\documentclass[rmp,twocolumn,unsortedaddress,showpacs,floatfix]{revtex4}
\usepackage{graphicx}
\usepackage{amsmath}
\usepackage{amssymb}
\usepackage[linktocpage]{hyperref}

\graphicspath{{.}}

\DeclareMathOperator{\re}{Re}
\DeclareMathOperator{\im}{Im}
\DeclareMathOperator{\trace}{Tr}
\DeclareMathOperator{\pv}{\mathcal{P}}
\DeclareMathOperator{\I}{i}
\DeclareMathOperator{\E}{e}

\newcommand{\llangle}{\ensuremath{\big\langle\!\!\big\langle}}
\newcommand{\rrangle}{\ensuremath{\big\rangle\!\!\big\rangle}}
\newcommand{\vopo}{(VO)$_2$P$_2$O$_7$}
\newcommand{\conline}{(Color in online edition) }

\begin{document}

\title{The Kernel Polynomial Method}

\author{Alexander Wei{\ss}e} 
\affiliation{School of Physics, The
  University of New South Wales, Sydney, NSW 2052, Australia}
\altaffiliation{New address: Institut f\"ur Physik,
  Ernst-Moritz-Arndt-Universit\"at Greifswald, 17487 Greifswald,
  Germany}

\author{Gerhard Wellein}
\affiliation{Regionales Rechenzentrum Erlangen, 
  Universit\"at Erlangen, 91058 Erlangen, Germany}

\author{Andreas Alvermann}

\author{Holger Fehske}
\affiliation{Institut f\"ur Physik, Ernst-Moritz-Arndt-Universit\"at 
  Greifswald, 17487 Greifswald, Germany}

\date{April 3, 2006}

\begin{abstract}
  Efficient and stable algorithms for the calculation of spectral
  quantities and correlation functions are some of the key tools in
  computational condensed matter physics. In this article we review
  basic properties and recent developments of Chebyshev expansion
  based algorithms and the Kernel Polynomial Method. Characterized by
  a resource consumption that scales linearly with the problem
  dimension these methods enjoyed growing popularity over the last
  decade and found broad application not only in physics.
  Representative examples from the fields of disordered systems,
  strongly correlated electrons, electron-phonon interaction, and
  quantum spin systems we discuss in detail. In addition, we
  illustrate how the Kernel Polynomial Method is successfully embedded
  into other numerical techniques, such as Cluster Perturbation Theory
  or Monte Carlo simulation.
\end{abstract}


\maketitle

\tableofcontents

\section{Introduction}\label{secintro}
In most areas of physics the fundamental interactions and the
equations of motion that govern the behavior of real systems on a
microscopic scale are very well known, but when it comes to solving
these equations they turn out to be exceedingly complicated. This
holds, in particular, if a large and realistic number of particles is
involved.  Inventing and developing suitable approximations and
analytical tools has therefore always been a cornerstone of
theoretical physics.  Recently, however, research continued to focus
on systems and materials, whose properties depend on the interplay of
many different degrees of freedom or on interactions that compete on
similar energy scales.  Analytical and approximate methods quite often
fail to describe the properties of such systems, so that the use of
numerical methods remains the only way to proceed.  On the other hand,
the available computer power increased tremendously over the last
decades, making direct simulations of the microscopic equations for
reasonable system sizes or particle numbers more and more feasible.
The success of such simulations, though, depends on the development
and improvement of efficient algorithms. Corresponding research
therefore plays an increasingly important role.

On a microscopic level the behavior of most physical systems, like
their thermodynamics or response to external probes, depends on the
distribution of the eigenvalues and the properties of the
eigenfunctions of a Hamilton operator or dynamical matrix. In
numerical approaches the latter correspond to Hermitian matrices of
finite dimension $D$, which can become huge already for a moderate
number of particles, lattice sites or grid points. The calculation of
all eigenvalues and eigenvectors then easily turns into an intractable
task, since for a $D$-dimensional matrix in general it requires memory
of the order of $D^2$, and the number of operations and the computation
time scale as $D^3$. Of course, this large resource consumption
severely restricts the size of the systems that can be studied by such
a ``naive'' approach.  For dense matrices the limit is currently of
the order of $D\approx 10^5$, and for sparse matrices the situation is
only slightly better.

Fortunately, alternatives are at hand: In the present article we
review basic properties and recent developments of numerical Chebyshev
expansion and of the Kernel Polynomial Method (KPM). As the most time
consuming step these iterative approaches require only multiplications
of the considered matrix with a small set of vectors, and therefore
allow for the calculation of spectral properties and dynamical
correlation functions with a resource consumption that scales
\emph{linearly} with~$D$ for sparse matrices, or like $D^2$ otherwise.
If the matrix is not stored but constructed on-the-fly dimensions of
the order of $D\approx 10^9$ or more are accessible.

The first step to achieve this favorable behavior is setting aside the
requirement for a complete and exact knowledge of the spectrum. A
natural approach, which has been considered from the early days of
quantum mechanics, is the characterization of the spectral density
$\rho(E)$ in terms of its moments $\mu_l = \int \rho(E) E^l dE$. By
iteration these moments can usually be calculated very efficiently,
but practical implementations in the context of Gaussian quadrature
showed that the reconstruction of $\rho(E)$ from ordinary power
moments is plagued by substantial numerical instabilities~\cite{Ga68}.
These occur mainly because the powers $E^l$ put too much weight to the
boundaries of the spectrum at the expense of poor precision for
intermediate energies.  The observation of this deficiency advanced
the development of modified moment approaches~\cite{Ga70,SD72}, where
$E^l$ is replaced by (preferably orthogonal) polynomials of $E$. With
studies of the spectral density of harmonic
solids~\cite{WB72,BW73,WPB74} and of autocorrelation
functions~\cite{Wh74}, which made use of Chebyshev polynomials of
second kind, these ideas soon found their way into physics
application. Later, similar Chebyshev expansion methods became popular
also in quantum chemistry, where the focus was on the time evolution
of quantum states~\cite{TK84,Ko88,MT97,CG99} and on Filter
Diagonalization~\cite{Ne90}.  The modified moment approach noticeably
improved when kernel polynomials were introduced to damp the Gibbs
oscillations, which for truncated polynomial series occur near
discontinuities of the expanded function~\cite{SR94,WZ94,Wa94,SRVK96}.
At this time also the name Kernel Polynomial Method was coined, and
applications then included high-resolution spectral densities, static
thermodynamic quantities as well as zero-temperature dynamical
correlations~\cite{SR94,WZ94,Wa94}.  Only recently this range was
extended to cover also dynamical correlation functions at
finite-temperature~\cite{We04}, and below we present some new
applications to complex-valued quantities, e.g. Green functions.
Being such a general tool for studying large matrix problems, KPM can
also be used as a core component of more involved numerical
techniques. As recent examples we discuss Monte Carlo~(MC) simulations
and Cluster Perturbation Theory~(CPT).

In parallel to Chebyshev expansion techniques and to KPM also the
Lanczos Recursion Method was
developed~\cite{HHK72,HHK75,LG82,BRP92,JP94,ADEL03}, which is based on
a recursive Lanczos tridiagonalization~\cite{La50} of the considered
matrix and the expression of the spectral density or of correlation
functions in terms of continued fractions. The approach, in general,
is applicable to the same problems as KPM and found wide application
in solid state physics~\cite{Pa78,Da94,Or98,JP00}.  It suffers,
however, from the shortcomings of the Lanczos algorithm, namely loss
of orthogonality and spurious degeneracies if extremal eigenstates
start to converge. We will compare the two methods in
Sec.~\ref{secalt} and explain, why we prefer to use Lanczos for the
calculation of extremal eigenstates and KPM for the calculation of
spectral properties and correlation functions. In addition, we will
comment on more specialized iterative schemes, such as projection
methods~\cite{GC94b,Go99,IE03} and Maximum Entropy
approaches~\cite{Sk88,SR97,BBBD05}.  Drawing more attention to KPM as
a potent alternative to all these techniques is one of the purposes of
the present work.

The outline of the article is as follows: In Sec.~\ref{seckpm} we
give a detailed introduction to Chebyshev expansion and the Kernel
Polynomial Method, its mathematical background, convergence properties
and practical aspects of its implementation. In Sec.~\ref{secapp}
we apply KPM to a variety of problems from solid state physics.
Thereby, we focus mainly on illustrating the types of quantities that
can be calculated with KPM, rather than on the physics of the
considered models. In Sec.~\ref{seccomp} we show how KPM can be
embedded into other numerical approaches that require knowledge of
spectral properties or correlation functions, namely Monte Carlo
simulation and Cluster Perturbation Theory. In Sec.~\ref{secalt} we
shortly discuss alternatives to KPM and compare their performance and
precision, before summarizing in Sec.~\ref{secsum}.

\section{Chebyshev expansion and the Kernel Polynomial Method (KPM)}\label{seckpm}
\subsection{Basic features of Chebyshev expansion}
\subsubsection{Chebyshev polynomials}
Let us first recall the basic properties of expansions in orthogonal
polynomials and of Chebyshev expansion in particular. Given a
positive weight function $w(x)$ defined on the interval $[a,b]$ we
can introduce a scalar product
\begin{equation}
  \langle f|g\rangle = 
  \int\limits_{a}^{b} w(x) f(x) g(x) \,dx
\end{equation}
between two integrable functions $f,g: [a,b]\rightarrow\mathbb{R}$.
With respect to each such scalar product there exists a complete set
of polynomials $p_n(x)$, which fulfil the orthogonality relations
\begin{equation}
  \langle p_n|p_m\rangle = \delta_{n,m} / h_n\,,
\end{equation}
where $h_n=1/\langle p_n|p_n\rangle$ denotes the inverse of the
squared norm of $p_n(x)$. These orthogonality relations allow for an
easy expansion of a given function $f(x)$ in terms of the $p_n(x)$,
since the expansion coefficients are proportional to the scalar
products of $f$ and $p_n$,
\begin{equation}
  f(x) = \sum_{n=0}^{\infty} \alpha_n\, p_n(x)
  \quad\text{with}\quad
  \alpha_n = \langle p_n|f\rangle\, h_n\,.
\end{equation}

In general, all types of orthogonal polynomials can be used for such
an expansion and for the Kernel Polynomial approach we discuss in this
article (see e.g.~\textcite{SR94}). However, as we frequently
observe whenever we work with polynomial expansions~\cite{Bo89},
Chebyshev polynomials~\cite{AS70,Ri90b} of first and second kind turn
out to be the best choice for most applications, mainly due to the
good convergence properties of the corresponding series and to the
close relation to Fourier transform~\cite{Ch66,Lo66}. The latter is
also an important prerequisite for the derivation of optimal kernels
(see Sec.~\ref{seckern}), which are required for the regularization
of finite-order expansions, and which so far have not been derived for
other sets of orthogonal polynomials.

Both sets of Chebyshev polynomials are defined on the interval $[a,b]
= [-1,1]$, where the weight function $w(x) = (\pi\sqrt{1-x^2})^{-1}$
yields the polynomials of first kind, $T_n$, and the weight function
$w(x) = \pi\sqrt{1-x^2}$ those of second kind, $U_n$. Based on the scalar
products 
\begin{align}
  \label{scalprod1}
  \langle f|g\rangle_1 & = 
  \int\limits_{-1}^{1} \frac{f(x)\, g(x)}{\pi\sqrt{1-x^2}}\,dx\,,\\
  \label{scalprod2}
  \langle f|g\rangle_2 & = 
  \int\limits_{-1}^{1} \pi\sqrt{1-x^2}\, f(x)\, g(x)\,dx\,,
\end{align}
the orthogonality relations thus read
\begin{align}
  \label{cheborth1}
  \langle T_n | T_m\rangle_1 & = 
  \tfrac{1 + \delta_{n,0}}{2}\ \delta_{n,m}\,, \\
  \label{cheborth2}
  \langle U_n | U_m\rangle_2 & = \tfrac{\pi^2}{2}\ \delta_{n,m}\,.
\end{align}
By substituting $x=\cos(\varphi)$ one can easily verify that they
correspond to the orthogonality relations of trigonometric functions,
and that in terms of those the Chebyshev polynomials can be expressed
in explicit form,
\begin{align}
  \label{chebdev1}
  T_n(x) & = \cos(n \arccos(x))\,,\\
  \label{chebdev2}
  U_n(x) & = \frac{\sin((n+1) \arccos(x))}{\sin(\arccos(x))}\,.
\end{align}
These expressions can then be used to prove the recursion relations,
\begin{equation}\label{chebrec1}
  \begin{gathered}
    T_0(x)=1\,,\quad T_{-1}(x) = T_1(x) = x\,,\\
    T_{m+1}(x)  = 2\, x\, T_m(x) - T_{m-1}(x)\,,
  \end{gathered}
\end{equation}
and
\begin{equation}\label{chebrec2}
  \begin{gathered}
    U_0(x)=1\,,\quad U_{-1}(x) = 0\,,\\
    U_{m+1}(x)  = 2\, x\, U_m(x) - U_{m-1}(x)\,,
  \end{gathered}
\end{equation}
which illustrate that Eqs.~\eqref{chebdev1} and~\eqref{chebdev2}
indeed describe polynomials, and which, moreover, are an integral part
of the iterative numerical scheme we develop later on. Two other
useful relations are
\begin{gather}
  \label{chebpro1}
  2\, T_m(x) T_n(x) = T_{m+n}(x) + T_{m-n}(x)\,,\\
  \label{chebpro2}
  2\, (x^2-1)\, U_{m-1}(x) U_{n-1}(x) = T_{m+n}(x) - T_{m-n}(x)\,.
\end{gather}
When calculating Green functions we also need Hilbert transforms
of the polynomials~\cite{AS70},
\begin{align}
  \label{chebpint1}
  \pv\int\limits_{-1}^{1} \frac{T_n(y)\,dy}{(y-x) \sqrt{1-y^2}} & = 
  \pi\,U_{n-1}(x)\,,\\
  \label{chebpint2}
  \pv\int\limits_{-1}^{1} \frac{\sqrt{1-y^2}\,U_{n-1}(y)\,dy}{(y-x)} & = 
  -\pi\,T_{n}(x)\,,
\end{align}
where $\pv$ denotes the principal value. Chebyshev polynomials have
many more interesting properties, for a detailed discussion we refer
the reader to text books such as~\cite{Ri90b}.

\subsubsection{Modified moments}
As sketched above, the standard way of expanding a function $f:
[-1,1]\rightarrow \mathbb{R}$ in terms of Chebyshev
polynomials of first kind is given by
\begin{equation}
  f(x) = \sum_{n=0}^{\infty} 
  \frac{\langle f|T_n\rangle_1}{\langle T_n|T_n\rangle_1}\ T_n(x)
  = \alpha_0 + 2 \sum_{n=1}^{\infty} \alpha_n \ T_n(x)
\end{equation}
with coefficients
\begin{equation}
  \alpha_n = \langle f|T_n\rangle_1 
  = \int\limits_{-1}^{1}\frac{f(x) T_n(x)}{\pi\sqrt{1-x^2}}\,dx\,.
\end{equation}
However, the calculation of these coefficients requires integrations
over the weight function $w(x)$, which in practical applications to
matrix problems prohibits a simple iterative scheme. The solution to
this problem follows from a slight rearrangement of the expansion,
namely
\begin{equation}\label{altser}
  f(x) = \frac{1}{\pi\sqrt{1-x^2}}\left[
      \mu_0 + 2 \sum_{n=1}^{\infty} \mu_n \ T_n(x)
    \right]
\end{equation}
with coefficients
\begin{equation}\label{defmom}
  \mu_n = \int\limits_{-1}^{1} f(x) T_n(x)\,dx\,.
\end{equation}
More formally this rearrangement of the Chebyshev series corresponds
to using the second scalar product $\langle.|.\rangle_2$ and expanding
in terms of the orthogonal functions
\begin{equation}
  \phi_n(x) = \frac{T_n(x)}{\pi\sqrt{1-x^2}}\,,
\end{equation}
which fulfil the orthogonality relations
\begin{equation}
  \langle\phi_n|\phi_m\rangle_2 = \tfrac{1+\delta_{n,0}}{2}\ \delta_{n,m}\,.
\end{equation}
The expansion in Eq.~\eqref{altser} is thus equivalent to
\begin{align}
  f(x) & = \sum_{n=0}^{\infty} 
  \frac{\langle f|\phi_n\rangle_2}{\langle \phi_n|\phi_n\rangle_2}
  \ \phi_n(x)\nonumber{}\\
  & = \frac{1}{\pi\sqrt{1-x^2}}\left[
    \mu_0 + 2 \sum_{n=1}^{\infty} \mu_n \ T_n(x)
  \right]\,\\
\intertext{with moments}
  \mu_n & = \langle f|\phi_n\rangle_2 
  = \int\limits_{-1}^{1} f(x) T_n(x)\,dx\,.
\end{align}

The $\mu_n$ now have the form of modified moments that we announced in
the introduction, and Eqs.~\eqref{altser} and~\eqref{defmom} represent
the elementary basis for the numerical method which we review in this
article. In the remaining sections we will explain how to translate
physical quantities into polynomial expansions of the form of
Eq.~\eqref{altser}, how to calculate the moments $\mu_n$ in practice,
and, most importantly, how to regularize expansions of finite order.

Naturally, the moments $\mu_n$ depend on the considered quantity
$f(x)$ and on the underlying model. We will specify these details when
discussing particular applications in Sec.~\ref{secapp}.
Nevertheless, there are features which are similar to all types of
applications, and we start with presenting these general aspects in
what follows.

\subsection{Calculation of moments}\label{secmom}
\subsubsection{General considerations}
A common feature of basically all Chebyshev expansions is the requirement
for a rescaling of the underlying matrix or Hamiltonian~$H$. As we
described above, the Chebyshev polynomials of both first and second
kind are defined on the real interval $[-1,1]$, whereas the quantities
we are interested in usually depend on the eigenvalues $\{E_k\}$ of
the considered (finite-dimensional) matrix.  To fit this spectrum into
the interval $[-1,1]$ we apply a simple linear transformation to the
Hamiltonian and all energy scales,
\begin{align}
  \tilde H &= (H - b) / a\,,\\
  \tilde E &= (E - b) / a\,,
\end{align}
and denote all rescaled quantities with a tilde hereafter.  Given the
extremal eigenvalues of the Hamiltonian, $E_\text{min}$ and
$E_\text{max}$, which can be calculated, e.g. with the Lanczos
algorithm~\cite{La50}, or for which bounds may be known analytically,
the scaling factors $a$ and $b$ read
\begin{align}
  a &= (E_\text{max} - E_\text{min}) / (2 - \epsilon)\,,\\
  b &= (E_\text{max} + E_\text{min}) / 2\,.
\end{align}
The parameter $\epsilon$ is a small cut-off introduced to avoid
stability problems that arise if the spectrum includes or exceeds the
boundaries of the interval $[-1,1]$. It can be fixed, e.g. to
$\epsilon=0.01$, or adapted to the resolution of the calculation,
which for an expansion of finite order~$N$ is proportional $1/N$ (see
below).

The next similarity of most Chebyshev expansions is the form of the
moments, namely their dependence on the matrix or Hamiltonian~$\tilde
H$. In general, we find two types of moments: Simple expectation
values of Chebyshev polynomials in $\tilde H$,
\begin{equation}
  \mu_n = \langle\beta| T_n(\tilde H) |\alpha\rangle\,,
\end{equation}
where $|\alpha\rangle$ and $|\beta\rangle$ are certain states of the
system, or traces over such polynomials and a given operator $A$,
\begin{equation}
  \mu_n = \trace[A\, T_n(\tilde H)]\,.
\end{equation}

Handling the first case is rather straightforward. Starting from the
state $|\alpha\rangle$ we can iteratively construct the states
$|\alpha_n\rangle=T_n(\tilde H)|\alpha\rangle$ by using the recursion
relations for the $T_n$, Eq.~\eqref{chebrec1},
\begin{align}
  \label{momreca}
  |\alpha_0\rangle &= |\alpha\rangle\,,\\
  |\alpha_1\rangle &= \tilde H |\alpha_0\rangle\,,\\
  |\alpha_{n+1}\rangle &= 2\tilde H |\alpha_n\rangle - |\alpha_{n-1}\rangle\,.
\end{align}
Scalar products with $|\beta\rangle$ then directly yield 
\begin{equation}
  \label{momrece}
  \mu_n = \langle\beta|\alpha_n\rangle\,.
\end{equation}
This iterative calculation of the moments, in particular the
application of $\tilde H$ to the state $|\alpha_n\rangle$, represents
the most time consuming part of the whole expansion approach and
determines its performance. If $\tilde H$ is a sparse matrix of
dimension~$D$ the matrix vector multiplication is an order~$O(D)$
process and the calculation of~$N$ moments therefore requires~$O(ND)$
operations and time. The memory consumption depends on the
implementation. For moderate problem dimension we can store the matrix
and, in addition, need memory for two vectors of dimension~$D$. For
very large~$D$ the matrix certainly does not fit into the memory and has
to be reconstructed on-the-fly in each iteration or retrieved from
disc.  The two vectors then determine the memory consumption of the
calculation. Overall, the resource consumption of the moment iteration
is similar or even slightly better than that of the Lanczos algorithm,
which requires a few more vector operations (see our comparison in
Sec.~\ref{secalt}). In contrast to Lanczos, Chebyshev iteration
is completely stable and can be carried out to arbitrary high order.

The moment iteration can be simplified even further, if $|\beta\rangle
= |\alpha\rangle$. In this case the product relation~\eqref{chebpro1}
allows for the calculation of two moments from each new
$|\alpha_n\rangle$,
\begin{align}
  \label{momdblg}
  \mu_{2n} & = 2\langle\alpha_n|\alpha_n\rangle - \mu_0\,,\\
  \label{momdblu}
  \mu_{2n+1} & = 2\langle\alpha_{n+1}|\alpha_n\rangle - \mu_1\,,
\end{align}
which is equivalent to two moments per matrix vector multiplication.
The numerical effort for $N$ moments is thus reduced by a factor of
two. In addition, like many other numerical approaches KPM benefits
considerably from the use of symmetries that reduce the Hilbert space
dimension.

\subsubsection{Stochastic evaluation of traces}\label{secstoch}
The second case where the moments depend on a trace over the whole
Hilbert space, at first glance, looks far more complicated. Based on
the previous considerations we would estimate the numerical effort to
be proportional to~$D^2$, because the iteration needs to be repeated
for all $D$ states of a given basis. It turns out, however, that
extremely good approximations of the moments can be obtained with a
much simpler approach: the stochastic evaluation of the
trace~\cite{Sk88,DS93,SR94}, i.e., an estimate of $\mu_n$ based on the
average over only a small number $R\ll D$ of randomly chosen
states~$|r\rangle$,
\begin{equation}\label{stomu}
  \mu_n = \trace[A\, T_n(\tilde H)] \approx \frac{1}{R} \sum_{r=0}^{R-1} 
  \langle r|A\, T_n(\tilde H)|r\rangle\,.
\end{equation}
The number of random states, $R$, does not scale with $D$. It can be
kept constant or even reduced with increasing $D$. To understand this,
let us consider the convergence properties of the above estimate.
Given an arbitrary basis $\{|i\rangle\}$ and a set of independent
identically distributed random variables~$\xi_{ri}\in\mathbb{C}$,
which in terms of the statistical average $\llangle\ldots\rrangle$
fulfil
\begin{align}
  \label{ecxi1}
  \llangle \xi_{ri}^{}\rrangle &= 0\,,\\
  \label{ecxi2}
  \llangle \xi_{ri}^{} \xi_{r'j}^{}\rrangle &= 0\,,\\
  \label{ecxi3}
  \llangle \xi_{ri}^{*}\xi_{r'j}^{}\rrangle &= \delta_{rr'}\delta_{ij}\,,
\end{align}
a random vector is defined through
\begin{equation}
  |r\rangle = \sum_{i=0}^{D-1} \xi_{ri}^{} |i\rangle\,.
\end{equation}
We can now calculate the statistical expectation value of the trace
estimate $\Theta = \frac{1}{R}\sum_{r=0}^{R-1}\langle r|B|r\rangle$
for some Hermitian operator $B$ with matrix elements $B_{ij} = \langle
i|B|j\rangle$, and indeed find,
\begin{align}
  \llangle \Theta \rrangle & =
  \llangle \frac{1}{R}\sum_{r=0}^{R-1}\langle r|B|r\rangle \rrangle  = 
  \frac{1}{R}\sum_{r=0}^{R-1}\sum_{i,j=0}^{D-1}
  \llangle\xi_{ri}^{*}\xi_{rj}^{}\rrangle B_{ij}\nonumber{}\\
  & = \sum_{i=0}^{D-1}B_{ii} = \trace(B)\,.
\end{align}
Of course, this only shows that we obtain the correct result on
average. To assess the associated error we also need to study the
fluctuation of $\Theta$, which is characterized by $(\delta\Theta)^2 =
\llangle \Theta^2\rrangle - \llangle\Theta\rrangle^2$. Evaluating
\begin{equation}
  \begin{aligned}
    \llangle \Theta^2\rrangle 
    &=\llangle\frac{1}{R^2}\sum_{r,r'=0}^{R-1}
    \langle r|B|r\rangle\langle r'|B|r'\rangle\rrangle\\
    &=\frac{1}{R^2}\sum_{r,r'=0}^{R-1}\sum_{i,j,i',j'=0}^{D-1}
    \llangle\xi_{ri}^{*}\xi_{rj}^{}\xi_{r'i'}^{*}\xi_{r'j'}^{}
    \rrangle B_{ij} B_{i'j'}\\
    &=\frac{1}{R^2}\Big(\sum_{\substack{r,r'=0 \\ r\ne r'}}^{R-1}
    \sum_{i,j,i',j'=0}^{D-1} \delta_{ij}\delta_{i'j'} B_{ij} B_{i'j'} \\
    & \quad + \sum_{r}^{R-1} \sum_{i,j,i',j'=0}^{D-1} 
    \llangle\xi_{ri}^{*}\xi_{rj}^{}\xi_{ri'}^{*}\xi_{rj'}^{}
    \rrangle B_{ij} B_{i'j'}\Big)\\
    & = \frac{R-1}{R} (\trace B)^2 
    + \frac{1}{R}\Big(\sum_{j=0}^{D-1}\llangle|\xi_{rj}^{}|^4
    \rrangle B_{jj}^2\\
    & \quad + \sum_{\substack{i,j=0\\i\ne j}}^{D-1} B_{ii} B_{jj} 
    + \sum_{\substack{i,j=0\\i\ne j}}^{D-1} B_{ij} B_{ji}\Big)\\
    &=(\trace B)^2 + \frac{1}{R} \Big(\trace(B^2) + 
    (\llangle|\xi_{ri}^{}|^4\rrangle -2)\sum_{j=0}^{D-1} B_{jj}^2\Big)
   \end{aligned}
\end{equation}
we get for the fluctuation
\begin{equation}\label{dtheta}
  (\delta\Theta)^2 = \frac{1}{R} \Big(\trace(B^2) + 
    (\llangle|\xi_{ri}^{}|^4\rrangle -2)\sum_{j=0}^{D-1} B_{jj}^2\Big)\,.
\end{equation}
The trace of $B^2$ will usually be of order $O(D)$, and the relative
error of the trace estimate, $\delta\Theta / \Theta$, is thus of order
$O(1/\sqrt{RD})$. It is this favorable behavior, which ensures the
convergence of the stochastic approach, and which was the basis for
our initial statement that the number of random states $R\ll D$ can be
kept small or even be reduced with the problem dimension~$D$.

Note also that the distribution of the elements of $|r\rangle$,
$p(\xi_{ri})$, has a slight influence on the precision of the
estimate, since it determines the expectation value
$\llangle|\xi_{ri}^{}|^4\rrangle$ that enters Eq.~\eqref{dtheta}. For
an optimal distribution $\llangle|\xi_{ri}^{}|^4\rrangle$ should be as
close as possible to its lower bound
$\llangle|\xi_{ri}^{}|^2\rrangle^2 = 1$, and indeed, we find this
result if we fix the amplitude of the $\xi_{ri}^{}$ and allow only for
a random phase $\phi\in[0,2\pi]$, $\xi_{ri}=\E^{i\phi}$. Moreover, if
we were working in the eigenbasis of $B$ this would cause
$\delta\Theta$ to vanish entirely, which led \textcite{IE04}
to conclude that random phase vectors are the optimal choice for
stochastic trace estimates. However, all these considerations depend
on the basis that we are working in, which in practice will never be
the eigenbasis of $B$ (in particular, if $B$ corresponds to something
like $A\, T_n(\tilde H)$, as in Eq.~\eqref{stomu}). A random phase
vector in one basis does not necessarily correspond to a random phase
vector in another basis, but the other basis may well lead to smaller
value of $\sum_{j=0}^{D-1} B_{jj}^2$, thus compensating for the larger
value of $\llangle|\xi_{ri}^{}|^4\rrangle$. Presumably, the most
natural choice are Gaussian distributed $\xi_{ri}$, which lead to
$\llangle|\xi_{ri}^{}|^4\rrangle = 2$ and thus a basis-independent
fluctuation $(\delta\Theta)^2$. To summarize this section, we think
that the actual choice of the distribution of $\xi_{ri}^{}$ is not of
high practical significance, as long as
Eqs.~\eqref{ecxi1}--\eqref{ecxi3} are fulfilled for
$\xi_{ri}^{}\in\mathbb{C}$, or
\begin{align}
  \label{erxi1}
  \llangle \xi_{ri}\rrangle &= 0\,,\\
  \label{erxi2}
  \llangle \xi_{ri}\xi_{r'j}\rrangle &= \delta_{rr'}\delta_{ij}\,,
\end{align}
hold for $\xi_{ri}^{}\in\mathbb{R}$. Typically, within this article we
will consider Gaussian~\cite{Sk88,SR94} or uniformly distributed
variables~$\xi_{ri}^{}\in\mathbb{R}$.

\subsection{Kernel polynomials and Gibbs oscillations}\label{seckern}

\subsubsection{Expansions of finite order \& simple kernels}
In the preceding sections we introduced the basic ideas underlying the
expansion of a function $f(x)$ in an \emph{infinite} series of
Chebyshev polynomials, and gave a few hints for the numerical
calculation of the expansion coefficients~$\mu_n$. As expected for a
numerical approach, however, the total number of these moments will
remain finite, and we thus arrive at a classical problem of
approximation theory.  Namely, we are looking for the best (uniform)
approximation to $f(x)$ by a polynomial of given maximal degree, which
in our case is equivalent to finding the best approximation to $f(x)$
given a \emph{finite} number $N$ of moments $\mu_n$. To our advantage,
such problems have been studied for at least 150 years and we can make
use of results by many renowned mathematicians, such as Chebyshev,
Weierstrass, Dirichlet, Fej\'er, Jackson, to name only a few. We will
also introduce the concept of kernels, which facilitates the study of
the convergence properties of the mapping $f(x)\to f_{\text{KPM}}(x)$
from the considered function $f(x)$ to our approximation
$f_{\text{KPM}}(x)$.

Experience shows that a simple truncation of an infinite series,
\begin{equation}\label{plain}
  f(x) \approx \frac{1}{\pi\sqrt{1-x^2}}\Big[
    \mu_0 + 2 \sum_{n=1}^{N-1} \mu_n\, T_n(x)
  \Big]\,,
\end{equation}
leads to poor precision and fluctuations --- also known as Gibbs
oscillations --- near points where the function $f(x)$ is not
continuously differentiable. The situation is even worse for
discontinuities or singularities of $f(x)$, as we illustrate below in
Figure~\ref{figkernels}. A common procedure to damp these oscillations
relies on an appropriate modification of the expansion coefficients,
$\mu_n\to g_n\mu_n$, which depends on the order of the
approximation~$N$,
\begin{equation}\label{fkpmdef}
  \begin{aligned}
    f_{\text{KPM}}(x) & = \sum_{n=0}^{N-1} 
    g_n \frac{\langle f|\phi_n\rangle_2}{\langle\phi_n|\phi_n\rangle_2}
    \ \phi_n(x)\\
    & = \frac{1}{\pi\sqrt{1-x^2}}\Big[
      g_0 \mu_0 + 2 \sum_{n=1}^{N-1} g_n \mu_n\, T_n(x)
    \Big].
  \end{aligned}
\end{equation}
In more abstract terms this truncation of the infinite series to
order~$N$ together with the corresponding modification of the
coefficients is equivalent to the convolution of $f(x)$ with a kernel
of the form
\begin{equation}\label{kerneldef}
  K_N(x,y) = g_0 \phi_0(x) \phi_0(y) 
  + 2\sum_{n=1}^{N-1} g_n \phi_n(x) \phi_n(y)\,,
\end{equation}
namely
\begin{equation}
  \begin{aligned}
    f_{\text{KPM}}(x) & = \int\limits_{-1}^{1} \pi\sqrt{1-y^2}\,K_N(x,y)\,f(y)\,dy\\
    & = \langle K_N(x,y)|f(y)\rangle_2\,.
  \end{aligned}
\end{equation}
The problem now translates into finding an optimal kernel $K_N(x,y)$,
i.e., coefficients $g_n$, where the notion of ``optimal'' partially
depends on the considered application.

The simplest kernel, which is usually attributed to Dirichlet, is
obtained by setting $g_n^D=1$ and evaluating the sum with the help of the 
Christoffel-Darboux identity~\cite{AS70},
\begin{equation}\label{dirichlet}
  \begin{aligned}
    K_N^D(x,y) & = \phi_0(x) \phi_0(y) 
    + 2\sum_{n=1}^{N-1} \phi_n(x) \phi_n(y)\\
    &=\frac{\phi_{N}(x)\phi_{N-1}(y)-\phi_{N-1}(x)\phi_{N}(y)}{x-y}\,.
  \end{aligned}
\end{equation}
Obviously, convolution of $K_N^D$ with an integrable function $f$
yields the above truncated series, Eq.~\eqref{plain}, which for
$N\to\infty$ converges to $f$ within the integral norm defined by the
scalar product Eq.~\eqref{scalprod2}, $||f||_2 = \sqrt{\langle
  f|f\rangle_2}$, i.e. we have
\begin{equation}
  ||f-f_{\text{KPM}}||_2 \xrightarrow{N\to\infty} 0\,.
\end{equation}
This is, of course, not particularly restrictive and leads to the 
disadvantages we mentioned earlier.

\subsubsection{Fej\'er kernel}

A first improvement is due to \textcite{Fe04} who showed that for
continuous functions an approximation based on the kernel
\begin{equation}\label{fejer}
  K_N^F(x,y) = \frac{1}{N} \sum_{\nu=1}^{N} K_\nu^D(x,y)\,,
  \quad\text{i.e.,}\quad g_n^F = 1 - \frac{n}{N}\,,
\end{equation}
converges uniformly in any restricted interval $[-1+\epsilon,1-\epsilon]$.
This means that now the absolute difference
between the function $f$ and the approximation $f_{\text{KPM}}$ goes
to zero,
\begin{equation}
  ||f-f_{\text{KPM}}||^\epsilon_\infty = \max_{-1+\epsilon<x<1-\epsilon} |f(x)-f_{\text{KPM}}(x)| \xrightarrow{N\to\infty} 0\,.
\end{equation}
Owing to the denominator in the expansion~\eqref{plain}
convergence is not uniform in the vicinity of the endpoints $x=\pm 1$,
which we accounted for by the choice of a small $\epsilon$ in the
rescaling of the Hamiltonian $H \to \tilde{H}$.

The more favorable uniform convergence is obtained under very general
conditions. Specifically, it suffices to demand that:
\begin{enumerate}
\item\label{kcond1} The kernel is positive: $K_N(x,y)>0\ \forall
  x,y\in [-1,1]$.
\item\label{kcond2} The kernel is normalized, $\int_{-1}^1 K(x,y)\, dx
  =\phi_0(y)$, which is equivalent to $g_0=1$.
\item\label{kcond3} The second coefficient $g_1$ approaches $1$ as
  $N\to \infty$.
\end{enumerate}
Then, as a corollary to Korovkin's theorem~\cite{Ko59}, an
approximation based on $K_N(x,y)$ converges uniformly in the sense
explicated for the Fej\'er kernel. The coefficients $g_n$, $n \ge 2$
are restricted only through the positivity of the kernel, the latter
one being equivalent to monotonicity of the mapping $f\to
f_{\text{KPM}}$, i.e.  $f \ge f' \Rightarrow \ f_{\text{KPM}} \ge
f'_{\text{KPM}}$. Note also that the conditions~\ref{kcond1}
and~\ref{kcond2} are very useful for practical applications: The first
ensures that approximations of positive quantities become positive,
the second conserves the integral of the expanded function, 
\begin{equation}
  \int\limits_{-1}^{1} f_{\text{KPM}}(x)\,dx 
    = \int\limits_{-1}^{1} f(x) \, dx\,.
\end{equation}
Applying the kernel, for example, to a density of states thus yields
an approximation which is strictly positive and normalized.

For a proof of the above theorem we refer the reader to the
literature~\cite{Ch66,Lo66}. Let us here only check that the Fej\'er
kernel indeed fulfils the conditions~\ref{kcond1} to~\ref{kcond3}: The
last two are obvious by inspection of Eq.~\eqref{fejer}. To prove the
positivity we start from the positive $2\pi$-periodic function
\begin{equation}
  p(\varphi) = \left|\sum_{\nu=0}^{N-1} a_{\nu} \E^{\I\nu\varphi}\right|^2
\end{equation} 
with arbitrary $a_\nu\in\mathbb{R}$. Straight-forward calculation
then shows
\begin{equation}
  \begin{split}
    p(\varphi) 
    & = \sum_{\nu,\mu=0}^{N-1} a_{\nu}a_{\mu} \E^{\I(\nu-\mu)\varphi}
    = \sum_{\nu,\mu=0}^{N-1} a_{\nu}a_{\mu} \cos(\nu-\mu)\varphi\\
    & = \sum_{\nu=0}^{N-1} a_{\nu}^2 
    + 2 \sum_{n=1}^{N-1}\sum_{\nu=0}^{N-1-n}a_{\nu}a_{\nu+n}\cos n\varphi\,.
  \end{split}
\end{equation} 
Hence, with
\begin{equation}\label{gofa}
  g_n = \sum_{\nu=0}^{N-1-n} a_{\nu}a_{\nu+n}
\end{equation}
the function 
\begin{equation}
  p(\varphi) = g_0 + 2\sum_{n=1}^{N-1} g_n \cos n\varphi
\end{equation}
is positive and periodic in $\varphi$. However, if $p(\varphi)$ is
positive, then the expression $\frac{1}{2}[p(\arccos x + \arccos y) +
p(\arccos x - \arccos y)]$ is positive $\forall\ x,y\in [-1,1]$.
Using Eq.~\eqref{chebdev1} and $\cos\alpha\cos\beta =
\frac{1}{2}[\cos(\alpha+\beta)+\cos(\alpha-\beta)]$, we immediately
observe that the general kernel $K_N(x,y)$ from Eq.~\eqref{kerneldef}
is positive $\forall\,x,y\in [-1,1]$, if the coefficients $g_n$
depend on arbitrary coefficients $a_\nu\in\mathbb{R}$ via
Eq.~\eqref{gofa}. Setting $a_\nu = 1/\sqrt{N}$ yields the Fej\'er
kernel $K_N^F(x,y)$, thus immediately proving its positivity.

In terms of its analytical properties and of the convergence in the
limit $N\to\infty$ the Fej\'er kernel is a major improvement over the
Dirichlet kernel. However, as yet we did not quantify the actual error
of an order-$N$ approximation: For continuous functions an appropriate 
scale is given by the modulus of continuity,
\begin{equation}
  w_f(\Delta) = \max_{|x-y|\le\Delta} |f(x)-f(y)|\,,
\end{equation}
in terms of which the Fej\'er approximation fulfils
\begin{equation}
  ||f-f_{\text{KPM}}||_\infty \sim w_f(1/\sqrt{N})\,.
\end{equation}
For sufficiently smooth functions this is equivalent to an error of
order $O(1/\sqrt{N})$. The latter is also an estimate for the
resolution or broadening that we will observe when expanding less
regular functions containing discontinuities or singularities, like
the examples in Figure~\ref{figkernels}.

\subsubsection{Jackson kernel}
With the coefficients $g_n^F$ of the Fej\'er kernel we have not fully
exhausted the freedom offered by the coefficients $a_\nu$ and
Eq.~\eqref{gofa}. We can hope to further improve the kernel by
optimizing the $a_\nu$ in some sense, which will lead us to recover
old results by~\textcite{Ja11,Ja12}.

In particular, let us tighten the third of the previously defined
conditions for uniform convergence by demanding that the kernel has
optimal resolution in the sense that
\begin{equation}\label{reqopt}
  Q := \int\limits_{-1}^{1}\int\limits_{-1}^{1} (x-y)^2 K_N(x,y) \, dx\, dy
\end{equation}
is minimal. Since $K_N(x,y)$ will be peaked at $x = y$, $Q$ is
basically the squared width of this peak. For sufficiently smooth
functions this more stringent condition will minimize the error
$||f-f_{\text{KPM}}||_\infty$, and in all other cases lead to optimal
resolution and smallest broadening of ``sharp'' features.

To  express the variance $Q$ of the kernel in terms of $g_n$ and
$a_\nu$, respectively, note that
\begin{multline}\label{xminusy}
  (x-y)^2 = (T_1(x) - T_1(y))^2\\
  = \tfrac{1}{2}(T_2(x)+T_0(x)) T_0(y) - 2 T_1(x)T_1(y)\\
  + \tfrac{1}{2}T_0(x)(T_2(y)+T_0(y))\,.
\end{multline}
Using the orthogonality of the Chebyshev polynomials and inserting
Eqs.~\eqref{kerneldef} and~\eqref{xminusy} into~\eqref{reqopt}, we
can thus rephrase the condition of optimal resolution as
\begin{equation}
  Q = g_0 - g_1 \stackrel{!}{=} \text{minimal w.r.t. } a_\nu\,.
\end{equation}
Hence, compared to the previous section, where we merely required
$g_0=1$ and $g_1\to 1$ for $N\to\infty$, our new condition tries to
optimize the rate at which $g_1$ approaches unity.

Minimizing $Q = g_0 - g_1$ under the constraint $C = g_0 - 1 = 0$
yields the condition
\begin{equation}
  \frac{\partial Q}{\partial a_\nu} 
  = \lambda \frac{\partial C}{\partial a_\nu}\,,
\end{equation}
where $\lambda$ is a Lagrange multiplier. Using Eq.~\eqref{gofa}
and setting $a_{-1} = a_{N} = 0$ we arrive at
\begin{equation}
  2a_\nu - a_{\nu-1} - a_{\nu+1} = \lambda a_\nu
\end{equation}
which the alert reader recognizes as the eigenvalue problem of a
harmonic chain with fixed boundary conditions. Its solution is given by
\begin{equation}
  \begin{aligned}
    a_\nu & = \bar{a} \sin\frac{\pi k (\nu+1)}{N+1}\,,\\
    \lambda & = 1 - \cos\frac{\pi k}{N+1}\,,
  \end{aligned}
\end{equation}
where $\nu = 0,\ldots,(N-1)$ and $k=1, 2, \ldots, N$. Given $a_\nu$
and the abbreviation $q=\pi k / (N+1)$ we can easily calculate
the $g_n$:
\begin{equation}
  \begin{split}
    g_n & = \sum_{\nu=0}^{N-1-n} a_\nu a_{\nu+n} = 
    \bar{a}^2 \sum_{\nu=1}^{N-n} \sin q\nu \sin q(\nu+n)\\
    & = \frac{\bar{a}^2}{2} \sum_{\nu=1}^{N-n}[\cos qn-\cos q(2\nu+n)]\\
    & = \frac{\bar{a}^2}{2} \left[(N-n) \cos qn 
      - \re \sum_{\nu=1}^{N-n} \E^{\I q (2\nu+n)} \right]\\
    & = \frac{\bar{a}^2}{2} \left[(N-n+1) \cos qn + \sin qn \cot q\right]\,.
  \end{split}
\end{equation}

The normalization $g_0=1$ is ensured through $\bar{a}^2 = 2/(N+1)$,
and with $g_1 = \cos q$ we can directly read off the optimal value
for
\begin{equation}
  Q = g_0 - g_1 = 1-\cos\frac{\pi k}{N+1}\,,
\end{equation}
which is obtained for $k=1$,
\begin{equation}\label{qmin}
  Q_{\text{min}} = 1 - \cos\frac{\pi}{N+1} 
  \simeq \frac{1}{2}\left(\frac{\pi}{N}\right)^2\,.
\end{equation}
The latter result shows that for large $N$ the resolution $\sqrt{Q}$
of the new kernel is proportional to $1/N$. Clearly, this is an
improvement over the Fej\'er kernel $K_N^F(x,y)$ which gives only
$\sqrt{Q}= 1/\sqrt{N}$.

With the above calculation we reproduced results
by~\textcite{Ja11,Ja12}, who showed that with a similar kernel a
continuous function $f$ can be approximated by a polynomial of degree
$N-1$ such that
\begin{equation}
  ||f-f_{\text{KPM}}||_\infty \sim w_f(1/N)\,,
\end{equation}
which we may interpret as an error of the order of $O(1/N)$.
Hereafter we are thus referring to the new optimal kernel as the
Jackson kernel $K_N^J(x,y)$, with
\begin{equation}\label{jackson}
  g_n^J  = \frac{(N-n+1)\cos\frac{\pi n}{N+1} 
    + \sin\frac{\pi n}{N+1}\cot\frac{\pi}{N+1}}{N+1}\,.
\end{equation}

\begin{figure}
  \includegraphics[width=\linewidth]{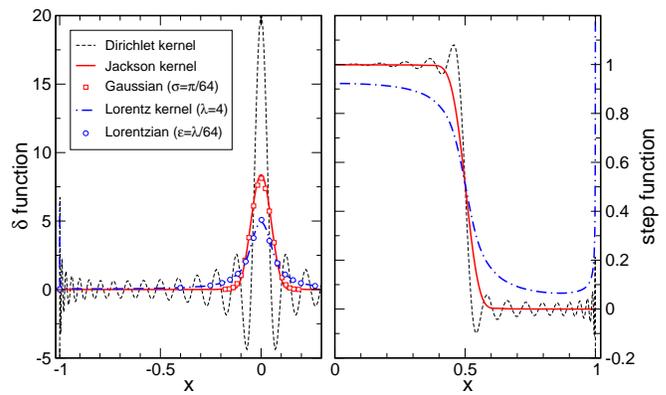}
  \caption{\conline Order $N=64$ expansions of $\delta(x)$ (left) and a 
    step function (right) based on different kernels. Whereas the
    truncated series (Dirichlet kernel) strongly oscillate, the
    Jackson results smoothly converge to the expanded functions. The
    Lorentz kernel leads to relatively poor convergence at the
    boundaries $x=\pm 1$, but otherwise yields perfect
    Lorentz-broadened approximations.}\label{figkernels}
\end{figure}

Before proceeding with other kernels let us add a few more details
on the resolution of the Jackson kernel: The quantity
$\sqrt{Q_{\text{min}}}$ obtained in Eq.~\eqref{qmin} is mainly a
measure for the spread of the kernel $K_N^J(x,y)$ in the
$x$-$y$-plane.  However, for practical calculations, which may also
involve singular functions, it is often reasonable to ask for the
broadening of a $\delta$-function under convolution with the kernel,
\begin{multline}
  \delta_{\text{KPM}}(x-a) = \langle K_N(x,y)|\delta(y-a)\rangle_2\\
  = g_0 \phi_0(x) T_0(a) + 2\sum_{n=1}^{N-1} g_n \phi_n(x) T_n(a)\,.
\end{multline}
It can be characterized by the variance $\sigma^2 = \llangle x^2
\rrangle - \llangle x\rrangle^2$, where we use $x = T_1(x)$ and $x^2 =
[T_2(x) + T_0(x)]/2$ to find
\begin{align}
  \llangle x\rrangle &= 
  \int\limits_{-1}^{1} x\, \delta_{\text{KPM}}(x-a)\, dx  = g_1 T_1(a)\,,\\
  \llangle x^2 \rrangle &= 
  \int\limits_{-1}^{1} x^2 \, \delta_{\text{KPM}}(x-a)\, dx
  = \frac{g_0 T_0(a) + g_2 T_2(a)}{2}\,.
\end{align}
Hence, for $K_N^J(x,y)$ the squared width of
$\delta_{\text{KPM}}(x-a)$ is given by
\begin{equation}
  \begin{split}
    \sigma^2 & = \llangle x^2 \rrangle - \llangle x\rrangle^2 = 
    a^2 (g_2^J - (g_1^J)^2) + (g_0^J - g_2^J)/2\\
    & = \frac{N - a^2(N-1)}{2 (N+1)}\left(1-\cos\frac{2\pi}{N+1}\right)\\
    & \simeq \left(\frac{\pi}{N}\right)^2 
    \left[1-a^2+\frac{3a^2-2}{N}\right] \,.
  \end{split}
\end{equation}
Using the Jackson kernel, an order~$N$ expansion of a
$\delta$-function at $x=0$ thus results in a broadened peak of width
$\sigma = \frac{\pi}{N}$, whereas close to the boundaries, $a=\pm 1$,
we find $\sigma = \frac{\pi}{N^{3/2}}$. It turns out that this peak is
a good approximation to a Gaussian,
\begin{equation}
  \delta_{\text{KPM}}^J(x) \approx \frac{1}{\sqrt{2\pi\sigma^2}}
  \exp\big(-\frac{x^2}{2\sigma^2}\big)\,,
\end{equation}
which we illustrate in Figure~\ref{figkernels}.

\subsubsection{Lorentz kernel}\label{seclorentz}
The Jackson kernel derived in the preceding sections is the best
choice for most of the applications we discuss below. In some
situations, however, special analytical properties of the expanded
functions become important, which only other kernels can account for.
The Green functions that appear in the Cluster Perturbation Theory,
Sec.~\ref{seccpt}, are an example.  Considering the imaginary part of
the Plemelj-Dirac formula which frequently occurs in connexion with
Green functions,
\begin{equation}
  \lim_{\epsilon\to 0}\frac{1}{x + \I\epsilon} 
  = \pv\Big(\frac{1}{x}\Big) - \I\pi\delta(x)\,,
\end{equation}
the $\delta$-function on the right hand side is approached in terms of
a Lorentz curve,
\begin{equation}
  \delta(x) = -\frac{1}{\pi}\lim_{\epsilon\to 0}\im\frac{1}{x + \I\epsilon} = 
  \lim_{\epsilon\to 0}\frac{\epsilon}{\pi(x^2+\epsilon^2)}\,,
\end{equation}
which has a different and broader shape compared to the approximations
of $\delta(x)$ we get with the Jackson kernel. There are attempts to
approximate Lorentzian like behavior in the framework of filter
diagonalization~\cite{VKH04}, but these solutions do not lead to a
positive kernel.  Note that positivity of the kernel is essential to
guarantee basic properties of Green functions, e.g.  that poles are
located in the lower (upper) half complex plane for a retarded
(advanced) Green function.  Since we know that the Fourier transform
of a Lorentz peak is given by $\exp(-\epsilon |k|)$, we can try to
construct an appropriate positive kernel assuming
$a_\nu=\E^{-\lambda\nu/N}$ in Eq.~\eqref{gofa}, and indeed, after
normalization, $g_0=1$, this yields what we call the Lorentz kernel
$K_N^L(x,y)$ hereafter,
\begin{equation}\label{lorentz}
  g_n^L = \frac{\sinh[\lambda(1-n/N)]}{\sinh(\lambda)}\,.
\end{equation}
The variable $\lambda$ is a free parameter of the kernel which as a
compromise between good resolution and sufficient damping of the Gibbs
oscillations we empirically choose to be of the order of $3\ldots 5$.
It is related to the $\epsilon$-parameter of the Lorentz curve, i.e.
to its resolution, via $\epsilon = \lambda/N$. Note also, that in the
limit $\lambda\to 0$ we recover the Fej\'er kernel $K_N^F(x,y)$ with
$g_n^F = 1 - n/N$, suggesting that both kernels share many of their
properties.

\begin{table*}
  \begin{ruledtabular}
    \begin{tabular}{lcccp{0.3\linewidth}}
      Name & $g_n$ & Parameters & positive? & Remarks\\
      \hline
      Jackson & $\frac{1}{N+1}[(N-n+1)\cos\frac{\pi n}{N+1} 
      + \sin\frac{\pi n}{N+1}\cot\frac{\pi}{N+1}]$ & none & yes & 
      best for most applications\\
      Lorentz & $\sinh[\lambda(1-n/N)]/\sinh(\lambda)$ & 
      $\lambda\in\mathbb{R}$ & yes &
      best for Green functions\\
      Fej\'er & $1-n/N$ & none & yes &
      mainly of academic interest\\
      Lanczos & $\left(\frac{\sin(\pi n/N)}{\pi n/N}\right)^M $ & 
      $M\in\mathbb{N}$ & no & 
      $M=3$ closely matches the Jackson kernel, but not strictly 
      positive~\cite{La66}\\
      Wang and Zunger & $\exp[-(\alpha\tfrac{n}{N})^\beta]$ & 
      $\alpha,\beta\in\mathbb{R}$ & no & 
      found empirically, not optimal~\cite{WZ94,Wa94}\\
      Dirichlet & $1$ & none & no & 
      least favorable choice\\
    \end{tabular}
  \end{ruledtabular}
  \caption{Summary of different integral kernels that can be used 
    to improve the quality of an order~$N$ Chebyshev series. 
    The coefficients $g_n$ refer to Eq.~\eqref{fkpmdef} or~\eqref{kerneldef}, 
    respectively.}\label{tabkern}
\end{table*}
In Figure~\ref{figkernels} we compare truncated Chebyshev expansions
--- equivalent to using the Dirichlet kernel --- to the approximations
obtained with the Jackson and Lorentz kernels, which we will later use
almost exclusively. Clearly, both kernels yield much better
approximations to the expanded functions and, in particular, the 
oscillations have disappeared almost completely. The comparison with a
Gaussian or Lorentzian, respectively, illustrates the nature of the
broadening of a $\delta$-function under convolution with the kernels,
which later on will facilitate the interpretation of our numerical
results.  With Table~\ref{tabkern} we conclude this section on
kernels, and, for the sake of completeness, also list two other
kernels that are occasionally used in the literature. Both have
certain disadvantages, in particular, they are not strictly positive.

\subsection{Implementational details and remarks}\label{secimp}
\subsubsection{Discrete cosine \& Fourier transforms}
Having discussed the theory behind Chebyshev expansion, the
calculation of moments, and the various kernel approximations, let us
now come to the practical issues of the implementation of KPM, namely
to the reconstruction of the expanded function~$f(x)$ from its
moments~$\mu_n$. Knowing a finite number $N$ of coefficients $\mu_n$
(see Sec.~\ref{secapp} for examples and details), we usually want
to reconstruct $f(x)$ on a finite set of
abscissas $x_k$. Naively we could sum up Eq.~\eqref{fkpmdef}
separately for each point, thereby making use of the recursion
relations for $T_n$, i.e.,
\begin{equation}
  f(x_k) = \frac{1}{\pi\sqrt{1-x_k^2}}\left[
    g_0 \mu_0 + 2 \sum_{n=1}^{N-1} g_n \mu_n\, T_n(x_k)
  \right].
\end{equation}
For a set $\{x_k\}$ containing $\tilde N$ points these summations
would require of the order of $N\tilde N$ operations.  We can do much
better, however, remembering the definition of the Chebyshev
polynomials $T_n$, Eq.~\eqref{chebdev1}, and the close relation
between KPM and Fourier expansion: First, we may introduce the
short-hand notation
\begin{equation}
  \tilde \mu_n = \mu_n g_n
\end{equation}
for the kernel improved moments.  Second and more important, we make a
special choice for our data points,
\begin{equation}
  x_k = \cos\frac{\pi (k+1/2)}{\tilde N}\quad\text{with}\quad 
  k=0,\ldots,(\tilde N-1)\,,
\end{equation}
which coincides with the abscissas of Chebyshev numerical
integration~\cite{AS70}. The number $\tilde N$ of points in the set
$\{x_k\}$ is not necessarily the same as the number of moments $N$.
Usually we will consider $\tilde N\ge N$ and a reasonable choice is,
e.g. $\tilde N = 2N$. All values $f(x_k)$ can now be obtained
through a discrete cosine transform,
\begin{equation}\label{defgamma}
  \begin{split}
    \gamma_k & = \pi \sqrt{1-x_k^2}\ f(x_k) \\
    & = \tilde\mu_0 + 2 \sum_{n=1}^{N-1} \tilde\mu_n 
    \cos\Big(\frac{\pi n (k+1/2)}{\tilde N}\Big)
  \end{split}
\end{equation}
which allows for the use of divide-and-conquer type algorithms that
require only $\tilde N\log \tilde N$ operations --- a clear advantage
over the above estimate $N\tilde N$.

Routines for fast discrete cosine transform are implemented in many
mathematical libraries or Fast Fourier Transform (FFT) packages, for
instance, in FFTW~\cite{FFTW05,FFTW} that ships with most Linux
distributions. If no direct implementation is at hand we may also use
fast discrete Fourier transform. With
\begin{equation}
  \lambda_n = 
  \begin{cases} 
    (2-\delta_{n,0})\,\tilde\mu_n\,\exp\left(\frac{\I\pi n}{2\tilde N}\right) &
    0<n<N\\
    0 & \text{otherwise}
  \end{cases}
\end{equation}
and the standard definition of discrete Fourier transform,
\begin{equation}
  \tilde\lambda_k = \sum_{n=0}^{\tilde N-1} 
  \lambda_n \exp\left(\frac{2 \pi \I n k}{\tilde N}\right)\,,
\end{equation}
after some reordering we find for an even number of data points
\begin{align}
  \gamma_{2j} & =  \re(\tilde\lambda_j)\,,\\
  \gamma_{2j+1} & =  \re(\tilde\lambda_{\tilde N-1-j})\,,
\end{align}
with $j = 0,\ldots,\tilde N/2-1$. If we need only a discrete cosine
transform this setup is not optimal, as it makes no use of the
imaginary part which the complex FFT calculates. It turns out,
however, that the ``wasted'' imaginary part is exactly what we need when
we later calculate Green functions and other complex quantities, i.e.,
we can use the setup
\begin{align}
  \label{cmplxkpm1}
  \gamma_{2j} & =  \tilde\lambda_j\,,\\
  \label{cmplxkpm2}
  \gamma_{2j+1} & = \tilde\lambda_{\tilde N-1-j}^{*}\,,
\end{align}
to evaluate Eq.~\eqref{eqcorr}.

\subsubsection{Integrals involving expanded functions}
We have already mentioned that our particular choice of $x_k$
corresponds to the abscissas of numerical Chebyshev integration.
Hence, Gauss-type numerical approximations~\cite{PFTV86} to integrals
of the form $\int_{-1}^{1} f(x) g(x) dx$ become simple sums,
\begin{multline}
  \int\limits_{-1}^{1} f(x) g(x)\, dx  = 
  \int\limits_{-1}^{1} \frac{\sqrt{1-x^2} f(x) g(x)}{\sqrt{1-x^2}}\, dx\\
  \simeq \frac{\pi}{\tilde N}\sum_{k=0}^{\tilde N-1} 
  \sqrt{1-x_k^2}\ f(x_k) g(x_k)
  = \frac{1}{\tilde N}\sum_{k=0}^{\tilde N-1} \gamma_k g(x_k)\,,
\end{multline}
where $\gamma_k$ denotes the raw output of the cosine or Fourier
transforms defined in Eq.~\eqref{defgamma}. We can use this feature,
for instance, to calculate partition functions, where $f(x)$
corresponds to the expansion of the spectral density $\rho(E)$ and
$g(x)$ to the Boltzmann or Fermi weight.

\subsection{Generalization to higher dimension}\label{secddim}
\subsubsection{Expansion of multivariate functions}
For the calculation of finite-temperature dynamical correlation
functions we will later need expansions of functions of two variables.
Let us therefore comment on the generalization of the previous
considerations to $d$-dimensional space, which is easily obtained by
extending the scalar product $\langle.|.\rangle_2$ to functions $f,g:
[-1,1]^d\rightarrow \mathbb{R}$,
\begin{equation}
  \langle f|g\rangle_2 = 
  \int\limits_{-1}^{1}\cdots\int\limits_{-1}^{1} f(\vec{x}) g(\vec{x})
  \Big(\prod_{j=1}^{d}\pi\sqrt{1-x_j^2}\Big) \,dx_1\ldots dx_d\,.
\end{equation}
Here $x_j$ denote the $d$ components of the vector $\vec{x}$.
Naturally, this scalar product leads to the expansion
\begin{equation}
  \begin{aligned}
    f(\vec{x}) & = \sum_{\vec{n}=\vec{0}}^{\infty} 
    \frac{\langle f|\phi_{\vec{n}}\rangle_2}{
      \langle \phi_{\vec{n}}|\phi_{\vec{n}}\rangle_2
      }\ \phi_{\vec{n}}(\vec{x})\\
    & = \frac{\sum_{\vec{n}=\vec{0}}^{\infty} 
      \mu_{\vec{n}} h_{\vec{n}}\ \prod_{j=1}^{d}T_{n_j}(x_j)
      }{\prod_{j=1}^{d}\pi\sqrt{1-x_j^2}}\,,
  \end{aligned}
\end{equation}
where we introduced a vector notation for indices, $\vec{n} =
\{n_1,\ldots,n_d\}$, and the following functions and coefficients
\begin{align}
  \phi_{\vec{n}}(\vec{x}) & = \prod_{j=1}^{d} \phi_{n_j}(x_j)\,,\\
  \mu_{\vec{n}} & = \langle f|\phi_{\vec{n}}\rangle_2\nonumber{}\\
  & = \int\limits_{-1}^{1}\cdots\int\limits_{-1}^{1} f(\vec{x})
  \Big(\prod_{j=1}^{d} T_{n_j}(x_j)\Big)\,dx_1\ldots dx_d\,,\\
  \label{defmulti}
  h_{\vec{n}} & = \frac{1}{\langle
    \phi_{\vec{n}}|\phi_{\vec{n}}\rangle_2} = \prod_{j=1}^{d}
  \frac{2}{1+\delta_{n_j,0}}\,.
\end{align}

\subsubsection{Kernels for multidimensional expansions}
As in the one-dimensional case, a simple truncation of the infinite
series will lead to Gibbs oscillations and poor convergence.
Fortunately, we can easily generalize our previous results for kernel
approximations. In particular, we find that the extended kernel
\begin{equation}
  K_N(\vec{x},\vec{y}) = \prod_{j=1}^{d} K_N(x_j,y_j)
\end{equation}
maps an infinite series onto an truncated series,
\begin{equation}
  \begin{aligned}
    f_{\text{KPM}}(\vec{x}) &= 
    \langle K_N(\vec{x},\vec{y})|f(\vec{y})\rangle_2\\
    & = \frac{\sum_{\vec{n}=\vec{0}}^{N-1} 
      \mu_{\vec{n}} h_{\vec{n}}\ 
      \prod_{j=1}^{d} g_{n_j}T_{n_j}(x_j)}{\prod_{j=1}^{d}\pi\sqrt{1-x_j^2}}\,,
  \end{aligned}
\end{equation}
where we can take the $g_n$ of any of the previously discussed
kernels. If we use the $g_n^J$ of the Jackson kernel,
$K_N^J(\vec{x},\vec{y})$ fulfils generalizations of our conditions for
an optimal kernel, namely
\begin{enumerate}
\item $K_N^J(\vec{x},\vec{y})$ is positive $\forall\ 
  \vec{x},\vec{y}\in[-1,1]^d$.
\item $K_N^J(\vec{x},\vec{y})$ is normalized with  
  \begin{equation}
    \int\limits_{-1}^{1}\cdots\int\limits_{-1}^{1} 
    f_{\text{KPM}}(\vec{x}) \, dx_1\ldots dx_d 
    = \int\limits_{-1}^{1}\cdots\int\limits_{-1}^{1} 
    f(\vec{x}) \, dx_1\ldots dx_d\,.
  \end{equation}
\item $K_N^J(\vec{x},\vec{y})$ has optimal resolution in the sense that 
  \begin{equation}
    \begin{aligned}
      Q &=\int\limits_{-1}^{1}\cdots\int\limits_{-1}^{1}
      (\vec{x}-\vec{y})^2 K_N(\vec{x},\vec{y}) 
      \, dx_1\ldots dx_d \, dy_1\ldots dy_d\\
      & = d(g_0-g_1)
    \end{aligned}
  \end{equation}
  is minimal.
\end{enumerate}
Note that for simplicity the order of the expansion, $N$, was chosen
to be the same for all spatial directions. Of course, we could also
define more general kernels,
\begin{equation}
  K_{\vec{N}}(\vec{x},\vec{y}) = \prod_{j=1}^{d} K_{N_j}(x_j,y_j)\,,
\end{equation}
where the vector $\vec{N}$ denotes the orders of expansion for the
different spatial directions.

\subsubsection{Reconstruction with cosine transforms}
Similar to the 1D case we may consider the function $f:[-1,1]^d
\rightarrow\mathbb{R}$ on a discrete grid $\vec{x}_{\vec{k}}$ with
\begin{align}
  x_{\vec{k},j} & = \cos(\varphi_{k_j})\,,\\
  \varphi_{k_j} & = \frac{\pi (k_j + 1/2)}{\tilde N}\,,\\
  k_j & = 0,\ldots,(\tilde N-1)\,.
\end{align}
Note again that we could define individual numbers of points for each
spatial direction, i.e., a vector $\vec{\tilde{N}}$ with elements
$\tilde N_j$ instead of a single $\tilde N$.  For all grid points
$\vec{x}_{\vec{k}}$ the function $f(\vec{x}_{\vec{k}})$ is obtained
through multidimensional discrete cosine transform, i.e., with
coefficients $\kappa_{\vec{n}} = \tilde\mu_{\vec{n}} h_{\vec{n}} =
\mu_{\vec{n}} g_{\vec{n}} h_{\vec{n}}$ we find
\begin{equation}
  \begin{aligned}
    \gamma_{\vec{k}} & = 
    f({\cos(\varphi_{k_1}),\ldots,\cos(\varphi_{k_d})})
    \prod_{j=1}^{d} \pi \sin(\varphi_{k_j})\\
    & = \sum_{\vec{n} = \vec{0}}^{N-1} \kappa_{\vec{n}} \prod_{j=1}^{d} 
    \cos(n_j \varphi_{k_j})\\
    & = \sum_{n_1=0}^{N-1} \cos(n_1 \varphi_{k_1}) \ldots 
    \sum_{n_d=0}^{N-1} \cos(n_d \varphi_{k_d}) \kappa_{\vec{n}}\,.
  \end{aligned}
\end{equation}
The last line shows that the multidimensional discrete cosine
transform is equivalent to a nesting of one-dimensional transforms in
every coordinate. With fast implementations the computational effort
is thus proportional to $d \tilde N^{d-1} \tilde N \log \tilde N$,
which equals the expected value for $\tilde N^d$ data points, $\tilde
N^d \log \tilde N^d$.  If we are not using libraries like FFTW, which
provide ready-to-use multidimensional routines, we may also resort to
one-dimensional cosine transform or the above translation into FFT to
obtain high-performance implementations of general $d$-dimensional
transforms.

\section{Applications of KPM}\label{secapp}
Having described the mathematical background and many details of the
implementation of the Kernel Polynomial Method, we are now in
the position to present practical applications of the approach.
Already in the introduction we have mentioned that KPM can be used
whenever we are interested in the spectral properties of large
matrices or in correlation functions that can be expressed through the
eigenstates of such matrices.  Apparently, this leads to a vast range
of applications. In what follows, we try to cover all types of
accessible quantities and for each give at least one example. We
thereby focus on lattice models from solid state physics.

\subsection{Densities of states}\label{secdos}
\subsubsection{General considerations}
The first and basic application of Chebyshev expansion and KPM is the
calculation of the spectral density of Hermitian matrices, which could
correspond to the densities of states of both interacting or
non-interacting quantum models~\cite{Wh74,Sk88,SR94,SRVK96}. To be
specific, let us consider a $D$-dimensional matrix $M$ with
eigenvalues $E_k$, whose spectral density is defined as
\begin{equation}
  \rho(E) = \frac{1}{D}\sum_{k=0}^{D-1} \delta(E-E_k)\,.
\end{equation}
As described earlier, the expansion of $\rho(E)$ in terms of Chebyshev
polynomials requires a rescaling of $M\rightarrow \tilde M$, such that
the spectrum of $\tilde M = (M - b)/a$ lies within the interval
$[-1,1]$.  Given the eigenvalues $\tilde E_k$ of $\tilde M$ the
rescaled density $\tilde\rho(\tilde E)$ reads
\begin{equation}
  \tilde\rho(\tilde E) = \frac{1}{D}\sum_{k=0}^{D-1} 
  \delta(\tilde E-\tilde E_k)\,,
\end{equation}
and according to Eq.~\eqref{defmom} the expansion coefficients
become
\begin{equation}\label{mu4dos}
  \begin{aligned}
    \mu_n & = \int\limits_{-1}^{1} 
    \tilde\rho(\tilde E)\,T_n(\tilde E)\,d\tilde E
    = \frac{1}{D}\sum_{k=0}^{D-1} T_n(\tilde E_k)\\
    & = \frac{1}{D}\sum_{k=0}^{D-1} \langle k|T_n(\tilde M)|k\rangle
    = \frac{1}{D}\,\trace(T_n(\tilde M))\,.
  \end{aligned}
\end{equation}
This is exactly the trace form that we introduced in
Sec.~\ref{secmom}, and we can immediately calculate the $\mu_n$
using the stochastic techniques described in Sec.~\ref{secstoch}.
Knowing the moments we can use the techniques of Sec.~\ref{secimp}
to reconstruct $\tilde\rho(\tilde E)$ for the whole range $[-1,1]$,
and a final rescaling yields $\rho(E)$.

\subsubsection{Non-interacting systems: Anderson model of disorder}

Applied to a generalized model of non-interacting fermions
$c_i^{(\dagger)}$,
\begin{equation}
  H = \sum_{i,j=0}^{D-1} c_i^{\dagger} M_{ij}^{} c_j^{} \,,
\end{equation}
the matrix of interest $M$ is formed by the coupling constants
$M_{ij}$. Knowing the spectrum of $M$, i.e. the single-particle
density of states $\rho(E)$, all thermodynamic quantities of the model
can be calculated. For example, the particle density is given by
\begin{equation}
  n = \int \frac{\rho(E)}{1 + \E^{\beta(E-\mu)}}\, dE
\end{equation}
and the free energy per site reads
\begin{equation}
  f = n \mu - \frac{1}{\beta}\int\rho(E)\, \log(1 + \E^{-\beta(E-\mu)})\, dE\,,
\end{equation}
where $\mu$ is the chemical potential and $\beta = 1/T$ the inverse
temperature. 

\begin{figure}
  \includegraphics[width=\linewidth]{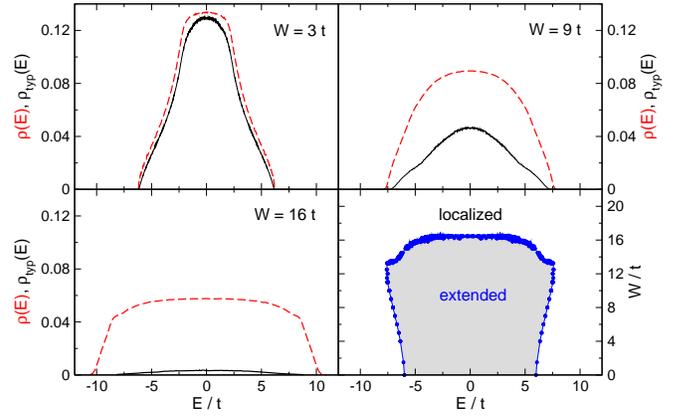}
  \caption{\conline Standard (dashed) and typical density of states
    (solid line), $\rho(E)$ and $\rho_{\text{typ}}(E)$ respectively,
    of the 3D Anderson model on a $50^3$ site cluster with periodic
    boundary conditions.  For $\rho(E)$ we calculated $N=2048$ moments
    with $R=10$ start vectors and $S=240$ realizations of disorder,
    for $\rho_{\text{typ}}(E)$ these numbers are $N=8192$, $R=32$ and
    $S=200$. The lower right panel shows the phase diagram of the
    model we obtained from $\rho_{\text{typ}}(E)/\rho(E)\to 0$
    (mobility edge).}\label{figanddos}
\end{figure}
As the first physical example let us consider the Anderson model of
non-interacting fermions moving in a random potential~\cite{An58},
\begin{equation}\label{hamand}
  H = -t\sum_{\langle ij\rangle} c_i^{\dagger}c_j^{} 
  + \sum_i \epsilon_i^{} c_i^{\dagger} c_i^{}\,.
\end{equation}
Here hopping occurs along nearest neighbor bonds $\langle ij\rangle$
on a simple cubic lattice and the local potential $\epsilon_i$ is
chosen randomly with uniform distribution in the interval
$[-W/2,W/2]$.  With increasing strength of disorder, $W$, the
single-particle eigenstates of the model tend to become localized in
the vicinity of a particular lattice site, which excludes these states
from contributing to electronic transport. Disorder can therefore
drive a transition from metallic behavior with delocalized fermions to
insulating behavior with localized fermions~\cite{Th74,LR85,KM93b}.
The disorder averaged density of states $\rho(E)$ of the model can be
obtained as described, but it contains no information about
localization. The KPM method, however, allows also for the calculation
of the local density of states,
\begin{equation}\label{defrhoi}
  \rho_i(E) = \frac{1}{D} \sum_{k=0}^{D-1} |\langle i|k\rangle|^2\, 
  \delta(E-E_k)\,,
\end{equation}
which is a measure for the contribution of a single lattice site
(denoted by the basis state $|i\rangle$) to the complete spectrum. For
delocalized states all sites contribute equally, whereas localized
states reside on just a few sites, or, equivalently, a certain site
contributes only to a few eigenstates. This property has a pronounced
effect on the distribution of $\rho_i(E)$, which at a fixed energy $E$
characterizes the variation of $\rho_i$ over different realizations of
disorder and sites $i$. For energies that correspond to localized
eigenstates the distribution is highly asymmetric and becomes singular
in the thermodynamic limit, whereas in the delocalized case the
distribution is regular and centered near its expectation value
$\rho(E)$.  Therefore a comparison of the geometric and the arithmetic
average of $\rho_i(E)$ over a set of realizations of disorder and over
lattice sites reveals the position of the Anderson
transition~\cite{DK97,DK98,SWWF05,SWF05}.  The expansion of $\rho_i(E)$ is
even simpler than the expansion of $\rho(E)$, since the moments have
the form of expectation values and do not involve a trace,
\begin{equation}
  \begin{aligned}
    \mu_n & = \int\limits_{-1}^{1} \tilde\rho_i(E)\, T_n(E)\, dE
    = \frac{1}{D}\sum_{k=0}^{D-1} |\langle i|k\rangle|^2 T_n(\tilde E_k)\\
    & = \frac{1}{D}\sum_{k=0}^{D-1}
    \langle i|T_n(\tilde M)|k\rangle\langle k|i\rangle
    = \frac{1}{D}\ \langle i|T_n(\tilde M)|i\rangle\,.
  \end{aligned}
\end{equation}
In Figure~\ref{figanddos} we show the standard density of states
$\rho(E)$, which coincides with the arithmetic mean of $\rho_i(E)$, in
comparison to the typical density of states $\rho_{\text{typ}}(E)$,
which is defined as the geometric mean of $\rho_i(E)$,
\begin{equation}
  \rho_{\text{typ}}(E) = \exp[\llangle\log(\rho_i(E))\rrangle]\,.
\end{equation}
With increasing disorder, starting from the boundaries of the
spectrum, $\rho_{\text{typ}}(E)$ is suppressed until it vanishes
completely for $W/t\gtrsim 16.5$, which is known as the critical
strength of disorder where the states in the band center become
localized~\cite{SO99}. The calculation yields the phase diagram shown
in the lower right corner of Figure~\ref{figanddos}, which compares
well to other numerical results.

Since the method requires storage only for the sparse Hamiltonian
matrix and for two vectors of the corresponding dimension, quite large
systems can be studied on standard desktop computers (of the order of
$100^3$ sites). The recursion is stable for arbitrarily high expansion
order. In the present case we calculated as many as $8192$ moments to
achieve maximum resolution in the local density of states.  The
standard density of states is usually far less demanding.

\subsubsection{Interacting systems: Double exchange}
Coming to interacting quantum systems, as a second example we study
the evolution of the quantum double-exchange model~\cite{AH55} for
large spin amplitude $S$, which in terms of spin-less fermions
$c_{i}^{(\dagger)}$ and Schwinger bosons $a_{i\sigma}^{(\dagger)}$
($\sigma=\uparrow,\downarrow$) is given by the Hamiltonian
\begin{equation}\label{hamqde}
    H = -\frac{t}{2S+1} \sum_{\langle ij\rangle,\sigma} a_{i\sigma}^{\dagger}
    a_{j\sigma}^{} c_{i}^{\dagger} c_{j}^{} 
\end{equation}
with the local constraint $\sum_{\sigma} a_{i\sigma}^{\dagger}
a_{i\sigma}^{} = 2S + c_{i}^{\dagger} c_{i}^{}$.  This model describes
itinerant electrons on a lattice whose spin is strongly coupled to
local spins of amplitude~$S$, so that the motion of the electrons
mediates an effective ferromagnetic interaction between these
localized spins. In the case of colossal magneto-resistant
manganites~\cite{CVM99}, for instance, cubic site symmetry leads to a
crystal field splitting of the manganese $d$-shell, and three
electrons in the resulting $t_{2g}$-shell form the local spins. The
remaining electrons occupy the $e_{g}$-shell and can become itinerant
upon doping, causing these materials to show ferromagnetic
order~\cite{Ze51b}. If the ferromagnetic (Hund's rule) coupling is
large, at each site only the high-spin states are relevant and we can
describe the total on-site spin in terms of Schwinger bosons
$a_{i\sigma}^{(\dagger)}$~\cite{Au94}.  From the electrons only the
charge degree of freedom remains, which is denoted by the spin-less
fermions $c_{i}^{(\dagger)}$ (see, e.g.~\cite{WLF01a} for more
details). The full quantum model, Eq.~\eqref{hamqde}, is rather complicated
for analytical or numerical studies, and we expect major
simplification by treating the spin background classically (remember
that $S$ is quite large for the systems of interest).  The limit of
classical spins, $S\rightarrow\infty$, is obtained by averaging
Eq.~\eqref{hamqde} over spin coherent states,
\begin{equation}
  |\Omega(S,\theta,\phi)\rangle 
  = \frac{\big(\cos(\frac{\theta}{2}) \E^{\I\phi/2} a_{\uparrow}^{\dagger} 
    + \sin(\frac{\theta}{2}) \E^{-\I\phi/2} 
    a_{\downarrow}^{\dagger}\big)^{2S}}{
    \sqrt{(2S)!}}\ |0\rangle\,,
\end{equation}
where $\theta$ and $\phi$ are the classical polar angles and
$|0\rangle$ the bosonic vacuum. The resulting non-interacting
Hamiltonian reads,
\begin{equation}\label{hamcde}
  H = -\sum_{\langle ij\rangle} 
  t_{ij}^{} c_{i}^{\dagger} c_{j}^{} + \textrm{H.c.}\,,
\end{equation}
with the matrix element~\cite{KA88}
\begin{multline}\label{tijcde}
  t_{ij}^{} = t \big[ \cos\tfrac{\theta_i}{2}
  \cos\tfrac{\theta_j}{2}\ \E^{-\I(\phi_i-\phi_j)/2}\\
  + \sin\tfrac{\theta_i}{2}
  \sin\tfrac{\theta_j}{2}\ \E^{\I(\phi_i-\phi_j)/2} \big] \,,
\end{multline}
i.e., spin-less fermions move in a background of random or ordered
classical spins which affect their hopping amplitude.

\begin{figure}
  \includegraphics[width=\linewidth]{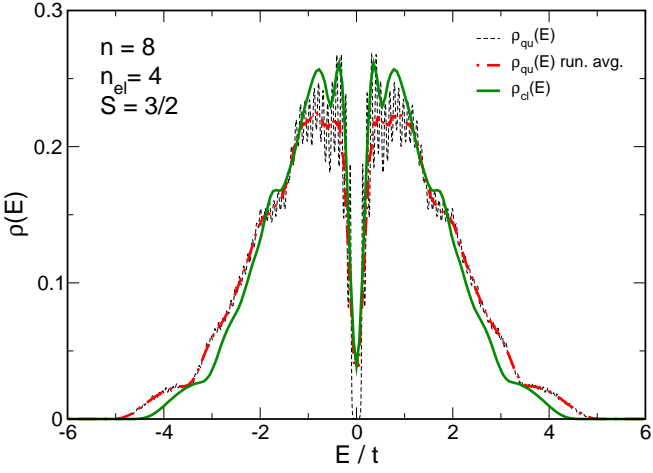}
  \caption{\conline Density of nonzero eigenvalues of the quantum 
    double-exchange model with $S=3/2$ (dashed line) and running
    average (red dot-dashed), calculated for 4 electrons on a 8-site
    ring, compared to the classical result $S\rightarrow\infty$ (green
    solid). Expansion parameters: $N=400$ moments and $R=100$ random
    vectors per $S^z$ sector.}\label{figdxdosn8}
\end{figure}
To assess the quality of this classical approximation we considered
four electrons moving on a ring of eight sites, and compared the
densities of states obtained for a background of $S=3/2$ quantum spins
and a background of classical spins. For the full quantum Hamiltonian,
Eq.~\eqref{hamqde}, the (canonical) density of states was calculated
on the basis of $400$ Chebyshev moments. To reduce the Hilbert space
dimension and to save resources we made use of the $SU(2)$
symmetry of the model: With the stochastic approach we calculated
separate moments $\mu_{n}^{S^z}$ for each $S^z$-sector,
\begin{equation}
  \mu_n^{S^z} = \trace^{S^z}[T_n(\tilde H)]\,,
\end{equation}
and used the dimensions $D^{S^z}$ of the sectors to obtain the total
normalized $\mu_n$ from the average
\begin{equation}
  \mu_n  = \frac{1}{D} \trace[T_n(\tilde H)]
  = \frac{\sum\limits_{S^z=-S^{\text{max}}}^{S^{\text{max}}}\mu_n^{S^z}}{
    \sum\limits_{S^z=-S^{\text{max}}}^{S^{\text{max}}} D^{S^z}}\,.
\end{equation}
Note, that such a setup can be used whenever the model under
consideration has certain symmetries.

On the other hand, we solved the effective non-interacting
model~\eqref{hamcde} and calculated the distributions of non-zero
energies for a background of fully disordered classical spins.  As
Figure~\ref{figdxdosn8} illustrates, the spectrum of the quantum model
with $S=3/2$ closely matches that of the system with classical spins,
providing good justification, e.g. for studies of colossal
magneto-resistive manganites that make use of a classical
approximation for the spin background. Since for the finite cluster
considered the spectrum of the quantum model is discrete, at the
present expansion order KPM starts to resolve distinct energy levels
(dashed line).  Therefore a running average (dot-dashed line) compares
better to the classical spin-averaged data (bold line).

\subsection{Static correlations at finite temperature}\label{secstatic}
Densities of states provide only the most basic information about a
given quantum system, and much more details can usually be learned
from the study of correlations and the response of the system to an
external probe or perturbation. Starting with static correlation
functions, let us now extend the application range of the expansion
techniques to such more involved quantities.

Given the eigenstates $|k\rangle$ of an interacting quantum system the
thermodynamic expectation value of an operator $A$ reads
\begin{align}
  \langle A\rangle & = \frac{1}{ZD}\trace(A \E^{-\beta H})
  = \frac{1}{ZD}\sum_{k=0}^{D-1} \langle k|A|k\rangle\, \E^{-\beta E_k}\,,\\
  Z & = \frac{1}{D}\trace(\E^{-\beta H}) 
  = \frac{1}{D}\sum_{k=0}^{D-1} \E^{-\beta E_k}\,,
\end{align}
where $H$ is the Hamiltonian of the system, $Z$ the partition
function, and $E_k$ the energy of the eigenstate $|k\rangle$. Using
the function
\begin{equation}
  a(E) = \frac{1}{D}\sum_{k=0}^{D-1} \langle k|A|k\rangle\, \delta(E - E_k)
\end{equation}
and the (canonical) density of states $\rho(E)$, we can express the
thermal expectation value in terms of integrals over the Boltzmann weight,
\begin{align}
  \label{thexpA}
  \langle A\rangle & = \frac{1}{Z}
  \int\limits_{-\infty}^{\infty} a(E)\, \E^{-\beta E}\,dE\,,\\
  Z & = \int\limits_{-\infty}^{\infty} \rho(E)\, \E^{-\beta E}\,dE\,.
\end{align}
Of course, similar relations hold also for non-interacting fermion
systems, where the Boltzmann weight $\E^{-\beta E}$ has to be replaced
by the Fermi function $f(E) = (1 + \E^{\beta (E-\mu)})^{-1}$ and the
single-electron wave functions play the role of~$|k\rangle$.

Again, the particular form of $a(E)$ suggests an expansion in
Chebyshev polynomials, and after rescaling we find
\begin{equation}
  \begin{aligned}
    \mu_n = \int\limits_{-1}^{1} \tilde a(E)\, T_n(E)\,dE
    & =  \frac{1}{D}\sum_{k=0}^{D-1} \langle k|A|k\rangle\, T_n(\tilde E_k)\\
    & =  \frac{1}{D}\trace(A T_n(\tilde H))\,,
  \end{aligned}
\end{equation}
which can be evaluated with the stochastic approach,
Sec.~\ref{secstoch}.

\begin{figure}
  \includegraphics[width=\linewidth]{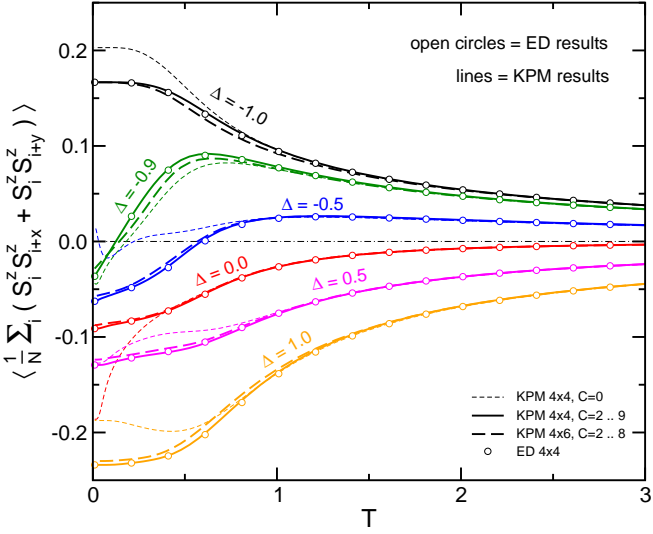}
  \caption{\conline Nearest-neighbor $S^z$-$S^z$ correlations of the
    XXZ model on a square lattice. Lines represent the KPM results
    with separation of low-lying eigenstates (bold solid and bold
    dashed) and without (thin dashed), open symbols denote exact
    results from a complete diagonalization of a $4\times4$
    system.}\label{figxxzcor}
\end{figure}
For interacting systems at low temperature the expression in
Eq.~\eqref{thexpA} is a bit problematic, since the Boltzmann factor
puts most of the weight on the lower end of the spectrum and heavily
amplifies small numerical errors in $\rho(E)$ and $a(E)$. We can avoid
these problems by calculating the ground state and some of the lowest
excitations exactly, using standard iterative diagonalization methods
like Lanczos or Jacobi-Davidson. Then we split the expectation value
of $A$ and the partition function~$Z$ into contributions from the
exactly known states and contributions from the rest of the spectrum,
\begin{align}
  \langle A\rangle & = 
  \frac{1}{ZD}\sum_{k=0}^{C-1} \langle k|A|k\rangle\, \E^{-\beta E_k} 
  + \frac{1}{Z}\int\limits_{-\infty}^{\infty} a_s(E) \E^{-\beta E}\,dE\,,\\
  Z & = \frac{1}{D}\sum_{k=0}^{C-1} \E^{-\beta E_k} 
  + \int\limits_{-\infty}^{\infty} \rho_s(E) \E^{-\beta E}\,dE
\end{align}
The functions
\begin{align}
  a_s(E) & = \frac{1}{D}\sum_{k=C}^{D-1} 
  \langle k|A|k\rangle\, \delta(E-E_k) \,, \\
  \rho_s(E) & = \frac{1}{D}\sum_{k=C}^{D-1} \delta(E-E_k)
\end{align}
describe the rest of the spectrum and can be expanded in Chebyshev
polynomials easily. Based on the known states we can introduce the
projection operator
\begin{equation}\label{projector}
  P = 1 - \sum_{k=0}^{C-1} |k\rangle\langle k|\,,
\end{equation}
and find for the expansion coefficients of $\tilde a_s(E)$
\begin{equation}
  \mu_n = \frac{1}{D}\trace(P A T_n(\tilde H)) 
  \approx \frac{1}{RD} \sum_{r=0}^{R-1} 
  \langle r|P A T_n(\tilde H) P|r\rangle\,,
\end{equation}
and similarly for those of $\tilde \rho_s(E)$
\begin{equation}
  \mu_n = \frac{1}{D}\trace(P T_n(\tilde H)) 
  \approx \frac{1}{RD} \sum_{r=0}^{R-1} \langle r|P T_n(\tilde H) P|r\rangle\,.
\end{equation}
Note, that in addition to the two vectors for the Chebyshev recursion
we now need memory also for the eigenstates $|k\rangle$. Otherwise the
resource consumption is the same as in the standard scheme.

We illustrate the accuracy of this approach in Figure~\ref{figxxzcor}
considering the nearest-neighbor $S^z$-$S^z$ correlations of the
square-lattice spin-$1/2$ XXZ model as an example,
\begin{equation}
  H = \sum_{i,\delta} (S_{i}^x S_{i+\delta}^x + S_{i}^y S_{i+\delta}^y
  + \Delta S_{i}^z S_{i+\delta}^z)\,.
\end{equation}
As a function of temperature and for an anisotropy $-1<\Delta<0$ this
model shows a quantum to classical crossover in the sense that the
correlations are anti-ferromagnetic at low temperature (quantum
effect) and ferromagnetic at high temperature (as expected for the
classical model).~\cite{FM99,SFBI00,FSWBI00} Comparing the KPM results
with the exact correlations of a $4\times 4$ system, which were
obtained from a complete diagonalization of the Hamiltonian, the
improvement due to the separation of only a few low-lying eigenstates
is obvious. Whereas for $C=0$ the data is more or less random below
$T\approx 1$, the agreement with the exact data is perfect, if the
ground state and one or two excitations are considered separately. The
numerical effort required for these calculations differs largely
between complete diagonalization and the KPM method.  For the former,
$18$ or $20$ sites are practically the limit, whereas the latter can
easily handle $30$ sites or more.

Note that for non-interacting systems the above separation of the
spectrum is not required, since for $T\to 0$ the Fermi function
converges to a simple step function without causing any numerical
problems.

\subsection{Dynamical correlations at zero temperature}\label{secgreen}

\subsubsection{General considerations}

Having discussed simple expectation values and static correlations,
the calculation of time dependent quantities is the natural next step
in the study of complex quantum models. This is motivated also by many
experimental setups, which probe the response of a physical system to
time dependent external perturbations. Examples are inelastic
scattering experiments or measurements of transport coefficients. In
the framework of linear response theory and the Kubo formalism the
system's response is expressed in terms of dynamical correlation
functions, which can also be calculated efficiently with Chebyshev
expansion and KPM. Technically though, we need to distinguish between
two different situations: For interacting many-particle systems at
zero temperature only matrix elements between the ground state and
excited states contribute to a dynamical correlation function, whereas
for interacting systems at finite temperature or for non-interacting
systems with a finite particle density transitions between all
eigenstates --- many-particle or single-particle, respectively ---
contribute. We therefore split the discussion of dynamical
correlations into two sections, starting here with interacting
many-particle systems at $T=0$.

Given two operators $A$ and $B$ a general dynamical correlation
function can be defined through
\begin{multline}\label{defcorrel}
  \langle A ; B \rangle_\omega^\pm  = \lim_{\epsilon\to 0}\ 
  \langle 0| A \frac{1}{\omega + \I\epsilon \mp H} B |0\rangle \\
  = \lim_{\epsilon\to 0}\sum\limits_{k=0}^{D-1} 
  \frac{\langle 0|A|k\rangle\langle k|B|0\rangle}{\omega+\I\epsilon \mp E_k} \,,
\end{multline}
where $E_k$ is the energy of the many-particle eigenstate $|k\rangle$
of the Hamiltonian~$H$, $|0\rangle$ its ground state, and
$\epsilon>0$.

If we assume that the product $\langle 0|A|k\rangle\langle
k|B|0\rangle$ is real the imaginary part
\begin{equation}
  \im \langle A ; B \rangle_\omega^\pm = 
  -\pi\sum\limits_{k=0}^{D-1}\langle 0|A|k\rangle\langle k|B|0\rangle
  \, \delta(\omega \mp E_k)
\end{equation}
has a similar structure as, e.g., the local density of states in
Eq.~\eqref{defrhoi}, and in fact, with $\rho_i(E)$ we already
calculated a dynamical correlation function. Hence, after rescaling
the Hamiltonian $H\to\tilde H$ and all energies
$\omega\to\tilde\omega$ we can proceed as usual and expand $\im
\langle A ; B \rangle_\omega^\pm$ in Chebyshev polynomials,
\begin{equation}\label{eqimcorr}
  \im \langle A ; B \rangle_{\tilde\omega}^\pm = 
  -\frac{1}{\sqrt{1-\tilde\omega^2}} \Big[\mu_0 + 2
  \sum\limits_{n=1}^{\infty} \mu_n\, T_n(\tilde\omega) \Big]\,.
\end{equation}
Again, the moments are obtained from expectation values
\begin{equation}\label{muimcorr}
  \mu_n = \frac{1}{\pi}\int\limits_{-1}^{1} 
  \im \langle A ; B \rangle_{\tilde\omega}^\pm\,T_n(\tilde\omega)\,d\tilde\omega 
  = \langle 0|A T_n(\mp\tilde H) B|0\rangle \,,
\end{equation}
and for $A\ne B^\dagger$ we can follow the scheme outlined in
Eqs.~\eqref{momreca} to~\eqref{momrece}. For $A=B^\dagger$ the
calculation simplifies to the one in Eqs.~\eqref{momdblg}
and~\eqref{momdblu}, now with $B |0\rangle$ as the starting vector.

In many cases, especially for the spectral functions and optical
conductivities studied below, only the imaginary part of $\langle A ;
B \rangle_\omega^\pm$ is of interest, and the above setup is all we need.
Sometimes however --- e.g., within the Cluster Perturbation Theory
discussed in Sec.~\ref{seccpt} --- also the real part of a general
correlation function $\langle A ; B \rangle_\omega^\pm$ is required.
Fortunately it can be calculated with almost no additional effort: The
analytical properties of $\langle A ; B \rangle_\omega^\pm$ arising from
causality imply that its real part is fully determined by the
imaginary part.  Indeed a Hilbert transform gives
\begin{multline}
  \re \langle A;B \rangle_{\tilde\omega}^\pm = \sum\limits_{k=0}^{D-1} 
  \langle 0|A|k\rangle\langle k|B|0\rangle\,
  \pv\left(\frac{1}{\tilde\omega \mp \tilde E_k}\right)\\
  = - \frac{1}{\pi} \pv\int\limits_{-1}^1 
  \frac{\im \langle A ; B \rangle_{\tilde\omega'}^\pm}{\tilde\omega-\tilde\omega'} 
  \,d\omega'
  = -2\sum\limits_{n=1}^{\infty} \mu_n\, U_{n-1}(\tilde\omega)\,,
\end{multline}
where we used Eq.~\eqref{chebpint1}. The full correlation function
\begin{equation}\label{eqcorr}
  \begin{aligned}
    \langle A ; B \rangle_{\tilde\omega}^\pm
    & = \frac{-\I\mu_0}{\sqrt{1-\tilde\omega^2}}
    - 2\sum\limits_{n=1}^{\infty} \mu_n 
    \Big[U_{n-1}(\tilde\omega)
    +\frac{\I\,T_n(\tilde\omega)}{\sqrt{1-\tilde\omega^2}} 
    \Big]\\
    & = \frac{-\I}{\sqrt{1-\tilde\omega^2}}
    \Big[\mu_0 + 2\sum\limits_{n=1}^{\infty}
    \mu_n\exp(-\I n\arccos\tilde\omega)\Big]
  \end{aligned}
\end{equation}
can thus be reconstructed from the same moments $\mu_n$ that we
derived for its imaginary part, Eq.~\eqref{muimcorr}. In contrast to
the real quantities we considered so far, the reconstruction merely
requires complex Fourier transform, see Eqs.~\eqref{cmplxkpm1}
and~\eqref{cmplxkpm2}.  If only the imaginary or real part of $\langle
A ; B \rangle_{\omega}^\pm$ is needed, a cosine or sine transform,
respectively, is sufficient.

Note again, that the calculation of dynamical correlation functions
for non-interacting electron systems is \emph{not} possible with the
scheme discussed in this section, not even at zero temperature. At
finite band filling (finite chemical potential) the ground state
consists of a sum over occupied single-electron states, and dynamical
correlation functions thus involve a double summation over matrix
elements between all single-particle eigenstates, weighted by the
Fermi function.  Clearly, this is more complicated than
Eq.~\eqref{defcorrel}, and we postpone the discussion of this case to
Sec.~\ref{secdyntgt0}, where we describe methods for dynamical
correlation functions at finite temperature and --- for the case of
non-interacting electrons --- finite density.

\subsubsection{One-particle spectral function}
An important example of a dynamical correlation function is the
(retarded) Green function in momentum space,
\begin{equation}
  G_{\sigma}(\vec{k},\omega) = 
  \langle c^{\phantom{\dagger}}_{\vec{k},\sigma} ; 
  c^{\dagger}_{\vec{k},\sigma}\rangle_{\omega}^{+}  + 
  \langle c^{\dagger}_{\vec{k},\sigma} ; 
  c^{\phantom{\dagger}}_{\vec{k},\sigma} \rangle_{\omega}^{-}\,,
\end{equation}
and the associated spectral function 
\begin{equation}
  \begin{split}
    A_{\sigma}(\vec{k},\omega) 
    & = -\frac{1}{\pi}\im\,G_{\sigma}(\vec{k},\omega)\\
    & = A^+_{\sigma}(\vec{k},\omega) + A^-_{\sigma}(\vec{k},\omega)\,,
  \end{split}
\end{equation}
which characterizes the electron absorption or emission of an
interacting system. For instance, $A^-_{\sigma}$ can be measured
experimentally in angle resolved photo-emission spectroscopy (ARPES).

As the first application, let us consider the one-dimensional Holstein 
model for spinless fermions~\cite{Ho59a,Ho59b},
\begin{multline}\label{hamhol}
  H =  -t\sum_{i}(c_{i}^{\dagger} c_{i+1}^{}+\text{H.c.})\\
  -g\omega_0\sum_{i,\sigma}(b_i^{\dagger}+b_i^{})n_{i}
  +\omega_0 \sum_{i} b_i^{\dagger} b_i^{}\,,
\end{multline}
which is one of the basic models for the study of electron-lattice
interaction. A single band of electrons is approximated by spinless
fermions $c_i^{(\dagger)}$, the density of which couples to the local
lattice distortion described by dispersionless phonons
$b_i^{(\dagger)}$. At low fermion density, with increasing electron
phonon interaction the charges get dressed by a surrounding lattice
distortion and form new, heavy quasi-particles known as polarons.
Eventually, for strong coupling the width of the corresponding band is
suppressed exponentially, leading to a process called self-trapping.
For a half-filled band, i.e., $0.5$ fermions per site, the model
allows for the study of quantum effects at the transition from a metal
to a band (or Peierls) insulator, marked by the opening of a gap at
the Fermi wave vector and the development of a matching lattice
distortion.

Since the Hamiltonian~\eqref{hamhol} involves bosonic degrees of
freedom, the Hilbert space of even a finite system has infinite
dimension.  In practice, nevertheless, the contribution of highly
excited phonon states is found to be negligible at low temperature or
for the ground-state, and the system is well approximated by a
truncated phonon space with $\sum_i b_i^{\dagger} b_i^{} \le M$
\cite{BWF98}. In addition, the translational symmetry of the model can
be used to reduce the Hilbert space dimension, and, moreover, the
symmetric phonon mode with momentum $q=0$ can be excluded from the
numerics: Since it couples to the total number of electrons, which is
a conserved quantity, its contribution can be handled
analytically~\cite{Ro97,SHBWF05}. Below we present results for a
cluster size of $L=8$ or $10$, where a cut-off $M=24$ or $15$,
respectively, leads to truncation errors $<10^{-6}$ for the
ground-state energy. Alternatively, for one or two fermionic particles
and low temperatures an optimized variational basis can be constructed
for infinite systems~\cite{BTB99}, which would also be suitable for
our numerical approach.

\begin{figure}
  \includegraphics[width=\linewidth]{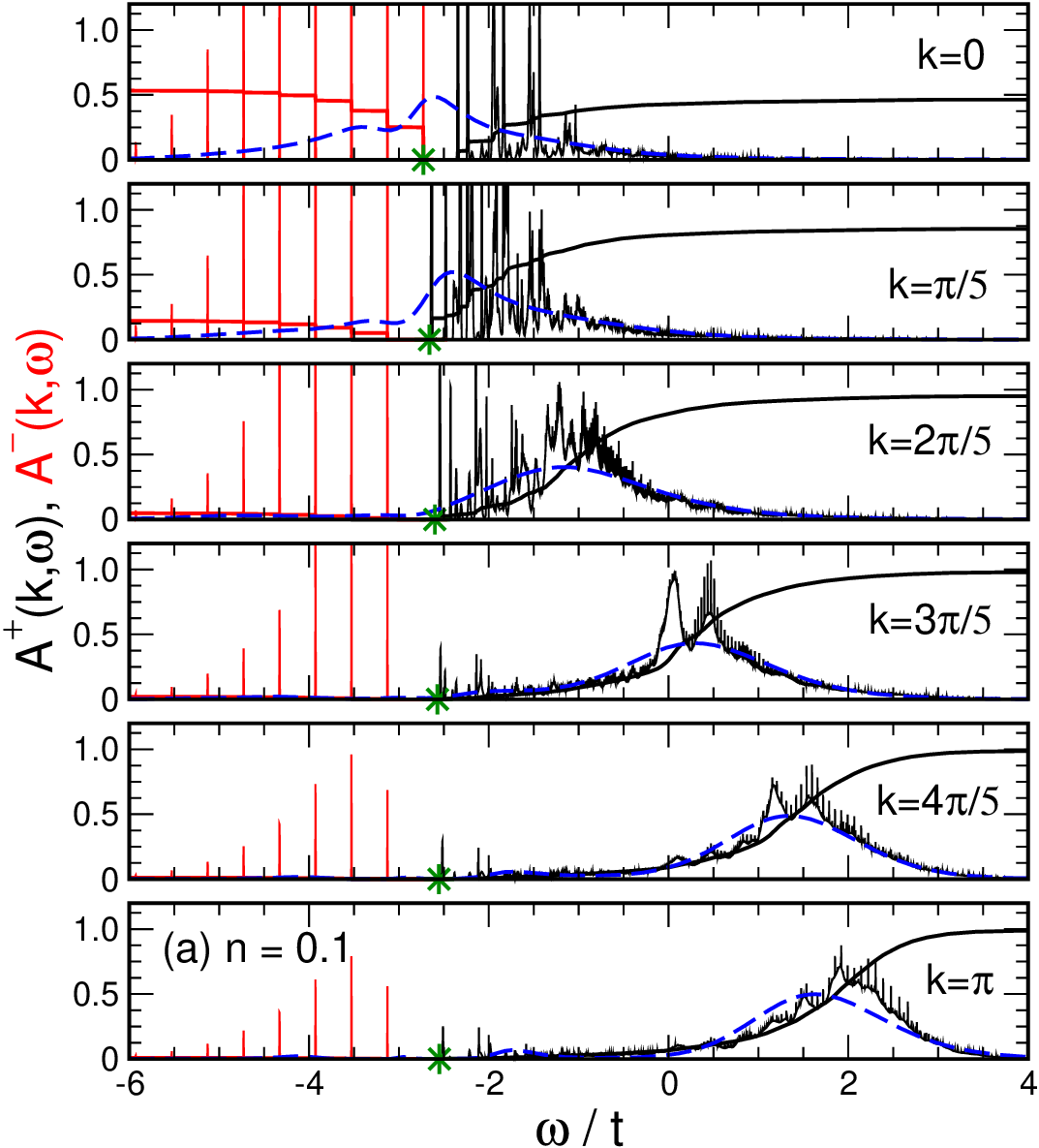}
  \includegraphics[width=\linewidth]{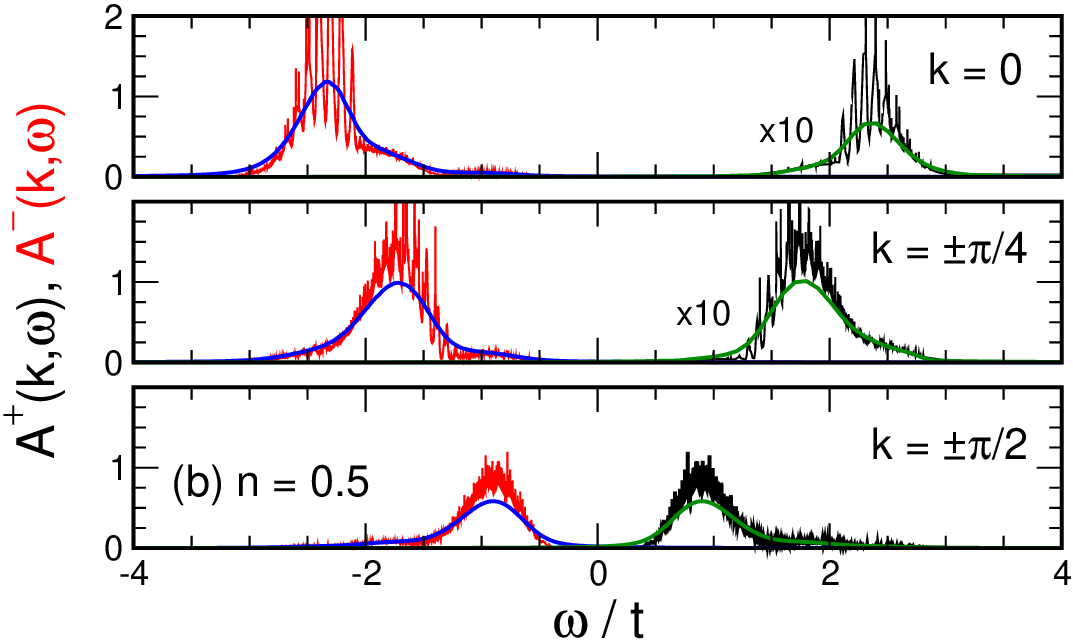}
  \caption{\conline One-particle spectral function and its integral for the
    Holstein model (a) on a 10-site ring with one electron,
    $\varepsilon_p = g^2\omega_0=2.0 t$, $\omega_0=0.4 t$, and (b) on
    a 8-site ring, band filling $n=0.5$, $\varepsilon_p =
    g^2\omega_0=1.6 t$, $\omega_0=0.1 t$. For comparison, in (a) the
    blue dashed lines represent Quantum Monte Carlo data at $\beta t =
    8$~\cite{Hoea05}, and green stars indicate the position of the
    polaron band in the infinite system~\cite{BTB99}. In (b) the blue
    and green curves denote results of Dynamical DMRG for the same
    lattice size and $T=0$~\cite{JF05p}.}\label{fighhspec}
\end{figure}
In Figure~\ref{fighhspec} we present KPM data for the spectral
function of the spinless-fermion Holstein model and assess its quality
by comparing with results from Quantum Monte Carlo (QMC) and Dynamical
Density Matrix Renormalization Group (DDMRG)~\cite{Je02b} calculations.
Starting with the case of a single electron on a ten-site ring,
Figure~\ref{fighhspec}~(a) illustrates the presence of a narrow
polaron band at the Fermi level and of a broad range of incoherent
contributions to the spectral function, which in the spinless case
reads
\begin{multline}\label{a-}
  A^-(k,\omega)  = 
  \sum_l |\langle l,N_\text{e}-1|\,c_{k}\,|0,N_\text{e}\rangle |^2 \\
  \times\delta[\omega + (E_{l,N_\text{e}-1}-E_{0,N_\text{e}})]
\end{multline}
and
\begin{multline}\label{a+}
  A^+(k,\omega)  = 
  \sum_l |\langle l,N_\text{e}+1|\,c_{k}^{\dagger}\,|0,N_\text{e}\rangle |^2 \\
  \times\delta[\omega - (E_{l,N_\text{e}+1}-E_{0,N_\text{e}})]\,.
\end{multline}
Here $|l,N_\text{e}\rangle$ denotes the $l$th eigenstate with
$N_\text{e}$ electrons and energy $E_{l,N_\text{e}}$. The
photo-emission part $A^-$ reflects the Poisson-like phonon
distribution of the polaron ground state, whereas $A^+$ has most of
its weight in the vicinity of the original free electron band.  In
terms of the overall shape and the integrated weight, both KPM and QMC
agree very well.  QMC, however, is not able to resolve all the narrow
features of the spectral function, and the polaron band is hardly
observable. Nevertheless, QMC has the advantage that larger systems
can be studied, in particular at finite temperature. As a guide to the
eye we also show the position of the polaron band in the infinite
system, which was calculated with the approach of~\textcite{BTB99}.
In Figure~\ref{fighhspec}~(b) we consider the case of a half-filled
band and strong electron-phonon coupling, where the system is in an
insulating phase with an excitation gap at the Fermi momentum $k=\pm
\pi/2$. Below and above the gap the spectrum is characterized by broad
multi-phonon absorption. Compared to DDMRG, again KPM offers the
better resolution and unfolds all the discrete phonon sidebands.
Concerning numerical performance DDMRG has the advantage of a small
optimized Hilbert space, which can be handled with standard
workstations. However, the basis optimization is rather time consuming
and, in addition, each frequency value $\omega$ requires a new
simulation. The KPM calculations, on the other hand, involved matrix
dimensions between $10^8$ and $10^{10}$, and we therefore used
high-performance computers such as Hitachi SR8000-F1 or IBM p690 for
the moment calculation. For the reconstruction of the spectra, of
course, a desktop computer is sufficient.

\subsubsection{Optical conductivity}\label{secsigT0}
\begin{figure}
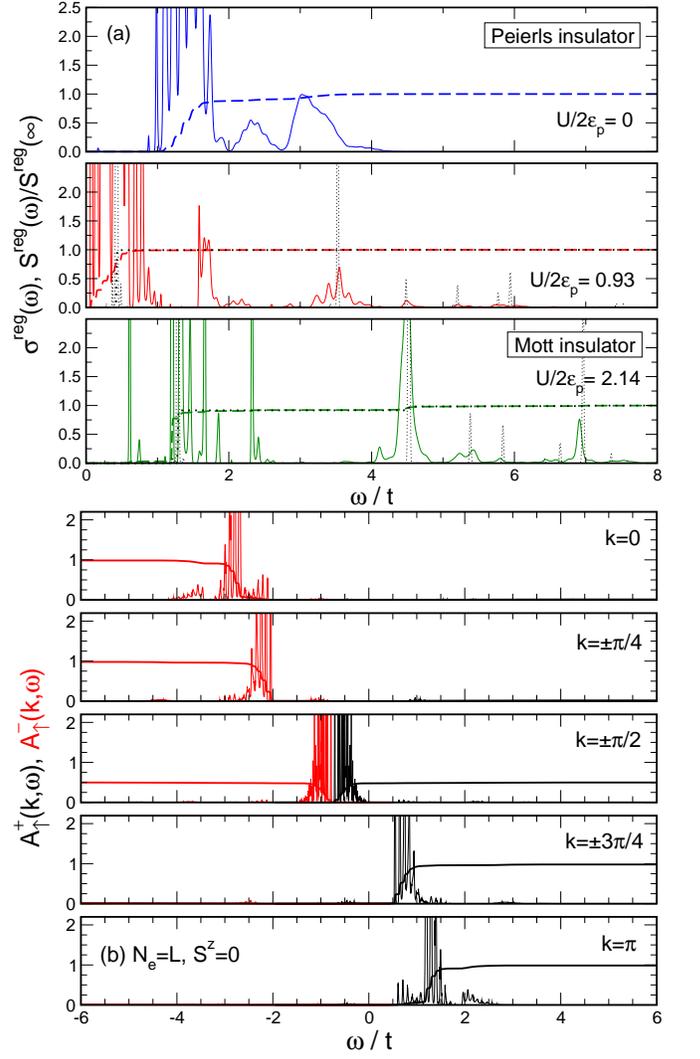

  \includegraphics[width=\linewidth]{figure6a.eps}
  \includegraphics[width=\linewidth]{figure6b.eps}
  \caption{\conline (a) The optical conductivity $\sigma^{\text{reg}}(\omega)$
    and its integral $S^{\text{reg}}(\omega)$ for the Holstein Hubbard
    model at half-filling with different ratios of the Coulomb
    interaction $U$ to the electron-lattice coupling
    $\varepsilon_p=g^2\omega_0$, $\omega_0=0.1t$, and $g^2=7$. Black
    dotted lines denote excitations of the pure Hubbard model.  (b)
    The one-particle spectral function at the transition point, i.e.,
    for the same parameters as in the middle panel of (a). The system
    size is $L=8$.}\label{fighhsigma}
\end{figure}
The next example of a dynamical correlation function is the optical
conductivity. Here the imaginary and real parts of our general
correlation functions $\langle A ; B \rangle_{\omega}$ change their
roles due to an additional frequency integration. The so-called
regular contribution to the real part of the optical conductivity is
thus given by,
\begin{equation}\label{sigmazero}
  \sigma^{\text{reg}}(\omega) = \frac{1}{\omega} \sum_{E_k>E_0} 
  |\langle k|J|0\rangle|^2\ \delta(\omega - (E_k - E_0))\,,
\end{equation}
where the operator
\begin{equation}\label{currentop}
  J = -\I q t\sum_{i,\sigma}(c_{i,\sigma}^{\dagger} 
  c_{i+1,\sigma}^{} - \text{H.c.})
\end{equation}
describes the current.  After rescaling the energy and shifting the
frequency, $\omega = \tilde\omega+\tilde E_0$, the sum can be expanded
as described earlier, now with $J|0\rangle$ as the initial state for
the Chebyshev recursion.  Back-scaling and dividing by $\omega$ then
yields the final result.

In Figure~\ref{fighhsigma} we apply this setup to the Holstein Hubbard
model, which is the generalization of the Holstein model to true,
spin-carrying electrons that interact via a screened Coulomb
interaction, modelled by a Hubbard $U$-term,
\begin{multline}\label{hamhh}
  H = -t\sum_{i,\sigma}(c_{i,\sigma}^{\dagger}
  c_{i+1,\sigma}^{}+\text{H.c.})
  +U\sum_i n_{i\uparrow}n_{i\downarrow}\\
  -g\omega_0\sum_{i,\sigma}(b_i^{\dagger}+b_i^{})n_{i\sigma} +\omega_0
  \sum_{i} b_i^{\dagger} b_i^{}\,.
\end{multline}
For a half-filled band, which now denotes a density of one electron
per site, the electronic properties of the model are governed by a
competition of two insulating phases: a Peierls (or band) insulator
caused by the electron-lattice interaction and a Mott (or correlated)
insulator caused by the electron-electron interaction. Within the
optical conductivity both phases are signalled by an excitation gap,
which closes at the transition between the two phases. We illustrate
this behavior in Figure~\ref{fighhsigma}~(a), showing
$\sigma^{\text{reg}}(\omega)$ at strong electron-phonon coupling and
for increasing $U$. The data for the one-particle spectral function in
Figure~\ref{fighhsigma}~(b) proves that simultaneously to the optical
gap also the charge gap vanishes at the quantum phase transition
point~\cite{Feea02b,FWHWB04}.

\subsubsection{Spin structure factor}
\begin{figure}
  \includegraphics[width=\linewidth]{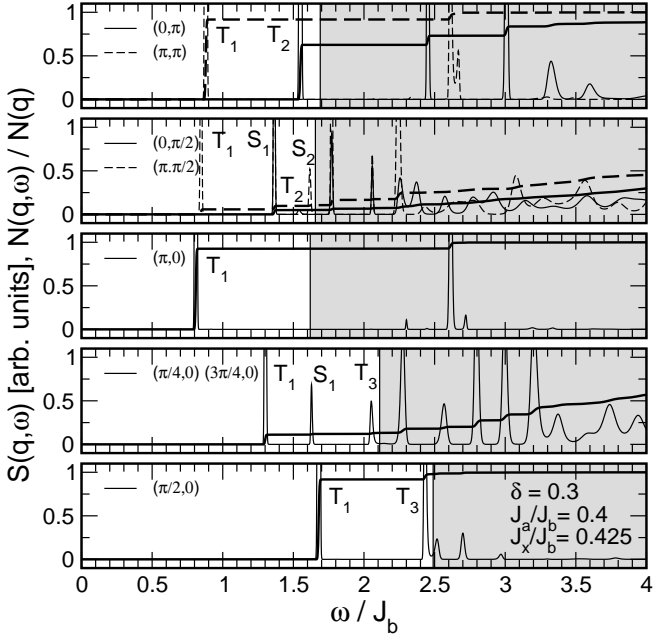}
  \caption{Spin structure factor at $T=0$ calculated for the 
    model~\eqref{hamvopo} which aims at describing the magnetic
    compound \vopo{}. For more details
    see~\cite{WBF99}.}\label{figvopo}
\end{figure}
Apart from electron systems, of course, the KPM approach works also
for other quantum problems such as pure spin systems. To describe the
excitation spectrum and the magnetic properties of the compound
\vopo{}, some years ago we proposed the 2D spin
Hamiltonian~\cite{WBF99}
\begin{multline}\label{hamvopo}
  H  = J_b \sum_{i, j} (1+\delta(-1)^i) 
  \vec{S}_{i,j}\cdot\vec{S}_{i+1,j} 
  + J_a \sum_{i,j} \vec{S}_{i,j}\cdot\vec{S}_{i,j+1}\\
  + J_{\times} \sum_{i,j} (\vec{S}_{2i,j}\cdot\vec{S}_{2i+1,j+1} +
  \vec{S}_{2i+1,j}\cdot\vec{S}_{2i,j+1})\,,
\end{multline}
where $\vec{S}_{i,j}$ denote spin-$1/2$ operators on a square lattice.
With this model we aimed at explaining the observation of two branches
of low-lying triplet excitations by neutron scattering~\cite{Gaea97},
which was inconsistent with the then prevailing picture of \vopo{}
being a spin-ladder or alternating chain compound.

Studying the low-energy physics of the model~\eqref{hamvopo} the KPM
approach can be used to calculate the spin structure factor and the
integrated spectral weight,
\begin{align}
  S(\vec{q},\omega) & = \sum_k |\langle k|\vec{S}^z(\vec{q})|0\rangle|^2 
  \delta(E_k - E_0 - \omega)\,,\\
  N(\vec{q},\omega) & = \int_0^{\omega} d\omega' S(\vec{q},\omega')\,,
\end{align}
where $\vec{S}^z(\vec{q}) = \sum_{i,j} \E^{\I
  \vec{q}\cdot\vec{r}_{i,j}} S^z_{i,j}$. Figure~\ref{figvopo} shows
these quantities for a $4\times 8$ cluster with periodic boundary
conditions. The dimension of the sector $S_z=0$, which contains the
ground state, is rather moderate here being of the order of $D\approx
4\cdot 10^7$ only. The expansion clearly resolves the lowest (massive)
triplet excitations $T_1$, a number of singlets and, in particular, a
second triplet branch $T_2$. The shaded region marks the two-particle
continuum obtained by exciting two of the elementary triplets $T_1$,
and illustrates that $T_2$ is lower in energy. Since the system is
finite in size, of course, the continuum appears only as a set of
broad discrete peaks, the density of which increases with the system
size.

\subsection{Dynamical correlations at finite temperature}\label{secdyntgt0}
\subsubsection{General considerations}\label{secdyngen}

In the preceding section we mentioned briefly that for non-interacting
electron systems or for interacting systems at finite temperature the
calculation of dynamical correlation functions is more involved, due
to the required double summation over all matrix elements of the
measured operators. Chebyshev expansion, nevertheless, offers an
efficient way for handling these problems.  To be specific, let us
derive all new ideas on the basis of the optical conductivity
$\sigma(\omega)$, which will be our primary application below.
Generalizations to other dynamical correlations can be derived without
much effort.

For an interacting system the extension of Eq.~\eqref{sigmazero} is
given by
\begin{equation}
  \sigma^{\text{reg}}(\omega)  
  = \sum_{k,q} \frac{|\langle k|J|q\rangle|^2 (\E^{-\beta E_k}-\E^{-\beta E_q})
  }{ZD\,\omega}\,\delta(\omega - \omega_{qk})\,,
\end{equation}
with $\omega_{qk} = E_q - E_k$.  Compared to Eq.~\eqref{sigmazero} a
straight-forward expansion of the finite temperature conductivity is
spoiled by the presence of the Boltzmann weighting factors. Some
authors~\cite{IE03} try to handle this problem by expanding these
factors in Chebyshev polynomials and performing a numerical time
evolution subsequently, which, however, requires a new simulation for
each temperature. A much simpler approach is based on the function
\begin{equation}
  j(x,y) = \frac{1}{D} \sum_{k,q} |\langle k|J|q \rangle|^2 
  \ \delta(x-E_k)\ \delta(y-E_q)
\end{equation}
which we may interpret as a matrix element density. Being a function
of two variables, $j(x,y)$ can be expanded with two-dimensional KPM,
\begin{equation}
  \tilde{j}(x,y) = \sum\limits_{n,m=0}^{N-1}\frac{
    \mu_{nm} h_{nm} g_n g_m T_n(x) T_m(y)
  }{\pi^2 \sqrt{(1-x^2)(1-y^2)}}
\end{equation}
where $\tilde{j}(x,y)$ refers to the rescaled $j(x,y)$, $g_n$ are the
usual kernel damping factors (see Eq.~\eqref{jackson}), and $h_{nm}$
account for the correct normalization (see Eq.~\eqref{defmulti}).  The
moments $\mu_{nm}$ are obtained from
\begin{equation}
  \begin{aligned}
    \mu_{nm} & = \int\limits_{-1}^{1}\int\limits_{-1}^{1} 
    \tilde{j}(x,y) T_n(x) T_m(y)\,dx\,dy\\
    & = \frac{1}{D} \sum_{k,q} |\langle k|J|q \rangle|^2
    \ T_n(\tilde E_k)\ T_m(\tilde E_q)\\
    & = \frac{1}{D} \sum_{k,q} \langle k|T_n(\tilde H) J|q\rangle
    \langle q|T_m(\tilde H) J|k \rangle\\
    & = \frac{1}{D} \trace\big(T_n(\tilde H) J T_m(\tilde H) J\big)\,,
  \end{aligned}
\end{equation}
and again the trace can be replaced by an average over a relatively
small number $R$ of random vectors $|r\rangle$. The numerical effort
for an expansion of order $n,m<N$ ranges between $2RDN$ and $RDN^2$
operations, depending on whether memory is available for up to $N$
vectors of the Hilbert space dimension $D$ or not. Given the operator
density $j(x,y)$ we find the optical conductivity by integrating over
Boltzmann factors,
\begin{equation}
  \begin{aligned}
    \sigma^{\text{reg}}(\omega) & = 
    \frac{1}{Z\omega} \int\limits_{-\infty}^{\infty}
    j(y+\omega,y) \big(\E^{-\beta y} - \E^{-\beta (y+\omega)}\big)\,dy\\
    & = \sum_{k,q} \frac{|\langle k|J|q\rangle|^2 
    (\E^{-\beta E_k}-\E^{-\beta E_q})}{ZD\omega}\,\delta(\omega-\omega_{qk})\,,
  \end{aligned}
\end{equation}
and, as above, we get the partition function $Z$ from an integral over
the density of states $\rho(E)$. The latter can be expanded in
parallel to $j(x,y)$. Note that the calculation of the conductivity at
different temperatures is based on the same operator density $j(x,y)$,
i.e., it needs to be expanded only once for all temperatures.

Surprisingly, the basic steps of this approach were suggested already
ten years ago~\cite{WZ94,Wa94}, but --- probably overlooking its
potential --- applied only to the zero-temperature response of
non-interacting electrons. A reason for the poor appreciation of these
old ideas may also lie in the use of non-optimal kernels, which did
not ensure the positivity of $j(x,y)$ and reduced the numerical
precision.  Only recently, one of the authors generalized the Jackson
kernel and obtained high resolution optical data for the Anderson
model~\cite{We04}. More results, in particular for interacting quantum
systems at finite temperature, we present hereafter.

\subsubsection{Optical conductivity of the Anderson model}\label{secoptand}
\begin{figure}
  \includegraphics[width=0.8\linewidth]{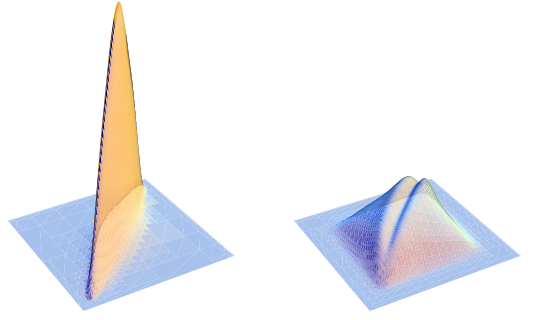}
  \caption{\conline The matrix element density $j(x,y)$ for the 3D
    Anderson model with disorder $W/t=2$ and $12$.}\label{figandopd}
\end{figure}
Since the Anderson model describes non-interacting fermions, the
eigenstates $|k\rangle$ occurring in $\sigma(\omega)$ now denote
single-particle wave functions and the Boltzmann weight has to be
replaced by the Fermi function,
\begin{equation}
  \begin{aligned}\label{sigfermi}
    \sigma^{\text{reg}}(\omega) & = 
    \frac{1}{\omega}\int\limits_{-\infty}^{\infty} 
    j(y+\omega,y) \big(f(y) - f(y+\omega)\big)\,dy\\
    & = {\sum_{k,q}}\frac{|\langle k|J|q\rangle|^2 
      (f(E_k) - f(E_q))}{\omega}\,\delta(\omega - \omega_{qk})\,.
  \end{aligned}
\end{equation}
Clearly, from a computational point of view this expression is of the
same complexity for both, zero and finite temperature, and indeed,
compared to Sec.~\ref{secgreen}, we need the more advanced 2D~KPM
approach. 

\begin{figure}
  \includegraphics[width=\linewidth]{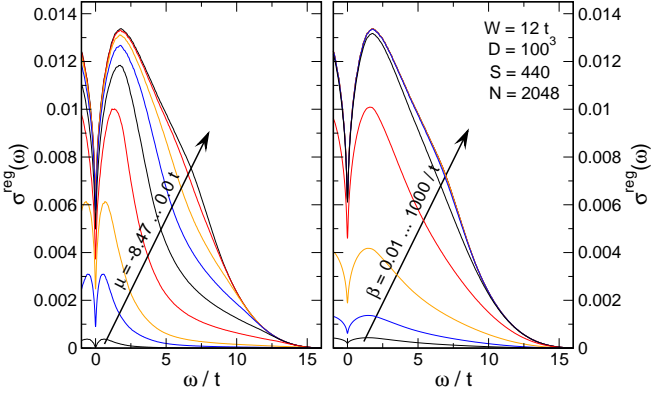}
  \caption{\conline Optical conductivity of the 3D Anderson model at
    disorder $W=12$ and for different chemical potentials $\mu$ and
    temperatures $\beta=1/T$.}\label{figandsigma}
\end{figure}

Figure~\ref{figandopd} shows the matrix element density $j(x,y)$
calculated for the 3D Anderson model on a $D=50^3$ site cluster. The
expansion order is $N=64$, and the moment data was averaged over
$S=10$ disorder samples and $R=10$ random start vectors each. Starting
from a ``shark fin'' at weak disorder, with increasing $W$ the density
$j(x,y)$ spreads in the entire energy plane, simultaneously developing
a sharp dip along $x=y$. A comparison with Eq.~\eqref{sigfermi}
reveals that this dip is responsible for the decreasing and finally
vanishing DC conductivity of the model~\cite{We04}. In
Figure~\ref{figandsigma} we show the resulting optical conductivity at
$W/t=12$ for different chemical potentials $\mu$ and temperatures
$\beta=1/T$. Note that all curves are derived from the same
matrix element density $j(x,y)$, which is now based on a $D=100^3$
site cluster, expansion order $N=2048$, an average over $S=440$
samples and only $R=1$ random start vectors each.

\subsubsection{Optical conductivity of the Holstein model}
Having discussed dynamical correlations for non-interacting electrons,
let us now come back to the case of interacting systems. The setup
described so far works well for high temperatures, but as soon as $T$
gets small we experience the same problems as with thermal expectation
values and static correlations. Again, the Boltzmann factors put most
of the weight to the margins of the domain of $j(x,y)$, thus
amplifying small numerical errors. To properly approach the limit
$T\to 0$ we therefore have to separate the ground state and a few
lowest excitations from the rest of the spectrum in a fashion similar
to the static correlations in Sec.~\ref{secstatic}. Since we start
from a 2D expansion, the correlation function (optical conductivity)
now splits into three parts: a contribution from the transitions (or
matrix elements) between the separated eigenstates, a sum of 1D
expansions for the transitions between the separated states and the
rest of the spectrum (see Sec.~\ref{secgreen}), and a 2D expansion
for all transitions within the rest of the spectrum,
\begin{equation}
  \sigma^{\text{reg}}(\omega) 
  = \underbrace{\sum_{k,q=0}^{C-1} \sigma_{k,q}}_{\sigma^{\text{reg}
    }_{\text{ED}}(\omega)}
  + \underbrace{\sum_{k=0}^{C-1}\sum_{q=C}^{D-1}(\sigma_{k,q}+\sigma_{q,k})
  }_{\sigma^{\text{reg}}_{\text{1D}}(\omega)}
  + \underbrace{\sum_{k,q=C}^{D-1}\sigma_{k,q}}_{\sigma^{\text{reg}
    }_{\text{2D}}(\omega)}\,,
\end{equation}
with
\begin{equation}
  \sigma_{k,q} = \frac{|\langle k|J|q\rangle|^2 
    (\E^{-\beta E_k}-\E^{-\beta E_q})\,\delta(\omega-\omega_{qk})}{ZD\omega}\,.
\end{equation}
The expansions required for $\sigma^{\text{reg}}_{\text{1D}}(\omega)$
are carried out in analogy to Sec.~\ref{secsigT0}, but the
resulting conductivities are weighted appropriately when all
contributions are combined to $\sigma^{\text{reg}}(\omega)$. Using the
projection operator defined in Eq.~\eqref{projector}, the
corresponding moments read
\begin{equation}
  \mu_n^k  = \langle k|JPT_n(\tilde H)PJ|k\rangle\,.
\end{equation}
For $\sigma^{\text{reg}}_{\text{2D}}(\omega)$ we follow the scheme
outlined in~\ref{secdyngen}, but use projected moments
\begin{equation}
  \mu_{nm}  = \trace(T_n(\tilde H) P J T_m(\tilde H) P J)/D \,.
\end{equation}

\begin{figure}
  \begin{center}
    \includegraphics[width=\linewidth]{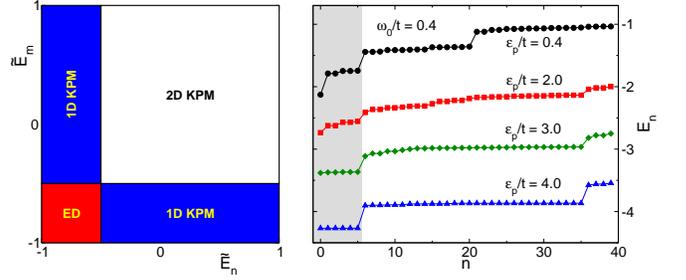}
  \end{center}
  \caption{\conline Left: Schematic setup for the calculation of
    finite-temperature dynamical correlations for interacting quantum
    systems, which requires a separation into parts handled by exact
    diagonalization (ED), 1D Chebyshev expansion and 2D Chebyshev
    expansion.  Right: The lowest eigenvalues of the Holstein model on
    a six site chain for different electron-phonon coupling
    $\varepsilon_p$. The shaded region marks the lowest polaron band,
    which was handled separately when calculating the spectra in
    Figure~\ref{figsigmaw}.}\label{figevals}
\end{figure}
\begin{figure}
  \begin{center}
    \includegraphics[width=\linewidth]{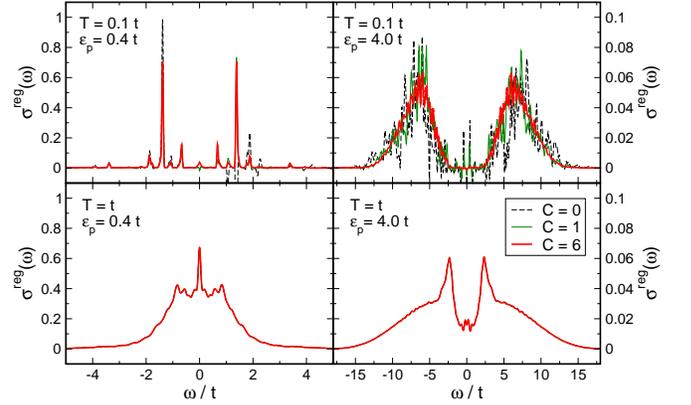}
  \end{center}
  \caption{\conline Finite temperature optical conductivity of a
    single electron coupled to the lattice via a Holstein type
    interaction. Different colors illustrate how, in particular, the
    low-temperature spectra benefit from a separation of $C=0, 1$ or
    $6$ low-energy states~\cite{SWWAF05}. The phonon frequency is
    $\omega_0/t = 0.4$.}\label{figsigmaw}
\end{figure}
In Figure~\ref{figevals} we illustrate our setup schematically and
show the lowest forty eigenvalues of the Holstein model,
Eq.~\eqref{hamhol}, with a band filling of one electron. Separating up
to six states from the rest of the spectrum we obtain the
finite-temperature optical conductivity of the system,
Figure~\ref{figsigmaw}. For high temperatures ($T=t$, see lower
panels) the separation of low-energy states is not necessary, the
conductivity curves for $C=0, 1$ and $6$ agree very well. For low
temperatures ($T=0.1 t$, see upper panels), the separation is crucial.
Without any separated states ($C=0$) the conductivity has substantial
numerical errors and can even become negative, if large Boltzmann
factors amplify infinitesimal numerical round-off errors of negative
sign. Splitting off the ground state ($C=1$) or the entire (narrow)
polaron band ($C=6$ for the present six-site cluster), we obtain
reliable, high-resolution spectra down to the lowest temperatures.
From a physics point of view, at strong electron phonon coupling
(right panels) the conductivity shows an interesting transfer of
spectral weight from high to low frequencies, if the temperature is
increased (see~\textcite{SWWAF05} for more details).

With this discussion of optical conductivity as a finite temperature
dynamical correlation function we conclude the section on direct
applications of KPM. Of course, the described techniques can be used
for the solution of many other interesting and numerically demanding
problems, but an equally important field of applications emerges, when
KPM is embedded into other numerical or analytical techniques, which
is the subject of the next section.

\section{KPM as a component of other methods}\label{seccomp}

\subsection{Monte Carlo simulations}
In condensed matter physics some of the most intensely studied
materials are affected by a complex interplay of many degrees of
freedom, and when deriving suitable approximate descriptions we
frequently arrive at models, where non-interacting fermions are
coupled to classical degrees of freedom. Examples are colossal
magneto-resistant manganites~\cite{Da03} or magnetic
semiconductors~\cite{SKM01}, where the classical variables correspond to
localized spin degrees of freedom.  We already introduced such a model
when we discussed the limit $S\to\infty$ of the double-exchange model,
Eq.~\eqref{hamcde}. The properties of these systems, e.g. a
ferromagnetic ordering as a function of temperature, can be studied by
standard MC procedures.  However, in contrast to purely classical
systems the energy of a given spin configuration, which enters the
transition probabilities, cannot be calculated directly, but requires
the solution of the corresponding non-interacting fermion problem.
This is usually the most time consuming part, and an efficient MC
algorithm should therefore evaluate the fermionic trace as fast and as
seldom as possible.

The first requirement can be matched by using KPM for calculating the
density of states of the fermion system, which by integration over the
Fermi function yields the energy of the underlying spin configuration.
Combined with standard Metropolis single-spin updates this led to the
first MC simulations of double-exchange systems~\cite{MF99,MF00,MF01}
on reasonably large clusters ($8^3$ sites), which were later improved
by replacing full traces by trace estimates and by increasing the
efficiency of the matrix vector multiplications~\cite{FM04,ASFMD05}.

To fulfil the second requirement it would be advantageous to replace
the above single-spin updates by updates of the whole spin background.
The first implementation of such ideas was given in terms of an hybrid
Monte Carlo algorithm~\cite{AFGLM01}, which combines an approximate
time evolution of the spin system with a diagonalization of the
fermionic problem by Legendre expansion, and requires a much smaller
number of MC accept-reject steps. However, this approach has the
drawback of involving a molecular dynamics type simulation of the
classical degrees of freedom, which is a bit complicated and which may
bias the system in the direction of the assumed approximate dynamics.

\begin{figure}
  \includegraphics[width=\linewidth]{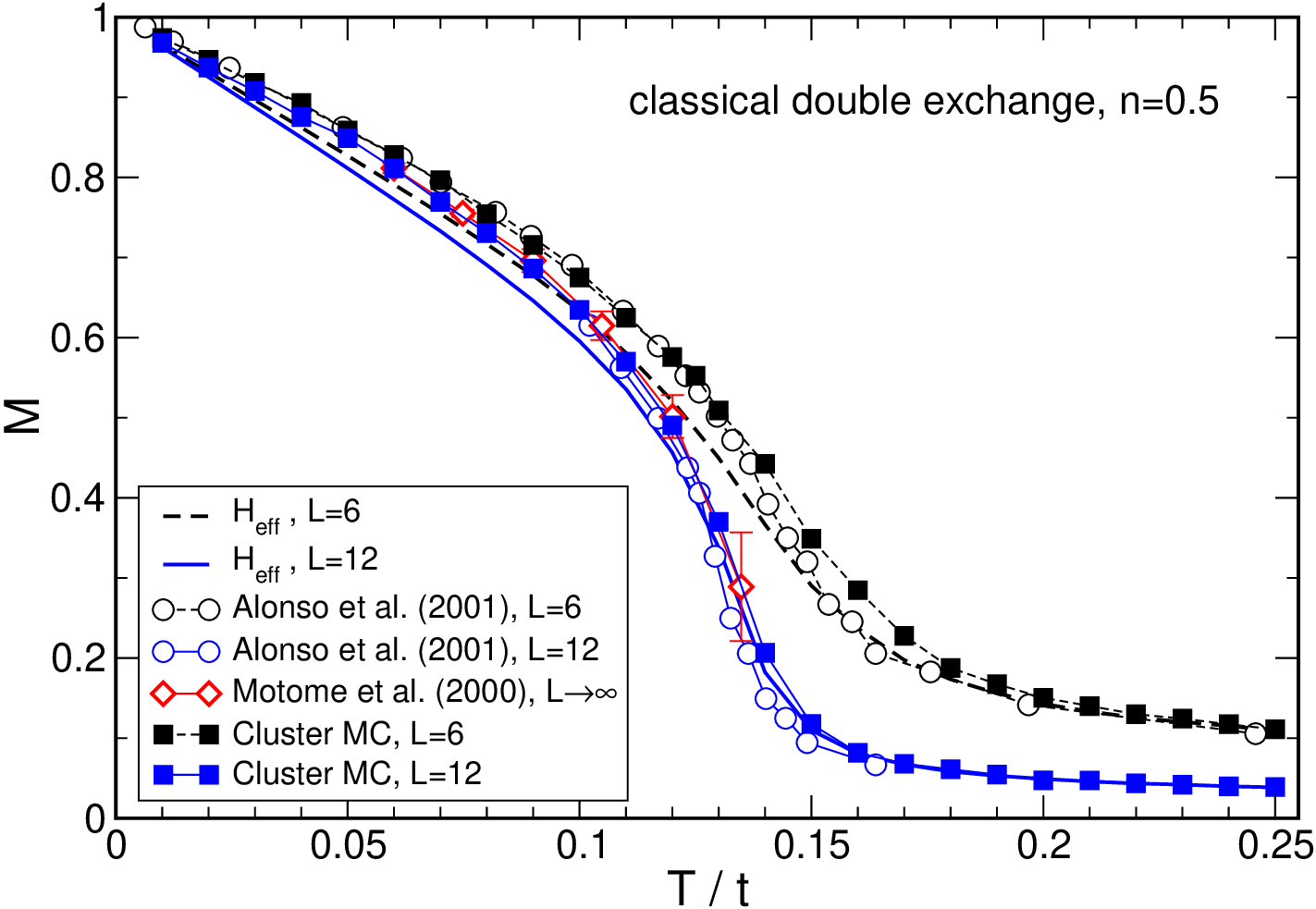}
  \caption{\conline Magnetization as a function of temperature for the
    classical double-exchange model at doping $n=0.5$. We compare data
    obtained from the effective model $H_\text{eff}$ (see text), from
    a hybrid Monte Carlo approach~\cite{AFGLM01}, the Truncated
    Polynomial Expansion Method~\cite{MF00,MF01}, and from a KPM
    based Cluster Monte Carlo technique~\cite{WFI05}. $L$ denotes the
    size of the underlying three-dimensional cluster, i.e., $D=L^3$ is
    the dimension of the fermionic problem.}\label{figdemc}
\end{figure}
Focussing on the problem of classical double exchange,
Eq.~\eqref{hamcde}, we therefore proposed a third
approach~\cite{WFI05}, which combines the advantages of KPM with the
highly efficient Cluster MC algorithms~\cite{Wo89,Ja98b,Kr04}.  In
general, for a classical MC algorithm the transition probability from
state $a$ to state $b$ can be written as
\begin{equation}
  P(a\to b) = A(a\to b) \tilde P(a\to b)\,,
\end{equation}
where $A(a\to b)$ is the probability of {\em considering} the move
$a\to b$, and $\tilde P(a\to b)$ is the probability of {\em accepting}
the move $a\to b$. Given the Boltzmann weights of the states $a$ and
$b$, $W(a)$ and $W(b)$, detailed balance requires that
\begin{equation}
  W(a) P(a\to b) = W(b) P(b\to a)\,,
\end{equation}
which can be fulfilled with a generalized Metropolis algorithm
\begin{equation}\label{metroprob}
  \tilde P(a\to b) =
  \min\left(1, \frac{W(b) A(b\to a)}{W(a) A(a\to b)}\right)\,.
\end{equation}
In the standard MC approach for spin systems only a single randomly
chosen spin is flipped. Hence, $A(a\to b)=A(b\to a)$ and the
probability $\tilde P(a\to b)$ is usually much smaller than $1$, since
it depends on temperature via the weights $W(a)$ and $W(b)$. This
disadvantage can be avoided by a clever construction of clusters of
spins, which are flipped simultaneously, such that the {\em a priori}
probabilities $A(a\to b)$ and $A(b\to a)$ soak up any difference in
the weights $W(a)$ and $W(b)$. We then arrive at the famous
rejection-free cluster MC algorithms~\cite{Wo89}, which are
characterized by $\tilde P(a\to b) = 1$.

For the double-exchange model~\eqref{hamcde} we cannot expect to find
an algorithm with $\tilde P(a\to b) = 1$, but even a method with
$\tilde P(a\to b) = 0.5$ would be highly efficient. The amplitude of
the hopping matrix element~\eqref{tijcde} is given by the cosine of
half the relative angle between neighboring spins, or $|t_{ij}|^2 = (1
+ \vec S_i\cdot \vec S_j)/2$.  Averaging over the fermionic degrees of
freedom, we thus arrive at an effective classical spin model
\begin{equation}
  H_{\text{eff}} = -J_{\text{eff}} \sum_{\langle ij \rangle}
  \sqrt{1 + \vec S_i\cdot \vec S_j}\,,
\end{equation}
where the particle density $n$ approximately defines the coupling,
$J_{\text{eff}} \approx n(1-n)/\sqrt{2}$. Similar to a classical
Heisenberg model, the Hamiltonian $H_{\text{eff}}$ is a sum over
contributions of single bonds, and we can therefore construct a
cluster algorithm with $\tilde P(a\to b) = 1$.  Surprisingly, the
simulation of this pure spin model yields magnetization data, which
almost perfectly matches the results for the full classical
double-exchange model at doping $n=0.5$, see Figure~\ref{figdemc}.

For simulating the coupled spin fermion model~\eqref{hamcde} we
suggested to apply the single cluster algorithm for $H_{\text{eff}}$
until approximately every spin in the system has been flipped once,
thereby keeping track of all {\em a priori} probabilities $A(a\to b)$
of subsequent cluster flips.  Then for the new spin configuration the
energy of the electron system is evaluated with the help of KPM. Note
however, that for a reliable discrimination of $H_{\text{eff}}$ and
the full fermionic model~\eqref{hamcde} the energy calculation needs
to be very precise.  For the moment calculation we therefore relied on
complete trace summations instead of stochastic estimates. The KPM
step is thus no longer linear in $D$, but still much faster than a
full diagonalization of the bilinear fermionic model. Based on the
resulting energy, the new spin configuration is accepted with the
probability~\eqref{metroprob}. Figure~\ref{figdemc} shows the
magnetization of the double-exchange model as a function of
temperature for $n=0.5$. Except for small deviations near the critical
temperature the data obtained with the new approach compares well with
the results of the hybrid MC approach~\cite{AFGLM01}, and due to the
low numerical effort rather large systems can be studied.

Of course, the combination of KPM and classical Monte Carlo not only
works for spin systems. We may also think of models involving the
coupling of electronic degrees of freedom to adiabatic lattice
distortions or other classical variables~\cite{ASFMD05}, and as yet
the potential of such combined approaches is certainly not fully
exhausted.

The next application, which makes use of KPM as a component of a more
general numerical approach, brings us back to interacting quantum
systems, in particular, correlated electron systems with strong local
interactions.

\subsection{Cluster Perturbation Theory (CPT)}\label{seccpt}
\subsubsection{General features of CPT}
Earlier in this review we have demonstrated the advantages of the
Chebyshev approach for the calculation of spectral functions, optical
conductivities and structure factors of complicated interacting
quantum systems. However, owing to the finite size of the considered
systems, quantities like the spectral function $A(\vec{k},\omega)$
could only be calculated for a finite set of independent momenta
$\vec{k}$.  The interpretation of this ``discrete'' data may sometimes
be less convenient, e.g. the $\vec{k}$-integrated one-electron density
$\rho(\omega)=\int dk^d \, A(\vec{k},\omega)$ does not show bands but only
discrete poles which are grouped to band-like structures.  Although
this does not substantially bias the interpretation it is desirable to
restore the translational symmetry of the lattice and reintroduce an
infinite momentum space.

With the Cluster Perturbation Theory (CPT)~\cite{GV94,SPP00,SPP02} a
straightforward way to perform this task approximatively has recently been
devised.  To describe it in a nutshell, let us consider a model of
interacting fermions on a one-dimensional chain 
\begin{equation}
  H= -t \sum_{i\sigma}
  ( c^\dagger_{i+1,\sigma} c^{}_{i,\sigma} + \text{H.c.}) + \sum_i U_i\,.
\end{equation}
Here $U_i$ denotes a local interaction, e.g. $U_i= U n_{i\uparrow}
n_{i\downarrow}$ for the Hubbard model.  CPT starts by breaking up the
infinite system into short finite chains of $L$ sites each (clusters),
which all are equivalent due to translational symmetry.  From the
Green function of a finite chain, $G^c_{ij}(\omega)$ with
$i,j=0,\dots,L-1$, which is calculated exactly by a suitable numerical
method, the Green function $G(k,\omega)$ of the infinite chain is
obtained by reintroducing the hopping between the segments.  This
inter-chain hopping is treated on the level of a random phase
approximation, which neglects correlations between different chains.
The Green function $G_{ij}^{nm}(\omega)$ is then given through a Dyson
equation
\begin{equation}
  G_{ij}^{nm}(\omega)=\delta_{nm} G^c_{ij}(\omega) + \sum\limits_{i',j',m'} 
  G^c_{ii'}(\omega)V_{i'j'}^{nm'}G_{j'j}^{m'm}(\omega)\,,
\end{equation}
where $V_{ij}^{nm} = -t (\delta_{n,m+1} \delta_{i0} \delta_{j,L-1} +
\delta_{n,m-1} \delta_{i,L-1} \delta_{j0})$ describes the inter-chain
hopping and upper indices number the different clusters. A partial
Fourier transform of the inter-chain hopping, $V_{ij}(Q)=-t(\E^{\I Q}
\delta_{i0} \delta_{j,L-1} + \E^{-\I Q} \delta_{i,L-1} \delta_{j0})$,
gives the infinite-lattice Green function in a mixed representation
\begin{equation}
  \hat{G}_{ij}(Q,\omega)=\left(
  \frac{G^c(\omega)}{1-V(Q)G^c(\omega)}\right)_{ij}
\end{equation}
for a momentum vector $Q$ of the super-lattice of finite chains and
cluster indices $i,j$.  Finally, from this mixed representation the
infinite lattice Green function in momentum space is recovered in the
CPT approximation as a simple Fourier transform
\begin{equation}
  G(k,\omega)=\frac{1}{L} \sum\limits_{i,j}
  \exp(\I (i-j)k)\,\hat{G}_{ij}(L k,\omega) \,.
\end{equation}

The reader should be aware that restoring translational symmetry in
the CPT sense is different from performing the thermodynamic limit of
the interacting system.  The CPT may be understood as a kind of
interpolation scheme from the discrete momentum space of a finite
cluster to the continuous $\vec{k}$-values of the infinite lattice.  The
amount of information attainable from the solution of a finite cluster
problem does however not increase.  Especially finite-size effects
affecting the interaction properties are by no means reduced, but
still determined through the size of the underlying cluster.
Nevertheless, CPT yields appealing presentations of the finite-cluster
data, which can ease its interpretation.

At present, all numerical studies within the CPT context use 
Lanczos recursion for the cluster diagonalization, thus suffering from
the shortcomings we discussed earlier. As an alternative, we prefer to
use the formalism introduced in Sec.~\ref{secgreen}, which is much
better suited for the calculation of spectral properties in a finite
energy interval.

\begin{figure}
  \includegraphics[width=\linewidth]{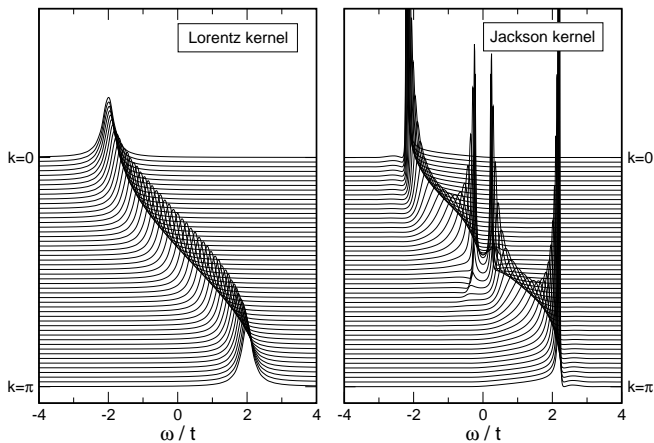}
  \caption{Spectral function for non-interacting tight-binding electrons. 
    Based on the Lorentz kernel CPT exactly reproduces the infinite
    system result (left), the Jackson kernel does not have the correct
    analytical properties, therefore CPT cannot close the finite size
    gap at $k=\pi/2$ (right). }\label{figcptkrnls}
\end{figure}
On applying the CPT crucial attention has to be paid to the kernel
used in the reconstruction of $G^c_{ij}(\omega)$.  As it turns out,
the Jackson kernel is an inadequate choice here, since already for the
non-interacting tight-binding model it introduces spurious structures
into the spectra. The failure can be attributed to the shape of the
Jackson kernel: Being optimized for high resolution, a pole in the
Green function will give a sharp peak with most of its weight
concentrated at the center, and rapidly decaying tails.  The
reconstructed (cluster) Green function therefore does not satisfy the
correct analytical properties required in the CPT step. To guarantee
these properties, instead, we use the Lorentz kernel, which we
constructed in Sec.~\ref{seclorentz} to mimic the effect of a
finite imaginary part in the energy argument of a Green function.
Using this kernel for the reconstruction of $G^c_{ij}(\omega)$ the CPT
works perfectly (cf. Figure~\ref{figcptkrnls}).

To provide further examples we present results for two different
interacting models where the cluster Green function $G^c_{ij}(\omega)$
has been calculated through a Chebyshev expansion as in
Eq.~\eqref{eqcorr}.  Using $G^c_{ij}(\omega) = G^c_{ji}(\omega)$ (no
magnetic field), for a $L$-site chain $L$ diagonal and $L(L-1)/2$
off-diagonal elements of $G^c_{ij}(\omega)$ have to be calculated.
The latter can be reduced to Chebyshev iterations for the operators
$c_i^{(\dagger)} + c_j^{(\dagger)}$, which allows application of the
``doubling trick'' (see the remark after Eq.~\eqref{muimcorr}).
However, the numerical effort can be further reduced by a factor
$1/L$: If we keep the ground state $|0\rangle$ of the system we can
calculate the moments $\mu^{ij}_n=\langle 0|c_i T_n(\tilde{H})
c^\dagger_j |0 \rangle$ for $L$ elements $i=1,\dots,L$ of
$G^c_{ij}(\omega)$ in a single Chebyshev iteration.  To achieve a
similar reduction within the Lanczos recursion we had to explicitly
construct the eigenstates to the Lanczos eigenvalues.  Then the factor
$1/L$ is exceeded by at least $ND$ additional operations for the
construction of $N$ eigenstates of a $D$-dimensional sparse matrix.
Hence using KPM for the CPT cluster diagonalization the numerical
effort can be reduced by a factor of $1/L$ in comparison to the
Lanczos recursion.

\subsubsection{CPT for the Hubbard model}
\begin{figure}
  \begin{center}
    \includegraphics[width=0.49\linewidth]{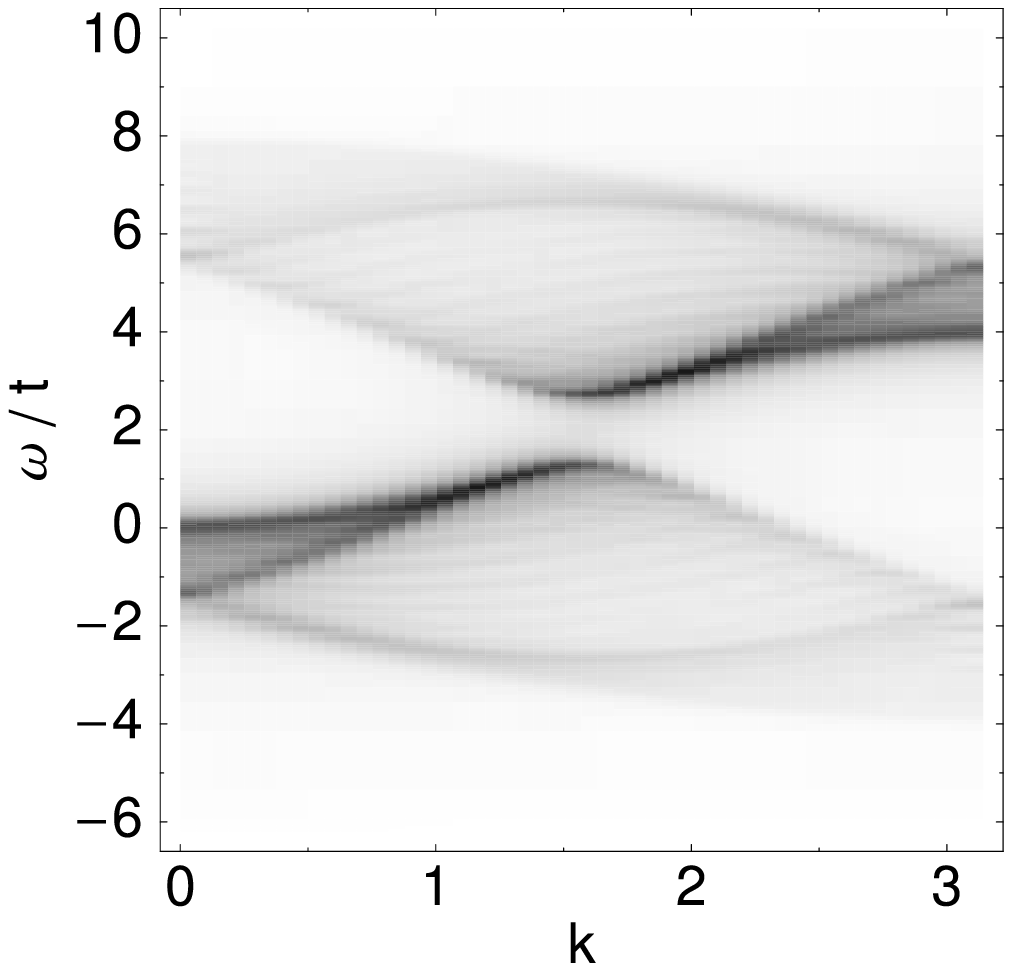}
    \includegraphics[width=0.49\linewidth]{figure14b.eps}
    \includegraphics[width=0.95\linewidth]{figure14c.eps}
  \end{center}
  \caption{\conline Spectral function of the 1D Hubbard model for half-filling
    and $U=4t$. Top left: CPT result with cluster size $L=16$ and
    expansion order $N=2048$. For similar data based on Lanczos
    recursion see~\textcite{SPP00}. Top right: Within the exact
    Bethe ansatz solution each electron separates into the sum of
    independent spinon (red dashed) and holon (green) excitations. The
    dots mark the energies of a 64-site chain.  Bottom: CPT data
    compared to selected DDMRG results for a system with $L=128$
    sites, open boundary conditions and a broadening of
    $\epsilon=0.0625 t$. Note that in DDMRG the momenta are
    approximate.}\label{figcpthub}
\end{figure}
As a first example we consider the 1D Hubbard model (Eq.~\eqref{hamhh}
with $g=\omega_0=0$), which is exactly solvable by Bethe
ansatz~\cite{EFGKK05} and was also extensively studied with
DDMRG~\cite{JGE00}. It thus provides the opportunity to assess the
precision of the KPM-based CPT. The top left panel of
Figure~\ref{figcpthub} shows the one-particle spectral function at
half-filling, calculated on the basis of $L=16$ site clusters and an
expansion order of $N=2048$. The matrix dimension is $D\approx
1.7\cdot 10^8$.  Remember that the cluster Green function is
calculated for a chain with open boundary conditions.  The reduced
symmetry compared to periodic boundary conditions results in a larger
dimension of the Hilbert space that has to be dealt with numerically.
In the top right panel the dots show the Bethe ansatz results for a
$L=64$ site chain, and the lines denote the $L\to\infty$ spinon and
holon excitations each electron separates into (spin-charge
separation).  So far the Bethe ansatz does not allow for a direct
calculation of the structure factor, the data thus represents only the
position and density of the eigenstates, but is not weighted with the
matrix elements of the operators $c_{k\sigma}^{(\dagger)}$. Although
for an infinite system we would expect a continuous response, the CPT
data shows some faint fine-structure. A comparison with the
finite-size Bethe ansatz data suggests that these features are an
artifact of the finite-cluster Greens function which the CPT spectral
function is based on. The fine-structure is also evident in the lower
panel of Figure~\ref{figcpthub}, where we compare with DDMRG data for
a $L=128$ site system. Otherwise the CPT nicely reproduces all
expected features, like the excitation gap, the two pronounced spinon
and holon branches, and the broad continuum. Note also, that CPT is
applicable to all spatial dimensions, whereas DDMRG works well only
for 1D models.

\subsubsection{CPT for the Holstein model}
\begin{figure}
  \includegraphics[width=\linewidth]{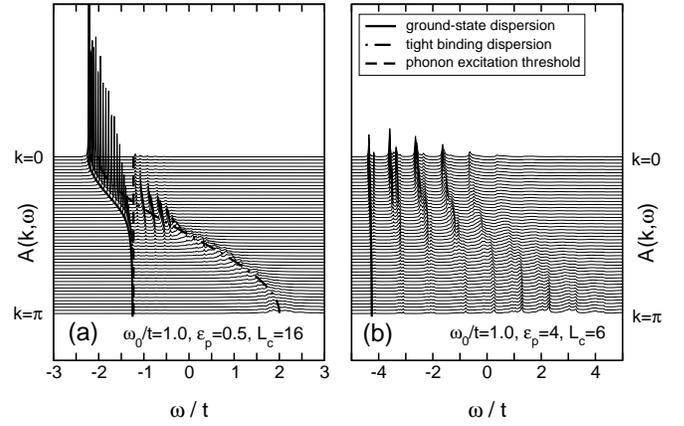}
  \caption{Spectral function $A^+(k,\omega)$ of a single electron 
    in the Holstein model (corresponding to $N_{\text{e}}=0$ in
    Eq.~\eqref{a+}). For weak electron-phonon coupling the original
    band is still very pronounced (left), for intermediate-to-strong
    coupling many narrow polaron bands develop (right). The cluster
    size is $L=16$ (left) or $L=6$ (right) and the expansion order
    $N=2048$. See~\textcite{HAL03} for similar data based on
    Lanczos recursion.}\label{figcpthol}
\end{figure}
Our second example is the spectral function of a single electron in
the Holstein model, i.e., Eq.~\eqref{hamhh} with $U=0$. Here, as a
function of the electron-phonon interaction, polaron formation sets in
and the band width of the resulting quasi particles becomes extremely
narrow at large coupling strength. Figure~\ref{figcpthol} illustrates
this behavior for two values of the electron-phonon coupling
$\varepsilon_p = g^2 \omega_0$. For weak coupling the original
one-electron band is still clearly visible (dot-dashed line), but the
dispersion-less phonon (dashed line) cuts in approximately at an
energy $\omega_0$ above the band minimum, causing the formation of a
polaron band (solid line; calculated with the approach of
\textcite{BTB99}), an avoided-crossing like gap and a number of
finite-size features. For strong coupling the spectral weight of the
electron is distributed over many narrow polaron bands separated
approximately by the bare phonon frequency $\omega_0$.

In all these cases, KPM works as a reliable high-resolution cluster
solver, and using the concepts from Sec.~\ref{secdyntgt0} we could
also extend these calculations to finite temperature. Probably, CPT is
not the only approximate technique that profits from the simplicity
and stability of KPM, and the range of its applications can certainly
be extended.

\section{KPM versus other numerical approaches}\label{secalt}

After we have given a very detailed description of the Kernel
Polynomial Method and presented a wide range of applications, let us
now classify the method in the context of numerical many-particle
techniques and comment on a number of other numerical approaches that
are closely related to KPM.

\subsection{KPM and dedicated many-particle techniques}

In the previous sections we already compared KPM data and results of
other numerical many-particle techniques. Nevertheless, it seems
appropriate to add a few comments about the general concept of such
calculations and the role KPM-like methods play in the field of
many-particle physics and complex quantum systems.  The numerical
study of interacting quantum many-particle systems is complicated by
the huge Hilbert space dimensions involved, which usually grow
exponentially with the number of particles or the system size. There
are different strategies to cope with this: In Monte Carlo approaches
only part of the Hilbert space is sampled stochastically, thereby
trying to capture the essential physics with an appropriate weighting
mechanism. On the other hand, variational methods, like
DMRG~\cite{Pe99,Sch05} or the specialized approach
of~\textcite{BTB99}, aim at reducing the Hilbert space dimension in an
intelligent way by discarding unimportant states, which, for instance,
contribute only at high temperature. Compared to such methods KPM is
much more basic: It is designed only for the fast and stable
calculation of the spectral properties of a given matrix and of
related correlations. Choosing a suitable Hilbert space or optimizing
the basis is the matter of the user or of external programs. It is
thus a more general approach, which can be used directly or embedded
into other methods, as we illustrated in the preceding section. Of
course, this simplicity and general applicability come at a certain
price: For interacting many-particle models the system sizes that can
be studied by using KPM directly are usually much smaller, compared to
DMRG and Monte Carlo.  Note however, that both of the latter methods
have limitations too: For many interesting models Monte Carlo methods
are plagued by the infamous sign problem, which is not present in KPM.
When it comes to the calculation of dynamical correlation functions
Monte Carlo approaches rely on power moments.  The reconstruction of
correlation functions from power moments is known to be an
ill-conditioned problem, in particular, if the moments are subject to
statistical noise. The resolution of Monte Carlo results is therefore
much smaller compared to the data obtained with KPM.  The DMRG method
develops its full potential only in one spatial dimension and for
short ranged interactions. In addition, the calculation of dynamical
correlations is limited to zero temperature, with only a few
exceptions~\cite{SK05}.  None of these restrictions apply to KPM.

\subsection{Close relatives of KPM}
Having compared KPM to specialized many-particle methods, let us now
discuss more direct competitors of KPM, i.e., methods that share the
broad application range and some of its general concepts.

\subsubsection{Chebyshev expansion and Maximum Entropy Methods}
\begin{figure}
  \includegraphics[width=\linewidth]{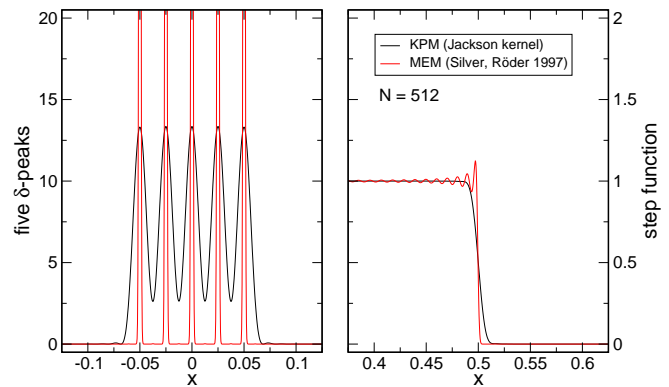}
  \caption{\conline Comparison of a KPM and a MEM approximation to a
    spectrum consisting of five isolated $\delta$-peaks, and to a step
    function. The expansion order is $N=512$. Clearly, for the
    $\delta$-peaks MEM yields a higher resolution, but for the step
    function the Gibbs oscillations return.}\label{figmem}
\end{figure}
The first of these approaches, the combination of Chebyshev expansion
and Maximum Entropy (MEM), is basically an alternative procedure to
transform moment data $\mu_n$ into convergent approximations of the
considered function $f(x)$. To achieve this, instead of (or in
addition to) applying kernel polynomials, an entropy
\begin{equation}
  S(f,f_0) = \int_{-1}^{1} (f(x) - f_0(x) - \log(f(x)/f_0(x)))\, dx
\end{equation}
is maximized under the constraint that the moments of the estimated
$f(x)$ agree with the given data. The function $f_0(x)$ describes our
initial knowledge about $f(x)$, and may in the worst case just be a
constant. Being related to Maximum Entropy approaches to the classical
moment problem~\cite{MP84,Tu88}, for the case of Chebyshev moments
different implementations of the method have been
suggested~\cite{Sk88,SR97,BBBD05}. Since for a given set of $N$
moments $\mu_n$ the approximation to the function $f(x)$ is usually
not restricted to a polynomial of degree $N-1$, compared to the KPM
with Jackson kernel the Maximum Entropy approach usually yields
estimates of higher resolution.  However, this higher resolution
results from adding a priori assumptions and not from a true
information gain (see also Figure~\ref{figmem}). The resource
consumption of Maximum Entropy is generally much higher than the
$N\log N$ behavior we found for KPM.  In addition, the approach is
non-linear in the moments and can occasionally become unstable for
large $N$. Note also that as yet Maximum Entropy methods have been
derived only for positive quantities, $f(x)>0$, such as densities of
states or strictly positive correlation functions.

Maximum Entropy, nevertheless, is a good alternative to KPM, if the
calculation of the $\mu_n$ is particularly time consuming. Based on
only a moderate number of moments it yields very detailed
approximations of $f(x)$, and we obtained very good results for some
computationally demanding problems~\cite{BWF98}.

\subsubsection{Lanczos recursion}

\begin{table*}
  \begin{ruledtabular}
  \begin{tabular}{p{0.35\linewidth}@{\hspace{0.03\linewidth}}c@{\hspace{0.03\linewidth}}p{0.35\linewidth}@{\hspace{0.03\linewidth}}c}
    Chebyshev / KPM & complexity & Lanczos recursion & complexity\\
    \hline
    Initialization: 
    \begin{gather*}
      \tilde H = (H-b)/a\\
      |\phi_0\rangle = A |0\rangle, \quad
      |\phi_1\rangle = \tilde H|\phi_0\rangle\\
      \mu_0 = \langle\phi_0|\phi_0\rangle, \quad
      \mu_1 = \langle\phi_1|\phi_0\rangle
    \end{gather*}
    & &
    Initialization: 
    \begin{gather*}
      \beta_0 = \sqrt{\langle 0| A^\dagger A |0\rangle}\\
      |\phi_0\rangle = A |0\rangle/\beta_0, \quad
      |\phi_{-1}\rangle = 0\\
    \end{gather*} 
    & \\
    \hline
    Recursion for $2N$ moments $\mu_n$:
    \begin{gather*}
      |\phi_{n+1}\rangle  = 2 \tilde H |\phi_{n}\rangle - |\phi_{n-1}\rangle \\
      \mu_{2n+2}  = 2\langle\phi_{n+1}|\phi_{n+1}\rangle - \mu_0 \\
      \mu_{2n+1} = 2\langle\phi_{n+1}|\phi_{n}\rangle - \mu_1
    \end{gather*}
    & $O(N D)$ &
    Recursion for $N$ coefficients $\alpha_n$, $\beta_n$:
    \begin{gather*}
      |\phi'\rangle = H|\phi_{n}\rangle-\beta_{n}|\phi_{n-1}\rangle,\quad
      \alpha_{n}  = \langle\phi_{n}|\phi'\rangle\\
      |\phi''\rangle = |\phi'\rangle-\alpha_{n}|\phi_{n}\rangle,\quad
      \beta_{n+1} = \sqrt{\langle\phi''|\phi''\rangle}\\
      |\phi_{n+1}\rangle  = |\phi''\rangle / \beta_{n+1}
    \end{gather*} 
    & $O(N D)$ \\
    $\to$ very stable & & 
    $\to$ tends to lose orthogonality\\
    \hline
    Reconstruction in three simple steps:
    \begin{description}
    \item[]Apply kernel: $\tilde\mu_n = g_n \mu_n$
    \item[]Fourier transform: $\tilde\mu_n \to \tilde f(\tilde\omega_i)$
    \item[]Rescale: $ f(\omega_i) = 
      \dfrac{\tilde f[(\omega_i-b)/a]}{\pi\sqrt{a^2-(\omega_i-b)^2}}$
    \end{description}
    & $O(M \log M)$ & 
    Reconstruction via continued fraction
    \begin{gather*}
      f(z) = -\frac{1}{\pi} \im 
      \cfrac{\beta_0^2}{z-\alpha_0-
        \cfrac{\beta_1^2}{z-\alpha_1-
          \cfrac{\beta_2^2}{z-\alpha_2-\ldots}}}\\
      \text{where } z = \omega_i + \I \epsilon
    \end{gather*} 
    & $O(NM)$ \\
    $\to$ procedure is linear in $\mu_n$ & & 
    $\to$ procedure is non-linear in $\alpha_n$, $\beta_n$\\
    $\to$ well defined resolution $\propto 1/N$ & & 
    $\to$ $\epsilon$ is somewhat arbitrary\\
  \end{tabular}
  \end{ruledtabular}
  \caption{Comparison of Chebyshev expansion and 
    Lanczos recursion for the calculation of a zero-temperature dynamical 
    correlation function $f(\omega) = \sum_{n} |\langle n|A|0\rangle|^2 
    \delta(\omega-\omega_n)$. We assume $N$ matrix vector multiplications with
    a $D$-dimensional sparse matrix $H$, and a reconstruction of 
    $f(\omega)$ at $M$ points $\omega_i$.}\label{tabkpmvslanc}
\end{table*}
The Lanczos Recursion Method is certainly the most capable competitor
of the Kernel Polynomial Method~\cite{Da94}. It is based on the Lanczos
algorithm~\cite{La50}, a method which was initially developed for the
tridiagonalization of Hermitian matrices and later evolved to one of
the most powerful methods for the calculation of extremal eigenstates
of sparse matrices~\cite{CW85}. Although ideas like the mapping of the
classical moment problem to tridiagonal matrices and continued
fractions have been suggested earlier~\cite{Go68}, the use of the
Lanczos algorithm for the characterization of spectral
densities~\cite{HHK72,HHK75} was first proposed at about the same time
as the Chebyshev expansion approaches, and in principle Lanczos
recursion is also a kind of modified moment
expansion~\cite{LG82,BRP92}. Its generalization from spectral
densities to zero temperature dynamical correlation functions was
first given in terms of continued fractions~\cite{GB87}, and later
also an approach based on the eigenstates of the tridiagonal matrix
was introduced and termed Spectral Decoding Method~\cite{ZSP94}.  This
technique was then generalized to finite temperature~\cite{JP94,JP00},
and, in addition, some variants of the approach for low
temperature~\cite{ADEL03} and based on the micro-canonical
ensemble~\cite{LPEKZ03} have been proposed recently.

To give an impression, in Table~\ref{tabkpmvslanc} we compare the
setup for the calculation of a zero temperature dynamical correlation
function within the Chebyshev and the Lanczos approach. The most time
consuming step for both methods is the recursive construction of a set
of vectors $|\phi_n\rangle$, which in terms of scalar products yield
the moments $\mu_n$ of the Chebyshev series or the elements
$\alpha_n$, $\beta_n$ of the Lanczos tridiagonal matrix. In terms of
the number of operations the Chebyshev recursion has a small
advantage, but, of course, the application of the Hamiltonian as the
dominant factor is the same for both methods. As a drawback, at high
expansion order the Lanczos iteration tends to lose the orthogonality
between the vectors $|\phi_n\rangle$, which it intends to establish by
construction. When the Lanczos algorithm is applied to eigenvalue
problems this loss of orthogonality usually signals the convergence of
extremal eigenstates, and the algorithm then starts to generate
artificial copies of the converged states. For the calculation of
spectral densities or correlation functions this means that the
information content of the $\alpha_n$ and $\beta_n$ does no longer
increase proportionally to the number of iterations. Unfortunately,
this deficiency can only be cured with more complex variants of the
algorithm, which also increase the resource consumption. Chebyshev
expansion is free from such defects, as there is a priori no
orthogonality between the $|\phi_n\rangle$.

The reconstruction of the considered function from its moments $\mu_n$
or coefficients $\alpha_n$, $\beta_n$, respectively, is also faster
and simpler within the KPM, as it makes use of Fast Fourier
Transformation. In addition, the KPM is a linear transformation of the
moments $\mu_n$, a property we used extensively above when averaging
moment data instead of the corresponding functions. Continued
fractions, in contrast, are non-linear in the coefficients $\alpha_n$,
$\beta_n$. A further advantage of KPM is our good understanding of its
convergence and resolution as a function of the expansion order~$N$.
For the Lanczos algorithm these issues have not been worked out with
the same rigor.

We therefore think that the Lanczos algorithm is an excellent tool for
the calculation of extremal eigenstates of large sparse matrices, but
for spectral densities and correlation functions the Kernel Polynomial
Method is the better choice. Of course, the advantages of both
algorithms can be combined, e.g. when the Chebyshev expansion starts
from an exact eigenstate that was calculated with the Lanczos
algorithm.

\subsubsection{Projection methods}
Projection methods were developed mainly in the context of electronic
structure calculations or tight-binding molecular dynamics, which both
require knowledge of the total energy of a non-interacting electron
system or of related expectation values~\cite{Or98,Go99}. The
starting point of these methods is the density matrix $F = f(H)$,
where $f(E)$ again represents the Fermi function. Thermal expectation
values, total energies and other quantities of interest are then 
expressed in terms of traces over $F$ and corresponding
operators~\cite{GC94b}.  For instance, the number of electrons and
their energy are given by $N_{\text{el}} = \trace(F)$ and $E =
\trace(FH)$, respectively. To obtain a numerical approach that is
linear in the dimension $D$ of $H$, $F$ is expanded as a series of
polynomials or other suitable functions in the Hamiltonian~$H$,
\begin{equation}\label{dmexp}
  F = \frac{1}{1+\E^{\beta(H-\mu)}} = \sum_{i=0}^{N-1} \alpha_i p_i(H)\,,
\end{equation}
and the above traces are replaced by averages over random vectors
$|r\rangle$. Chebyshev polynomials are a good basis for such an
expansion of $F$~\cite{GT95}, and the corresponding approaches are
thus closely related to the KPM setup we described in
Sec.~\ref{secdos}.  Note however, that the expansion in
Eq.~\eqref{dmexp} has to be repeated whenever the temperature
$1/\beta$ or the chemical potential $\mu$ is modified. This is
particularly inconvenient, if $\mu$ needs to be adjusted to fix the
electron density of the system. To compensate for this drawback, at
least partially, we can make use of the fact that in Eq.~\eqref{dmexp}
the expanded function and its expansion coefficients are known in
advance: Using implicit methods~\cite{Ni03} the order $N$
approximation of $F$ can be calculated with only $O(\log N)$ matrix
vector operations involving the Hamiltonian $H$. The total computation
time for one expansion is thus proportional to $D\log N$, compared to
$DN$ if the sum in Eq.~\eqref{dmexp} is evaluated iteratively, e.g.,
on the basis of the recursion relation Eq.~\eqref{chebrec1}.

\begin{figure}
  \includegraphics[width=\linewidth]{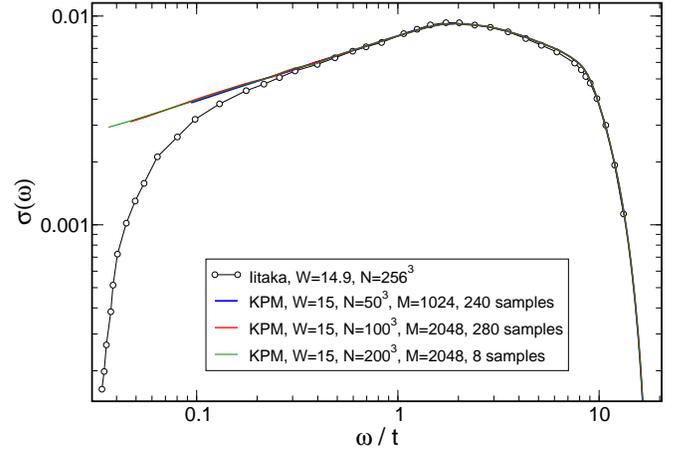}
  \caption{\conline The optical conductivity of the Anderson model,
    Eq.~\eqref{hamand}, calculated with KPM and a projection
    method~\cite{Ii98}. The disorder is $W\approx 15$; temperature and
    chemical potential read $T=0$ and $\mu=0$.}\label{figpromet}
\end{figure}
Projection methods can also be used for the calculation of dynamical
correlation functions. In this case the expansion of the density
matrix, which accounts for the thermodynamics, is supplemented by a
numerical time evolution. Hence, a general correlation function is
written as
\begin{equation}
  \langle A;B\rangle_\omega = 
  \lim_{\epsilon\to 0} \int\limits_{0}^{\infty} \E^{\I (\omega+\I\epsilon) t} 
  \trace(\E^{\I H t} A \E^{-\I H t} B F)\,dt\,,
\end{equation}
and the $\E^{\pm\I H t}$ terms are handled by standard methods, such
as Crank-Nicolson~\cite{PFTV86}, Suzuki-Trotter~\cite{VR93}, and,
very efficiently, Chebyshev expansion~\cite{DR03}. Of course, not only
the fermionic density matrix $F$ but also its interacting counterpart,
$\exp(-\beta H)$, can be expanded in polynomials, which leads to
similar methods for interacting quantum systems~\cite{IE03}.

To give an impression, in Figure~\ref{figpromet} we compare the
optical conductivity of the Anderson model calculated with KPM (see
Sec.~\ref{secoptand}) and with a projection approach~\cite{Ii98}.
Over a wide frequency range the data agrees very well, but at low
frequency the projection results deviate from both KPM and the
analytically expected power law $\sigma(\omega) - \sigma_0 \sim
\omega^\alpha$. Presumably this discrepancy is due to an insufficient
resolution or a too short time-integration interval. There is no
fundamental reason for the projection approach to fail here.

In summary, the projection methods have a similarly broad application
range as KPM, and can also compete in terms of numerical effort and
computation time. For finite-temperature dynamical correlations the
projection methods are characterized by a smaller memory consumption.
However, in contrast to KPM they require a new simulation for each
change in temperature or chemical potential, which represents their
major disadvantage.

\section{Conclusions \& Outlook}\label{secsum}
In this review we gave a detailed introduction to the Kernel
Polynomial Method, a numerical approach that on the basis of Chebyshev
expansion allows for an efficient calculation of the spectral
properties of large matrices and of the static and dynamic correlation
functions, which depend on them. The method has a wide range of
applications in different areas of physics and quantum chemistry, and
we illustrated its capability with numerous examples from solid state
physics, which covered such diverse topics as non-interacting
electrons in disordered media, quantum spin models, or strongly
correlated electron-phonon systems. Many of the considered quantities
are hardly accessible with other methods, or could previously be
studied only on smaller systems.  Comparing with alternative numerical
approaches, we demonstrated the advantages of KPM measured in terms of
general applicability, speed, resource consumption, algorithmic
simplicity and accuracy of the results.

Apart from further direct applications of the KPM outside the fields
of solid state physics and quantum chemistry, we think that the
combination of KPM with other numerical techniques will become one of
the major future research directions. Certainly not only classical MC
simulations and CPT, but potentially also other cluster
approaches~\cite{MJPH05} or quantum MC can profit from the concepts
outlined in this review.

\section*{Acknowledgements}
We thank A.~Basermann, B.~B\"auml, G.~Hager, M.~Hohenadler,
E.~Jeckelmann, M.~Kinateder, G.~Schubert, and in particular
R.N.~Silver for fruitful discussions and technical support. Most of
the calculations could only be performed with the generous grant of
resources by the John von Neumann-Institute for Computing (NIC
J\"ulich), the Leibniz-Rechenzentrum M\"unchen (LRZ), the High
Performance Computing Center Stuttgart (HLRS), the Norddeutscher
Verbund f\"ur Hoch- und H\"ochstleistungsrechnen (HLRN), the
Australian Partnership for Advanced Computing (APAC) and the
Australian Centre for Advanced Computing and Communications (ac3). In
addition, we are grateful for support by the Australian Research
Council, the Gordon Godfrey Bequest, and the Deutsche
Forschungsgemeinschaft through SFB~652.


\begin{thebibliography}{116}
\expandafter\ifx\csname natexlab\endcsname\relax\def\natexlab#1{#1}\fi
\expandafter\ifx\csname bibnamefont\endcsname\relax
  \def\bibnamefont#1{#1}\fi
\expandafter\ifx\csname bibfnamefont\endcsname\relax
  \def\bibfnamefont#1{#1}\fi
\expandafter\ifx\csname citenamefont\endcsname\relax
  \def\citenamefont#1{#1}\fi
\expandafter\ifx\csname url\endcsname\relax
  \def\url#1{\texttt{#1}}\fi
\expandafter\ifx\csname urlprefix\endcsname\relax\def\urlprefix{URL }\fi
\providecommand{\bibinfo}[2]{#2}
\providecommand{\eprint}[2][]{\url{#2}}

\bibitem[{\citenamefont{Abramowitz and Stegun}(1970)}]{AS70}
\bibinfo{editor}{\bibnamefont{Abramowitz}, \bibfnamefont{M.}}, and
  \bibinfo{editor}{\bibfnamefont{I.~A.} \bibnamefont{Stegun}} (eds.),
  \bibinfo{year}{1970}, \emph{\bibinfo{title}{Handbook of Mathematical
  Functions with formulas, graphs, and mathematical tables}}
  (\bibinfo{publisher}{Dover}, \bibinfo{address}{New York}).

\bibitem[{\citenamefont{Aichhorn} \emph{et~al.}(2003)\citenamefont{Aichhorn,
  Daghofer, Evertz, and von~der Linden}}]{ADEL03}
\bibinfo{author}{\bibnamefont{Aichhorn}, \bibfnamefont{M.}},
  \bibinfo{author}{\bibfnamefont{M.}~\bibnamefont{Daghofer}},
  \bibinfo{author}{\bibfnamefont{H.~G.} \bibnamefont{Evertz}}, and
  \bibinfo{author}{\bibfnamefont{W.}~\bibnamefont{von~der Linden}},
  \bibinfo{year}{2003}, \bibinfo{journal}{Phys. Rev. B}
  \textbf{\bibinfo{volume}{67}}, \bibinfo{pages}{161103}.

\bibitem[{\citenamefont{Alonso} \emph{et~al.}(2001)\citenamefont{Alonso,
  Fern{\'a}ndez, Guinea, Laliena, and Mart{\'i}n-Mayor}}]{AFGLM01}
\bibinfo{author}{\bibnamefont{Alonso}, \bibfnamefont{J.~L.}},
  \bibinfo{author}{\bibfnamefont{L.~A.} \bibnamefont{Fern{\'a}ndez}},
  \bibinfo{author}{\bibfnamefont{F.}~\bibnamefont{Guinea}},
  \bibinfo{author}{\bibfnamefont{V.}~\bibnamefont{Laliena}}, and
  \bibinfo{author}{\bibfnamefont{V.}~\bibnamefont{Mart{\'i}n-Mayor}},
  \bibinfo{year}{2001}, \bibinfo{journal}{Nucl. Phys. B}
  \textbf{\bibinfo{volume}{596}}, \bibinfo{pages}{587}.

\bibitem[{\citenamefont{Alvarez} \emph{et~al.}(2005)\citenamefont{Alvarez, Sen,
  Furukawa, Motome, and Dagotto}}]{ASFMD05}
\bibinfo{author}{\bibnamefont{Alvarez}, \bibfnamefont{G.}},
  \bibinfo{author}{\bibfnamefont{C.}~\bibnamefont{Sen}},
  \bibinfo{author}{\bibfnamefont{N.}~\bibnamefont{Furukawa}},
  \bibinfo{author}{\bibfnamefont{Y.}~\bibnamefont{Motome}}, and
  \bibinfo{author}{\bibfnamefont{E.}~\bibnamefont{Dagotto}},
  \bibinfo{year}{2005}, \bibinfo{journal}{Comp. Phys. Comm.}
  \textbf{\bibinfo{volume}{168}}, \bibinfo{pages}{32}.

\bibitem[{\citenamefont{Anderson}(1958)}]{An58}
\bibinfo{author}{\bibnamefont{Anderson}, \bibfnamefont{P.~W.}},
  \bibinfo{year}{1958}, \bibinfo{journal}{Phys. Rev.}
  \textbf{\bibinfo{volume}{109}}, \bibinfo{pages}{1492}.

\bibitem[{\citenamefont{Anderson and Hasegawa}(1955)}]{AH55}
\bibinfo{author}{\bibnamefont{Anderson}, \bibfnamefont{P.~W.}}, and
  \bibinfo{author}{\bibfnamefont{H.}~\bibnamefont{Hasegawa}},
  \bibinfo{year}{1955}, \bibinfo{journal}{Phys. Rev.}
  \textbf{\bibinfo{volume}{100}}, \bibinfo{pages}{675}.

\bibitem[{\citenamefont{Auerbach}(1994)}]{Au94}
\bibinfo{author}{\bibnamefont{Auerbach}, \bibfnamefont{A.}},
  \bibinfo{year}{1994}, \emph{\bibinfo{title}{Interacting Electrons and Quantum
  Magnetism}}, Graduate Texts in Contemporary Physics
  (\bibinfo{publisher}{Springer-Verlag}, \bibinfo{address}{Heidelberg}).

\bibitem[{\citenamefont{Bandyopadhyay}
  \emph{et~al.}(2005)\citenamefont{Bandyopadhyay, Bhattacharya, Biswas, and
  Drabold}}]{BBBD05}
\bibinfo{author}{\bibnamefont{Bandyopadhyay}, \bibfnamefont{K.}},
  \bibinfo{author}{\bibfnamefont{A.~K.} \bibnamefont{Bhattacharya}},
  \bibinfo{author}{\bibfnamefont{P.}~\bibnamefont{Biswas}}, and
  \bibinfo{author}{\bibfnamefont{D.~A.} \bibnamefont{Drabold}},
  \bibinfo{year}{2005}, \bibinfo{journal}{Phys. Rev. E}
  \textbf{\bibinfo{volume}{71}}, \bibinfo{pages}{057701}.

\bibitem[{\citenamefont{B{\"a}uml} \emph{et~al.}(1998)\citenamefont{B{\"a}uml,
  Wellein, and Fehske}}]{BWF98}
\bibinfo{author}{\bibnamefont{B{\"a}uml}, \bibfnamefont{B.}},
  \bibinfo{author}{\bibfnamefont{G.}~\bibnamefont{Wellein}}, and
  \bibinfo{author}{\bibfnamefont{H.}~\bibnamefont{Fehske}},
  \bibinfo{year}{1998}, \bibinfo{journal}{Phys. Rev. B}
  \textbf{\bibinfo{volume}{58}}, \bibinfo{pages}{3663}.

\bibitem[{\citenamefont{Benoit} \emph{et~al.}(1992)\citenamefont{Benoit, Royer,
  and Poussigue}}]{BRP92}
\bibinfo{author}{\bibnamefont{Benoit}, \bibfnamefont{C.}},
  \bibinfo{author}{\bibfnamefont{E.}~\bibnamefont{Royer}}, and
  \bibinfo{author}{\bibfnamefont{G.}~\bibnamefont{Poussigue}},
  \bibinfo{year}{1992}, \bibinfo{journal}{J. Phys. Condens. Matter}
  \textbf{\bibinfo{volume}{4}}, \bibinfo{pages}{3125}.

\bibitem[{\citenamefont{Blumstein and Wheeler}(1973)}]{BW73}
\bibinfo{author}{\bibnamefont{Blumstein}, \bibfnamefont{C.}}, and
  \bibinfo{author}{\bibfnamefont{J.~C.} \bibnamefont{Wheeler}},
  \bibinfo{year}{1973}, \bibinfo{journal}{Phys. Rev. B}
  \textbf{\bibinfo{volume}{8}}, \bibinfo{pages}{1764}.

\bibitem[{\citenamefont{Bon\v{c}a} \emph{et~al.}(1999)\citenamefont{Bon\v{c}a,
  Trugman, and Batisti\'{c}}}]{BTB99}
\bibinfo{author}{\bibnamefont{Bon\v{c}a}, \bibfnamefont{J.}},
  \bibinfo{author}{\bibfnamefont{S.~A.} \bibnamefont{Trugman}}, and
  \bibinfo{author}{\bibfnamefont{I.}~\bibnamefont{Batisti\'{c}}},
  \bibinfo{year}{1999}, \bibinfo{journal}{Phys. Rev. B}
  \textbf{\bibinfo{volume}{60}}, \bibinfo{pages}{1633}.

\bibitem[{\citenamefont{Boyd}(1989)}]{Bo89}
\bibinfo{author}{\bibnamefont{Boyd}, \bibfnamefont{J.~P.}},
  \bibinfo{year}{1989}, \emph{\bibinfo{title}{Chebyshev and Fourier Spectral
  Methods}}, number~\bibinfo{number}{49} in \bibinfo{series}{Lecture Notes in
  Engineering} (\bibinfo{publisher}{Springer-Verlag},
  \bibinfo{address}{Berlin}).

\bibitem[{\citenamefont{Chen and Guo}(1999)}]{CG99}
\bibinfo{author}{\bibnamefont{Chen}, \bibfnamefont{R.}}, and
  \bibinfo{author}{\bibfnamefont{H.}~\bibnamefont{Guo}}, \bibinfo{year}{1999},
  \bibinfo{journal}{Comp. Phys. Comm.} \textbf{\bibinfo{volume}{119}},
  \bibinfo{pages}{19}.

\bibitem[{\citenamefont{Cheney}(1966)}]{Ch66}
\bibinfo{author}{\bibnamefont{Cheney}, \bibfnamefont{E.~W.}},
  \bibinfo{year}{1966}, \emph{\bibinfo{title}{Introduction to Approximation
  Theory}} (\bibinfo{publisher}{McGraw-Hill}, \bibinfo{address}{New York}).

\bibitem[{\citenamefont{Coey} \emph{et~al.}(1999)\citenamefont{Coey, Viret, and
  von Moln{\'a}r}}]{CVM99}
\bibinfo{author}{\bibnamefont{Coey}, \bibfnamefont{J.~M.~D.}},
  \bibinfo{author}{\bibfnamefont{M.}~\bibnamefont{Viret}}, and
  \bibinfo{author}{\bibfnamefont{S.}~\bibnamefont{von Moln{\'a}r}},
  \bibinfo{year}{1999}, \bibinfo{journal}{Adv. Phys.}
  \textbf{\bibinfo{volume}{48}}, \bibinfo{pages}{167}.

\bibitem[{\citenamefont{Cullum and Willoughby}(1985)}]{CW85}
\bibinfo{author}{\bibnamefont{Cullum}, \bibfnamefont{J.~K.}}, and
  \bibinfo{author}{\bibfnamefont{R.~A.} \bibnamefont{Willoughby}},
  \bibinfo{year}{1985}, \emph{\bibinfo{title}{Lanczos Algorithms for Large
  Symmetric Eigenvalue Computations}}, volume \bibinfo{volume}{I \& II}
  (\bibinfo{publisher}{Birkh\"auser}, \bibinfo{address}{Boston}).

\bibitem[{\citenamefont{Dagotto}(1994)}]{Da94}
\bibinfo{author}{\bibnamefont{Dagotto}, \bibfnamefont{E.}},
  \bibinfo{year}{1994}, \bibinfo{journal}{Rev. Mod. Phys.}
  \textbf{\bibinfo{volume}{66}}, \bibinfo{pages}{763}.

\bibitem[{\citenamefont{Dagotto}(2003)}]{Da03}
\bibinfo{author}{\bibnamefont{Dagotto}, \bibfnamefont{E.}},
  \bibinfo{year}{2003}, \emph{\bibinfo{title}{Nanoscale Phase Separation and
  Colossal Magnetoresistance: The Physics of Manganites and Related
  Compounds}}, volume \bibinfo{volume}{136} of \emph{\bibinfo{series}{Springer
  Series in Solid-State Sciences}} (\bibinfo{publisher}{Springer},
  \bibinfo{address}{Heidelberg}).

\bibitem[{\citenamefont{Dobrosavljevi\'{c} and Kotliar}(1997)}]{DK97}
\bibinfo{author}{\bibnamefont{Dobrosavljevi\'{c}}, \bibfnamefont{V.}}, and
  \bibinfo{author}{\bibfnamefont{G.}~\bibnamefont{Kotliar}},
  \bibinfo{year}{1997}, \bibinfo{journal}{Phys. Rev. Lett.}
  \textbf{\bibinfo{volume}{78}}, \bibinfo{pages}{3943}.

\bibitem[{\citenamefont{Dobrosavljevi\'{c} and Kotliar}(1998)}]{DK98}
\bibinfo{author}{\bibnamefont{Dobrosavljevi\'{c}}, \bibfnamefont{V.}}, and
  \bibinfo{author}{\bibfnamefont{G.}~\bibnamefont{Kotliar}},
  \bibinfo{year}{1998}, \bibinfo{journal}{Philos. Trans. Roy. Soc. Lond., Ser.
  A} \textbf{\bibinfo{volume}{356}}, \bibinfo{pages}{57}.

\bibitem[{\citenamefont{Dobrovitski and De~Raedt}(2003)}]{DR03}
\bibinfo{author}{\bibnamefont{Dobrovitski}, \bibfnamefont{V.~V.}}, and
  \bibinfo{author}{\bibfnamefont{H.}~\bibnamefont{De~Raedt}},
  \bibinfo{year}{2003}, \bibinfo{journal}{Phys. Rev. E}
  \textbf{\bibinfo{volume}{67}}, \bibinfo{pages}{056702}.

\bibitem[{\citenamefont{Drabold and Sankey}(1993)}]{DS93}
\bibinfo{author}{\bibnamefont{Drabold}, \bibfnamefont{D.~A.}}, and
  \bibinfo{author}{\bibfnamefont{O.~F.} \bibnamefont{Sankey}},
  \bibinfo{year}{1993}, \bibinfo{journal}{Phys. Rev. Lett.}
  \textbf{\bibinfo{volume}{70}}, \bibinfo{pages}{3631}.

\bibitem[{\citenamefont{Essler} \emph{et~al.}(2005)\citenamefont{Essler, Frahm,
  G{\"o}hmann, Kl{\"u}mper, and Korepin}}]{EFGKK05}
\bibinfo{author}{\bibnamefont{Essler}, \bibfnamefont{F.~H.~L.}},
  \bibinfo{author}{\bibfnamefont{H.}~\bibnamefont{Frahm}},
  \bibinfo{author}{\bibfnamefont{F.}~\bibnamefont{G{\"o}hmann}},
  \bibinfo{author}{\bibfnamefont{A.}~\bibnamefont{Kl{\"u}mper}}, and
  \bibinfo{author}{\bibfnamefont{V.~E.} \bibnamefont{Korepin}},
  \bibinfo{year}{2005}, \emph{\bibinfo{title}{The One-Dimensional Hubbard
  Model}} (\bibinfo{publisher}{Cambridge University Press},
  \bibinfo{address}{Cambridge}).

\bibitem[{\citenamefont{Fabricius and McCoy}(1999)}]{FM99}
\bibinfo{author}{\bibnamefont{Fabricius}, \bibfnamefont{K.}}, and
  \bibinfo{author}{\bibfnamefont{B.~M.} \bibnamefont{McCoy}},
  \bibinfo{year}{1999}, \bibinfo{journal}{Phys. Rev. B}
  \textbf{\bibinfo{volume}{59}}, \bibinfo{pages}{381}.

\bibitem[{\citenamefont{Fehske} \emph{et~al.}(2000)\citenamefont{Fehske,
  Schindelin, Wei{\ss}e, B{\"u}ttner, and Ihle}}]{FSWBI00}
\bibinfo{author}{\bibnamefont{Fehske}, \bibfnamefont{H.}},
  \bibinfo{author}{\bibfnamefont{C.}~\bibnamefont{Schindelin}},
  \bibinfo{author}{\bibfnamefont{A.}~\bibnamefont{Wei{\ss}e}},
  \bibinfo{author}{\bibfnamefont{H.}~\bibnamefont{B{\"u}ttner}}, and
  \bibinfo{author}{\bibfnamefont{D.}~\bibnamefont{Ihle}}, \bibinfo{year}{2000},
  \bibinfo{journal}{Brazil. Jour. Phys.} \textbf{\bibinfo{volume}{30}},
  \bibinfo{pages}{720}.

\bibitem[{\citenamefont{Fehske} \emph{et~al.}(2004)\citenamefont{Fehske,
  Wellein, Hager, Wei{\ss}e, and Bishop}}]{FWHWB04}
\bibinfo{author}{\bibnamefont{Fehske}, \bibfnamefont{H.}},
  \bibinfo{author}{\bibfnamefont{G.}~\bibnamefont{Wellein}},
  \bibinfo{author}{\bibfnamefont{G.}~\bibnamefont{Hager}},
  \bibinfo{author}{\bibfnamefont{A.}~\bibnamefont{Wei{\ss}e}}, and
  \bibinfo{author}{\bibfnamefont{A.~R.} \bibnamefont{Bishop}},
  \bibinfo{year}{2004}, \bibinfo{journal}{Phys. Rev. B}
  \textbf{\bibinfo{volume}{69}}, \bibinfo{pages}{165115}.

\bibitem[{\citenamefont{Fehske} \emph{et~al.}(2002)\citenamefont{Fehske,
  Wellein, Kampf, Sekania, Hager, Wei{\ss}e, B{\"u}ttner, and
  Bishop}}]{Feea02b}
\bibinfo{author}{\bibnamefont{Fehske}, \bibfnamefont{H.}},
  \bibinfo{author}{\bibfnamefont{G.}~\bibnamefont{Wellein}},
  \bibinfo{author}{\bibfnamefont{A.~P.} \bibnamefont{Kampf}},
  \bibinfo{author}{\bibfnamefont{M.}~\bibnamefont{Sekania}},
  \bibinfo{author}{\bibfnamefont{G.}~\bibnamefont{Hager}},
  \bibinfo{author}{\bibfnamefont{A.}~\bibnamefont{Wei{\ss}e}},
  \bibinfo{author}{\bibfnamefont{H.}~\bibnamefont{B{\"u}ttner}}, and
  \bibinfo{author}{\bibfnamefont{A.~R.} \bibnamefont{Bishop}},
  \bibinfo{year}{2002}, in \emph{\bibinfo{booktitle}{High Performance Computing
  in Science and Engineering, Munich 2002}}, edited by
  \bibinfo{editor}{\bibfnamefont{S.}~\bibnamefont{Wagner}},
  \bibinfo{editor}{\bibfnamefont{W.}~\bibnamefont{Hanke}},
  \bibinfo{editor}{\bibfnamefont{A.}~\bibnamefont{Bode}}, and
  \bibinfo{editor}{\bibfnamefont{F.}~\bibnamefont{Durst}}
  (\bibinfo{publisher}{Springer-Verlag}, \bibinfo{address}{Heidelberg}), pp.
  \bibinfo{pages}{339--350}.

\bibitem[{\citenamefont{Fej{\'e}r}(1904)}]{Fe04}
\bibinfo{author}{\bibnamefont{Fej{\'e}r}, \bibfnamefont{L.}},
  \bibinfo{year}{1904}, \bibinfo{journal}{Math. Ann.}
  \textbf{\bibinfo{volume}{58}}, \bibinfo{pages}{51}.

\bibitem[{\citenamefont{Frigo and Johnson}(2005{\natexlab{a}})}]{FFTW05}
\bibinfo{author}{\bibnamefont{Frigo}, \bibfnamefont{M.}}, and
  \bibinfo{author}{\bibfnamefont{S.~G.} \bibnamefont{Johnson}},
  \bibinfo{year}{2005}{\natexlab{a}}, \bibinfo{journal}{Proceedings of the
  IEEE} \textbf{\bibinfo{volume}{93}}(\bibinfo{number}{2}),
  \bibinfo{pages}{216}, \bibinfo{note}{special issue on "Program Generation,
  Optimization, and Platform Adaptation"}.

\bibitem[{\citenamefont{Frigo and Johnson}(2005{\natexlab{b}})}]{FFTW}
\bibinfo{author}{\bibnamefont{Frigo}, \bibfnamefont{M.}}, and
  \bibinfo{author}{\bibfnamefont{S.~G.} \bibnamefont{Johnson}},
  \bibinfo{year}{2005}{\natexlab{b}}, \bibinfo{title}{{FFTW} fast fourier
  transform library}, \urlprefix\url{http://www.fftw.org/}.

\bibitem[{\citenamefont{Furukawa and Motome}(2004)}]{FM04}
\bibinfo{author}{\bibnamefont{Furukawa}, \bibfnamefont{N.}}, and
  \bibinfo{author}{\bibfnamefont{Y.}~\bibnamefont{Motome}},
  \bibinfo{year}{2004}, \bibinfo{journal}{J. Phys. Soc. Jpn.}
  \textbf{\bibinfo{volume}{73}}, \bibinfo{pages}{1482}.

\bibitem[{\citenamefont{Gagliano and Balseiro}(1987)}]{GB87}
\bibinfo{author}{\bibnamefont{Gagliano}, \bibfnamefont{E.}}, and
  \bibinfo{author}{\bibfnamefont{C.}~\bibnamefont{Balseiro}},
  \bibinfo{year}{1987}, \bibinfo{journal}{Phys. Rev. Lett.}
  \textbf{\bibinfo{volume}{59}}, \bibinfo{pages}{2999}.

\bibitem[{\citenamefont{Garrett} \emph{et~al.}(1997)\citenamefont{Garrett,
  Nagler, Tennant, Sales, and Barnes}}]{Gaea97}
\bibinfo{author}{\bibnamefont{Garrett}, \bibfnamefont{A.~W.}},
  \bibinfo{author}{\bibfnamefont{S.~E.} \bibnamefont{Nagler}},
  \bibinfo{author}{\bibfnamefont{D.}~\bibnamefont{Tennant}},
  \bibinfo{author}{\bibfnamefont{B.~C.} \bibnamefont{Sales}}, and
  \bibinfo{author}{\bibfnamefont{T.}~\bibnamefont{Barnes}},
  \bibinfo{year}{1997}, \bibinfo{journal}{Phys. Rev. Lett.}
  \textbf{\bibinfo{volume}{79}}, \bibinfo{pages}{745}.

\bibitem[{\citenamefont{Gautschi}(1968)}]{Ga68}
\bibinfo{author}{\bibnamefont{Gautschi}, \bibfnamefont{W.}},
  \bibinfo{year}{1968}, \bibinfo{journal}{Math. Comp.}
  \textbf{\bibinfo{volume}{22}}, \bibinfo{pages}{251}.

\bibitem[{\citenamefont{Gautschi}(1970)}]{Ga70}
\bibinfo{author}{\bibnamefont{Gautschi}, \bibfnamefont{W.}},
  \bibinfo{year}{1970}, \bibinfo{journal}{Math. Comp.}
  \textbf{\bibinfo{volume}{24}}, \bibinfo{pages}{245}.

\bibitem[{\citenamefont{Goedecker}(1999)}]{Go99}
\bibinfo{author}{\bibnamefont{Goedecker}, \bibfnamefont{S.}},
  \bibinfo{year}{1999}, \bibinfo{journal}{Rev. Mod. Phys.}
  \textbf{\bibinfo{volume}{71}}, \bibinfo{pages}{1085}.

\bibitem[{\citenamefont{Goedecker and Colombo}(1994)}]{GC94b}
\bibinfo{author}{\bibnamefont{Goedecker}, \bibfnamefont{S.}}, and
  \bibinfo{author}{\bibfnamefont{L.}~\bibnamefont{Colombo}},
  \bibinfo{year}{1994}, \bibinfo{journal}{Phys. Rev. Lett.}
  \textbf{\bibinfo{volume}{73}}, \bibinfo{pages}{122}.

\bibitem[{\citenamefont{Goedecker and Teter}(1995)}]{GT95}
\bibinfo{author}{\bibnamefont{Goedecker}, \bibfnamefont{S.}}, and
  \bibinfo{author}{\bibfnamefont{M.}~\bibnamefont{Teter}},
  \bibinfo{year}{1995}, \bibinfo{journal}{Phys. Rev. B}
  \textbf{\bibinfo{volume}{51}}, \bibinfo{pages}{9455}.

\bibitem[{\citenamefont{Gordon}(1968)}]{Go68}
\bibinfo{author}{\bibnamefont{Gordon}, \bibfnamefont{R.~G.}},
  \bibinfo{year}{1968}, \bibinfo{journal}{J. Math. Phys.}
  \textbf{\bibinfo{volume}{9}}, \bibinfo{pages}{655}.

\bibitem[{\citenamefont{Gros and Valent{\'i}}(1994)}]{GV94}
\bibinfo{author}{\bibnamefont{Gros}, \bibfnamefont{C.}}, and
  \bibinfo{author}{\bibfnamefont{R.}~\bibnamefont{Valent{\'i}}},
  \bibinfo{year}{1994}, \bibinfo{journal}{Ann. Phys. (Leipzig)}
  \textbf{\bibinfo{volume}{3}}, \bibinfo{pages}{460}.

\bibitem[{\citenamefont{Haydock} \emph{et~al.}(1972)\citenamefont{Haydock,
  Heine, and Kelly}}]{HHK72}
\bibinfo{author}{\bibnamefont{Haydock}, \bibfnamefont{R.}},
  \bibinfo{author}{\bibfnamefont{V.}~\bibnamefont{Heine}}, and
  \bibinfo{author}{\bibfnamefont{M.~J.} \bibnamefont{Kelly}},
  \bibinfo{year}{1972}, \bibinfo{journal}{J. Phys. C}
  \textbf{\bibinfo{volume}{5}}, \bibinfo{pages}{2845}.

\bibitem[{\citenamefont{Haydock} \emph{et~al.}(1975)\citenamefont{Haydock,
  Heine, and Kelly}}]{HHK75}
\bibinfo{author}{\bibnamefont{Haydock}, \bibfnamefont{R.}},
  \bibinfo{author}{\bibfnamefont{V.}~\bibnamefont{Heine}}, and
  \bibinfo{author}{\bibfnamefont{M.~J.} \bibnamefont{Kelly}},
  \bibinfo{year}{1975}, \bibinfo{journal}{J. Phys. C}
  \textbf{\bibinfo{volume}{8}}, \bibinfo{pages}{2591}.

\bibitem[{\citenamefont{Hohenadler}
  \emph{et~al.}(2003)\citenamefont{Hohenadler, Aichhorn, and von~der
  Linden}}]{HAL03}
\bibinfo{author}{\bibnamefont{Hohenadler}, \bibfnamefont{M.}},
  \bibinfo{author}{\bibfnamefont{M.}~\bibnamefont{Aichhorn}}, and
  \bibinfo{author}{\bibfnamefont{W.}~\bibnamefont{von~der Linden}},
  \bibinfo{year}{2003}, \bibinfo{journal}{Phys. Rev. B}
  \textbf{\bibinfo{volume}{68}}, \bibinfo{pages}{184304}.

\bibitem[{\citenamefont{Hohenadler}
  \emph{et~al.}(2005)\citenamefont{Hohenadler, Neuber, von~der Linden, Wellein,
  Loos, and Fehske}}]{Hoea05}
\bibinfo{author}{\bibnamefont{Hohenadler}, \bibfnamefont{M.}},
  \bibinfo{author}{\bibfnamefont{D.}~\bibnamefont{Neuber}},
  \bibinfo{author}{\bibfnamefont{W.}~\bibnamefont{von~der Linden}},
  \bibinfo{author}{\bibfnamefont{G.}~\bibnamefont{Wellein}},
  \bibinfo{author}{\bibfnamefont{J.}~\bibnamefont{Loos}}, and
  \bibinfo{author}{\bibfnamefont{H.}~\bibnamefont{Fehske}},
  \bibinfo{year}{2005}, \bibinfo{journal}{Phys. Rev. B}
  \textbf{\bibinfo{volume}{71}}, \bibinfo{pages}{245111}.

\bibitem[{\citenamefont{Holstein}(1959{\natexlab{a}})}]{Ho59a}
\bibinfo{author}{\bibnamefont{Holstein}, \bibfnamefont{T.}},
  \bibinfo{year}{1959}{\natexlab{a}}, \bibinfo{journal}{Ann. Phys. (N.Y.)}
  \textbf{\bibinfo{volume}{8}}, \bibinfo{pages}{325}.

\bibitem[{\citenamefont{Holstein}(1959{\natexlab{b}})}]{Ho59b}
\bibinfo{author}{\bibnamefont{Holstein}, \bibfnamefont{T.}},
  \bibinfo{year}{1959}{\natexlab{b}}, \bibinfo{journal}{Ann. Phys. (N.Y.)}
  \textbf{\bibinfo{volume}{8}}, \bibinfo{pages}{343}.

\bibitem[{\citenamefont{Iitaka}(1998)}]{Ii98}
\bibinfo{author}{\bibnamefont{Iitaka}, \bibfnamefont{T.}},
  \bibinfo{year}{1998}, in \emph{\bibinfo{booktitle}{High Performance Computing
  in {RIKEN} 1997}} (\bibinfo{publisher}{Inst. Phys. Chem. Res. ({RIKEN})},
  \bibinfo{address}{Japan}), volume~\bibinfo{volume}{19} of
  \emph{\bibinfo{series}{{RIKEN} Review}}, pp. \bibinfo{pages}{136--143}.

\bibitem[{\citenamefont{Iitaka and Ebisuzaki}(2003)}]{IE03}
\bibinfo{author}{\bibnamefont{Iitaka}, \bibfnamefont{T.}}, and
  \bibinfo{author}{\bibfnamefont{T.}~\bibnamefont{Ebisuzaki}},
  \bibinfo{year}{2003}, \bibinfo{journal}{Phys. Rev. Lett.}
  \textbf{\bibinfo{volume}{90}}, \bibinfo{pages}{047203}.

\bibitem[{\citenamefont{Iitaka and Ebisuzaki}(2004)}]{IE04}
\bibinfo{author}{\bibnamefont{Iitaka}, \bibfnamefont{T.}}, and
  \bibinfo{author}{\bibfnamefont{T.}~\bibnamefont{Ebisuzaki}},
  \bibinfo{year}{2004}, \bibinfo{journal}{Phys. Rev. E}
  \textbf{\bibinfo{volume}{69}}, \bibinfo{pages}{057701}.

\bibitem[{\citenamefont{Jackson}(1911)}]{Ja11}
\bibinfo{author}{\bibnamefont{Jackson}, \bibfnamefont{D.}},
  \bibinfo{year}{1911}, \emph{\bibinfo{title}{{{\"U}ber die Genauigkeit der
  Ann{\"a}herung stetiger Funktionen durch ganze rationale Funktionen gegebenen
  Grades und trigonometrische Summen gegebener Ordnung}}}, Ph.D. thesis,
  \bibinfo{school}{Georg-August-Universit{\"a}t G{\"o}ttingen}.

\bibitem[{\citenamefont{Jackson}(1912)}]{Ja12}
\bibinfo{author}{\bibnamefont{Jackson}, \bibfnamefont{D.}},
  \bibinfo{year}{1912}, \bibinfo{journal}{Trans. Amer. Math. Soc.}
  \textbf{\bibinfo{volume}{13}}, \bibinfo{pages}{491}.

\bibitem[{\citenamefont{Jakli\v{c} and Prelov\v{s}ek}(1994)}]{JP94}
\bibinfo{author}{\bibnamefont{Jakli\v{c}}, \bibfnamefont{J.}}, and
  \bibinfo{author}{\bibfnamefont{P.}~\bibnamefont{Prelov\v{s}ek}},
  \bibinfo{year}{1994}, \bibinfo{journal}{Phys. Rev. B}
  \textbf{\bibinfo{volume}{49}}, \bibinfo{pages}{5065}.

\bibitem[{\citenamefont{Jakli\v{c} and Prelov\v{s}ek}(2000)}]{JP00}
\bibinfo{author}{\bibnamefont{Jakli\v{c}}, \bibfnamefont{J.}}, and
  \bibinfo{author}{\bibfnamefont{P.}~\bibnamefont{Prelov\v{s}ek}},
  \bibinfo{year}{2000}, \bibinfo{journal}{Adv. Phys.}
  \textbf{\bibinfo{volume}{49}}, \bibinfo{pages}{1}.

\bibitem[{\citenamefont{Janke}(1998)}]{Ja98b}
\bibinfo{author}{\bibnamefont{Janke}, \bibfnamefont{W.}}, \bibinfo{year}{1998},
  \bibinfo{journal}{Math. and Comput. in Simul.} \textbf{\bibinfo{volume}{47}},
  \bibinfo{pages}{329}.

\bibitem[{\citenamefont{Jeckelmann}(2002)}]{Je02b}
\bibinfo{author}{\bibnamefont{Jeckelmann}, \bibfnamefont{E.}},
  \bibinfo{year}{2002}, \bibinfo{journal}{Phys. Rev. B}
  \textbf{\bibinfo{volume}{66}}, \bibinfo{pages}{045114}.

\bibitem[{\citenamefont{Jeckelmann and Fehske}(2006)}]{JF05p}
\bibinfo{author}{\bibnamefont{Jeckelmann}, \bibfnamefont{E.}}, and
  \bibinfo{author}{\bibfnamefont{H.}~\bibnamefont{Fehske}},
  \bibinfo{year}{2006}, in \emph{\bibinfo{booktitle}{Polarons in Bulk Materials
  and Systems with Reduced Dimensionality}}, edited by
  \bibinfo{editor}{\bibfnamefont{G.}~\bibnamefont{Iadonisi}},
  \bibinfo{editor}{\bibfnamefont{J.}~\bibnamefont{Ranninger}}, and
  \bibinfo{editor}{\bibfnamefont{G.~D.} \bibnamefont{Filippis}}
  (\bibinfo{publisher}{IOS Press}, \bibinfo{address}{Amsterdam}), volume
  \bibinfo{volume}{161} of \emph{\bibinfo{series}{International School of
  Phyics Enrico Fermi}}, p.~\bibinfo{pages}{?}, \bibinfo{note}{in press, see
  also \url{http://arXiv.org/abs/cond-mat/0510637}}.

\bibitem[{\citenamefont{Jeckelmann}
  \emph{et~al.}(2000)\citenamefont{Jeckelmann, Gebhard, and Essler}}]{JGE00}
\bibinfo{author}{\bibnamefont{Jeckelmann}, \bibfnamefont{E.}},
  \bibinfo{author}{\bibfnamefont{F.}~\bibnamefont{Gebhard}}, and
  \bibinfo{author}{\bibfnamefont{F.~H.~L.} \bibnamefont{Essler}},
  \bibinfo{year}{2000}, \bibinfo{journal}{Phys. Rev. Lett.}
  \textbf{\bibinfo{volume}{85}}, \bibinfo{pages}{3910}.

\bibitem[{\citenamefont{Kogan and Auslender}(1988)}]{KA88}
\bibinfo{author}{\bibnamefont{Kogan}, \bibfnamefont{E.~M.}}, and
  \bibinfo{author}{\bibfnamefont{M.~I.} \bibnamefont{Auslender}},
  \bibinfo{year}{1988}, \bibinfo{journal}{Phys. Status Solidi B}
  \textbf{\bibinfo{volume}{147}}, \bibinfo{pages}{613}.

\bibitem[{\citenamefont{Korovkin}(1959)}]{Ko59}
\bibinfo{author}{\bibnamefont{Korovkin}, \bibfnamefont{P.~P.}},
  \bibinfo{year}{1959}, \emph{\bibinfo{title}{Linejnye Operatory i teorija
  priblizenij}} (\bibinfo{publisher}{Gos. Izd. Fiziko-Matematiceskoj
  Literatury}, \bibinfo{address}{Moscow}).

\bibitem[{\citenamefont{Kosloff}(1988)}]{Ko88}
\bibinfo{author}{\bibnamefont{Kosloff}, \bibfnamefont{R.}},
  \bibinfo{year}{1988}, \bibinfo{journal}{J. Phys. Chem.}
  \textbf{\bibinfo{volume}{92}}, \bibinfo{pages}{2087}.

\bibitem[{\citenamefont{Kramer and Mac~Kinnon}(1993)}]{KM93b}
\bibinfo{author}{\bibnamefont{Kramer}, \bibfnamefont{B.}}, and
  \bibinfo{author}{\bibfnamefont{A.}~\bibnamefont{Mac~Kinnon}},
  \bibinfo{year}{1993}, \bibinfo{journal}{Rep. Prog. Phys.}
  \textbf{\bibinfo{volume}{56}}, \bibinfo{pages}{1469}.

\bibitem[{\citenamefont{Krauth}(2004)}]{Kr04}
\bibinfo{author}{\bibnamefont{Krauth}, \bibfnamefont{W.}},
  \bibinfo{year}{2004}, in \emph{\bibinfo{booktitle}{New Optimization
  Algorithms in Physics}}, edited by \bibinfo{editor}{\bibfnamefont{A.~K.}
  \bibnamefont{Hartmann}} and
  \bibinfo{editor}{\bibfnamefont{H.}~\bibnamefont{Rieger}}
  (\bibinfo{publisher}{Wiley-VCH}, \bibinfo{address}{Berlin}),
  chapter~\bibinfo{chapter}{2}, pp. \bibinfo{pages}{7--22}.

\bibitem[{\citenamefont{Lambin and Gaspard}(1982)}]{LG82}
\bibinfo{author}{\bibnamefont{Lambin}, \bibfnamefont{P.}}, and
  \bibinfo{author}{\bibfnamefont{J.-P.} \bibnamefont{Gaspard}},
  \bibinfo{year}{1982}, \bibinfo{journal}{Phys. Rev. B}
  \textbf{\bibinfo{volume}{26}}, \bibinfo{pages}{4356}.

\bibitem[{\citenamefont{Lanczos}(1950)}]{La50}
\bibinfo{author}{\bibnamefont{Lanczos}, \bibfnamefont{C.}},
  \bibinfo{year}{1950}, \bibinfo{journal}{J. Res. Nat. Bur. Stand.}
  \textbf{\bibinfo{volume}{45}}, \bibinfo{pages}{255}.

\bibitem[{\citenamefont{Lanczos}(1966)}]{La66}
\bibinfo{author}{\bibnamefont{Lanczos}, \bibfnamefont{C.}},
  \bibinfo{year}{1966}, \emph{\bibinfo{title}{Discourse on Fourier series}}
  (\bibinfo{publisher}{Hafner}, \bibinfo{address}{New York}).

\bibitem[{\citenamefont{Lee and Ramakrishnan}(1985)}]{LR85}
\bibinfo{author}{\bibnamefont{Lee}, \bibfnamefont{P.~A.}}, and
  \bibinfo{author}{\bibfnamefont{T.~V.} \bibnamefont{Ramakrishnan}},
  \bibinfo{year}{1985}, \bibinfo{journal}{Rev. Mod. Phys.}
  \textbf{\bibinfo{volume}{57}}, \bibinfo{pages}{287}.

\bibitem[{\citenamefont{Long} \emph{et~al.}(2003)\citenamefont{Long,
  Prelov\v{s}ek, El~Shawish, Karadamoglou, and Zotos}}]{LPEKZ03}
\bibinfo{author}{\bibnamefont{Long}, \bibfnamefont{M.~W.}},
  \bibinfo{author}{\bibfnamefont{P.}~\bibnamefont{Prelov\v{s}ek}},
  \bibinfo{author}{\bibfnamefont{S.}~\bibnamefont{El~Shawish}},
  \bibinfo{author}{\bibfnamefont{J.}~\bibnamefont{Karadamoglou}}, and
  \bibinfo{author}{\bibfnamefont{X.}~\bibnamefont{Zotos}},
  \bibinfo{year}{2003}, \bibinfo{journal}{Phys. Rev. B}
  \textbf{\bibinfo{volume}{68}}, \bibinfo{pages}{235106}.

\bibitem[{\citenamefont{Lorentz}(1966)}]{Lo66}
\bibinfo{author}{\bibnamefont{Lorentz}, \bibfnamefont{G.~G.}},
  \bibinfo{year}{1966}, \emph{\bibinfo{title}{Approximation of Functions}}
  (\bibinfo{publisher}{Holt, Rinehart and Winston}, \bibinfo{address}{New
  York}).

\bibitem[{\citenamefont{Maier} \emph{et~al.}(2005)\citenamefont{Maier, Jarrell,
  Pruschke, and Hettler}}]{MJPH05}
\bibinfo{author}{\bibnamefont{Maier}, \bibfnamefont{T.}},
  \bibinfo{author}{\bibfnamefont{M.}~\bibnamefont{Jarrell}},
  \bibinfo{author}{\bibfnamefont{T.}~\bibnamefont{Pruschke}}, and
  \bibinfo{author}{\bibfnamefont{M.}~\bibnamefont{Hettler}},
  \bibinfo{year}{2005}, \bibinfo{journal}{Rev. Mod. Phys.}
  \textbf{\bibinfo{volume}{77}}, \bibinfo{pages}{1027}.

\bibitem[{\citenamefont{Mandelshtam and Taylor}(1997)}]{MT97}
\bibinfo{author}{\bibnamefont{Mandelshtam}, \bibfnamefont{V.~A.}}, and
  \bibinfo{author}{\bibfnamefont{H.~S.} \bibnamefont{Taylor}},
  \bibinfo{year}{1997}, \bibinfo{journal}{J. Chem. Phys.}
  \textbf{\bibinfo{volume}{107}}, \bibinfo{pages}{6756}.

\bibitem[{\citenamefont{Mead and Papanicolaou}(1984)}]{MP84}
\bibinfo{author}{\bibnamefont{Mead}, \bibfnamefont{L.~R.}}, and
  \bibinfo{author}{\bibfnamefont{N.}~\bibnamefont{Papanicolaou}},
  \bibinfo{year}{1984}, \bibinfo{journal}{J. Math. Phys.}
  \textbf{\bibinfo{volume}{25}}, \bibinfo{pages}{2404}.

\bibitem[{\citenamefont{Motome and Furukawa}(1999)}]{MF99}
\bibinfo{author}{\bibnamefont{Motome}, \bibfnamefont{Y.}}, and
  \bibinfo{author}{\bibfnamefont{N.}~\bibnamefont{Furukawa}},
  \bibinfo{year}{1999}, \bibinfo{journal}{J. Phys. Soc. Jpn.}
  \textbf{\bibinfo{volume}{68}}, \bibinfo{pages}{3853}.

\bibitem[{\citenamefont{Motome and Furukawa}(2000)}]{MF00}
\bibinfo{author}{\bibnamefont{Motome}, \bibfnamefont{Y.}}, and
  \bibinfo{author}{\bibfnamefont{N.}~\bibnamefont{Furukawa}},
  \bibinfo{year}{2000}, \bibinfo{journal}{J. Phys. Soc. Jpn.}
  \textbf{\bibinfo{volume}{69}}, \bibinfo{pages}{3785}.

\bibitem[{\citenamefont{Motome and Furukawa}(2001)}]{MF01}
\bibinfo{author}{\bibnamefont{Motome}, \bibfnamefont{Y.}}, and
  \bibinfo{author}{\bibfnamefont{N.}~\bibnamefont{Furukawa}},
  \bibinfo{year}{2001}, \bibinfo{journal}{J. Phys. Soc. Jpn.}
  \textbf{\bibinfo{volume}{70}}, \bibinfo{pages}{3186},
  \bibinfo{note}{erratum}.

\bibitem[{\citenamefont{Neuhauser}(1990)}]{Ne90}
\bibinfo{author}{\bibnamefont{Neuhauser}, \bibfnamefont{D.}},
  \bibinfo{year}{1990}, \bibinfo{journal}{J. Chem. Phys.}
  \textbf{\bibinfo{volume}{93}}, \bibinfo{pages}{2611}.

\bibitem[{\citenamefont{Niklasson}(2003)}]{Ni03}
\bibinfo{author}{\bibnamefont{Niklasson}, \bibfnamefont{A.~M.~N.}},
  \bibinfo{year}{2003}, \bibinfo{journal}{Phys. Rev. B}
  \textbf{\bibinfo{volume}{68}}, \bibinfo{pages}{233104}.

\bibitem[{\citenamefont{Ordej{\'o}n}(1998)}]{Or98}
\bibinfo{author}{\bibnamefont{Ordej{\'o}n}, \bibfnamefont{P.}},
  \bibinfo{year}{1998}, \bibinfo{journal}{Comp. Mater. Sci.}
  \textbf{\bibinfo{volume}{12}}, \bibinfo{pages}{157}.

\bibitem[{\citenamefont{Pantelides}(1978)}]{Pa78}
\bibinfo{author}{\bibnamefont{Pantelides}, \bibfnamefont{S.~T.}},
  \bibinfo{year}{1978}, \bibinfo{journal}{Rev. Mod. Phys.}
  \textbf{\bibinfo{volume}{50}}, \bibinfo{pages}{797}.

\bibitem[{\citenamefont{Peschel} \emph{et~al.}(1999)\citenamefont{Peschel,
  Wang, Kaulke, and Hallberg}}]{Pe99}
\bibinfo{editor}{\bibnamefont{Peschel}, \bibfnamefont{I.}},
  \bibinfo{editor}{\bibfnamefont{X.}~\bibnamefont{Wang}},
  \bibinfo{editor}{\bibfnamefont{M.}~\bibnamefont{Kaulke}}, and
  \bibinfo{editor}{\bibfnamefont{K.}~\bibnamefont{Hallberg}} (eds.),
  \bibinfo{year}{1999}, \emph{\bibinfo{title}{Density-Matrix Renormalization. A
  New Numerical Method in Physics.}}, number \bibinfo{number}{528} in
  \bibinfo{series}{Lecture Notes in Physics}
  (\bibinfo{publisher}{Springer-Verlag}, \bibinfo{address}{Heidelberg}).

\bibitem[{\citenamefont{Press} \emph{et~al.}(1986)\citenamefont{Press,
  Flannery, Teukolsky, and Vetterling}}]{PFTV86}
\bibinfo{author}{\bibnamefont{Press}, \bibfnamefont{W.~H.}},
  \bibinfo{author}{\bibfnamefont{B.~P.} \bibnamefont{Flannery}},
  \bibinfo{author}{\bibfnamefont{S.~A.} \bibnamefont{Teukolsky}}, and
  \bibinfo{author}{\bibfnamefont{W.~T.} \bibnamefont{Vetterling}},
  \bibinfo{year}{1986}, \emph{\bibinfo{title}{Numerical Recipes}}
  (\bibinfo{publisher}{Cambridge University Press},
  \bibinfo{address}{Cambridge}).

\bibitem[{\citenamefont{Rivlin}(1990)}]{Ri90b}
\bibinfo{author}{\bibnamefont{Rivlin}, \bibfnamefont{T.~J.}},
  \bibinfo{year}{1990}, \emph{\bibinfo{title}{Chebyshev polynomials: From
  Approximation Theory to Algebra and Number Theory}}, Pure and Applied
  Mathematics (\bibinfo{publisher}{John Wiley \& Sons}, \bibinfo{address}{New
  York}), \bibinfo{edition}{2} edition.

\bibitem[{\citenamefont{Robin}(1997)}]{Ro97}
\bibinfo{author}{\bibnamefont{Robin}, \bibfnamefont{J.~M.}},
  \bibinfo{year}{1997}, \bibinfo{journal}{Phys. Rev. B}
  \textbf{\bibinfo{volume}{56}}, \bibinfo{pages}{13634}.

\bibitem[{\citenamefont{Sack and Donovan}(1972)}]{SD72}
\bibinfo{author}{\bibnamefont{Sack}, \bibfnamefont{R.~A.}}, and
  \bibinfo{author}{\bibfnamefont{A.~F.} \bibnamefont{Donovan}},
  \bibinfo{year}{1972}, \bibinfo{journal}{Numer. Math.}
  \textbf{\bibinfo{volume}{18}}, \bibinfo{pages}{465}.

\bibitem[{\citenamefont{Schindelin}
  \emph{et~al.}(2000)\citenamefont{Schindelin, Fehske, B\"uttner, and
  Ihle}}]{SFBI00}
\bibinfo{author}{\bibnamefont{Schindelin}, \bibfnamefont{C.}},
  \bibinfo{author}{\bibfnamefont{H.}~\bibnamefont{Fehske}},
  \bibinfo{author}{\bibfnamefont{H.}~\bibnamefont{B\"uttner}}, and
  \bibinfo{author}{\bibfnamefont{D.}~\bibnamefont{Ihle}}, \bibinfo{year}{2000},
  \bibinfo{journal}{Phys. Rev. B} \textbf{\bibinfo{volume}{62}},
  \bibinfo{pages}{12141}.

\bibitem[{\citenamefont{Schliemann}
  \emph{et~al.}(2001)\citenamefont{Schliemann, K{\"o}nig, and
  {MacDonald}}}]{SKM01}
\bibinfo{author}{\bibnamefont{Schliemann}, \bibfnamefont{J.}},
  \bibinfo{author}{\bibfnamefont{J.}~\bibnamefont{K{\"o}nig}}, and
  \bibinfo{author}{\bibfnamefont{A.~H.} \bibnamefont{{MacDonald}}},
  \bibinfo{year}{2001}, \bibinfo{journal}{Phys. Rev. B}
  \textbf{\bibinfo{volume}{64}}, \bibinfo{pages}{165201}.

\bibitem[{\citenamefont{Schollw{\"o}ck}(2005)}]{Sch05}
\bibinfo{author}{\bibnamefont{Schollw{\"o}ck}, \bibfnamefont{U.}},
  \bibinfo{year}{2005}, \bibinfo{journal}{Rev. Mod. Phys.}
  \textbf{\bibinfo{volume}{77}}, \bibinfo{pages}{259}.

\bibitem[{\citenamefont{Schubert}
  \emph{et~al.}(2005{\natexlab{a}})\citenamefont{Schubert, Wei{\ss}e, and
  Fehske}}]{SWF05}
\bibinfo{author}{\bibnamefont{Schubert}, \bibfnamefont{G.}},
  \bibinfo{author}{\bibfnamefont{A.}~\bibnamefont{Wei{\ss}e}}, and
  \bibinfo{author}{\bibfnamefont{H.}~\bibnamefont{Fehske}},
  \bibinfo{year}{2005}{\natexlab{a}}, \bibinfo{journal}{Phys. Rev. B}
  \textbf{\bibinfo{volume}{71}}, \bibinfo{pages}{045126}.

\bibitem[{\citenamefont{Schubert}
  \emph{et~al.}(2005{\natexlab{b}})\citenamefont{Schubert, Wei{\ss}e, Wellein,
  and Fehske}}]{SWWF05}
\bibinfo{author}{\bibnamefont{Schubert}, \bibfnamefont{G.}},
  \bibinfo{author}{\bibfnamefont{A.}~\bibnamefont{Wei{\ss}e}},
  \bibinfo{author}{\bibfnamefont{G.}~\bibnamefont{Wellein}}, and
  \bibinfo{author}{\bibfnamefont{H.}~\bibnamefont{Fehske}},
  \bibinfo{year}{2005}{\natexlab{b}}, in \emph{\bibinfo{booktitle}{High
  Performance Computing in Science and Engineering, Garching 2004}}, edited by
  \bibinfo{editor}{\bibfnamefont{A.}~\bibnamefont{Bode}} and
  \bibinfo{editor}{\bibfnamefont{F.}~\bibnamefont{Durst}}
  (\bibinfo{publisher}{Springer-Verlag}, \bibinfo{address}{Heidelberg}), pp.
  \bibinfo{pages}{237--250}.

\bibitem[{\citenamefont{Schubert}
  \emph{et~al.}(2005{\natexlab{c}})\citenamefont{Schubert, Wellein, Wei{\ss}e,
  Alvermann, and Fehske}}]{SWWAF05}
\bibinfo{author}{\bibnamefont{Schubert}, \bibfnamefont{G.}},
  \bibinfo{author}{\bibfnamefont{G.}~\bibnamefont{Wellein}},
  \bibinfo{author}{\bibfnamefont{A.}~\bibnamefont{Wei{\ss}e}},
  \bibinfo{author}{\bibfnamefont{A.}~\bibnamefont{Alvermann}}, and
  \bibinfo{author}{\bibfnamefont{H.}~\bibnamefont{Fehske}},
  \bibinfo{year}{2005}{\natexlab{c}}, \bibinfo{journal}{Phys. Rev. B}
  \textbf{\bibinfo{volume}{72}}, \bibinfo{pages}{104304}.

\bibitem[{\citenamefont{S\'en\'echal}
  \emph{et~al.}(2000)\citenamefont{S\'en\'echal, Perez, and
  Pioro-Ladri\`ere}}]{SPP00}
\bibinfo{author}{\bibnamefont{S\'en\'echal}, \bibfnamefont{D.}},
  \bibinfo{author}{\bibfnamefont{D.}~\bibnamefont{Perez}}, and
  \bibinfo{author}{\bibfnamefont{M.}~\bibnamefont{Pioro-Ladri\`ere}},
  \bibinfo{year}{2000}, \bibinfo{journal}{Phys. Rev. Lett.}
  \textbf{\bibinfo{volume}{84}}, \bibinfo{pages}{522}.

\bibitem[{\citenamefont{S\'{e}n\'{e}chal}
  \emph{et~al.}(2002)\citenamefont{S\'{e}n\'{e}chal, Perez, and
  Plouffe}}]{SPP02}
\bibinfo{author}{\bibnamefont{S\'{e}n\'{e}chal}, \bibfnamefont{D.}},
  \bibinfo{author}{\bibfnamefont{D.}~\bibnamefont{Perez}}, and
  \bibinfo{author}{\bibfnamefont{D.}~\bibnamefont{Plouffe}},
  \bibinfo{year}{2002}, \bibinfo{journal}{Phys. Rev. B}
  \textbf{\bibinfo{volume}{66}}, \bibinfo{pages}{075129}.

\bibitem[{\citenamefont{Silver and R\"oder}(1994)}]{SR94}
\bibinfo{author}{\bibnamefont{Silver}, \bibfnamefont{R.~N.}}, and
  \bibinfo{author}{\bibfnamefont{H.}~\bibnamefont{R\"oder}},
  \bibinfo{year}{1994}, \bibinfo{journal}{Int. J. Mod. Phys. C}
  \textbf{\bibinfo{volume}{5}}, \bibinfo{pages}{935}.

\bibitem[{\citenamefont{Silver and R\"oder}(1997)}]{SR97}
\bibinfo{author}{\bibnamefont{Silver}, \bibfnamefont{R.~N.}}, and
  \bibinfo{author}{\bibfnamefont{H.}~\bibnamefont{R\"oder}},
  \bibinfo{year}{1997}, \bibinfo{journal}{Phys. Rev. E}
  \textbf{\bibinfo{volume}{56}}, \bibinfo{pages}{4822}.

\bibitem[{\citenamefont{Silver} \emph{et~al.}(1996)\citenamefont{Silver,
  R\"oder, Voter, and Kress}}]{SRVK96}
\bibinfo{author}{\bibnamefont{Silver}, \bibfnamefont{R.~N.}},
  \bibinfo{author}{\bibfnamefont{H.}~\bibnamefont{R\"oder}},
  \bibinfo{author}{\bibfnamefont{A.~F.} \bibnamefont{Voter}}, and
  \bibinfo{author}{\bibfnamefont{D.~J.} \bibnamefont{Kress}},
  \bibinfo{year}{1996}, \bibinfo{journal}{J. of Comp. Phys.}
  \textbf{\bibinfo{volume}{124}}, \bibinfo{pages}{115}.

\bibitem[{\citenamefont{Sirker and Kl{\"u}mper}(2005)}]{SK05}
\bibinfo{author}{\bibnamefont{Sirker}, \bibfnamefont{J.}}, and
  \bibinfo{author}{\bibfnamefont{A.}~\bibnamefont{Kl{\"u}mper}},
  \bibinfo{year}{2005}, \bibinfo{journal}{Phys. Rev. B}
  \textbf{\bibinfo{volume}{71}}, \bibinfo{pages}{241101(R)}.

\bibitem[{\citenamefont{Skilling}(1988)}]{Sk88}
\bibinfo{author}{\bibnamefont{Skilling}, \bibfnamefont{J.}},
  \bibinfo{year}{1988}, in \emph{\bibinfo{booktitle}{Maximum Entropy and
  Bayesian Methods}}, edited by
  \bibinfo{editor}{\bibfnamefont{J.}~\bibnamefont{Skilling}}
  (\bibinfo{publisher}{Kluwer}, \bibinfo{address}{Dordrecht}), Fundamental
  Theories of Physics, pp. \bibinfo{pages}{455--466}.

\bibitem[{\citenamefont{Slevin and Ohtsuki}(1999)}]{SO99}
\bibinfo{author}{\bibnamefont{Slevin}, \bibfnamefont{K.}}, and
  \bibinfo{author}{\bibfnamefont{T.}~\bibnamefont{Ohtsuki}},
  \bibinfo{year}{1999}, \bibinfo{journal}{Phys. Rev. Lett.}
  \textbf{\bibinfo{volume}{82}}, \bibinfo{pages}{382}.

\bibitem[{\citenamefont{Sykora} \emph{et~al.}(2005)\citenamefont{Sykora,
  H{\"u}bsch, Becker, Wellein, and Fehske}}]{SHBWF05}
\bibinfo{author}{\bibnamefont{Sykora}, \bibfnamefont{S.}},
  \bibinfo{author}{\bibfnamefont{A.}~\bibnamefont{H{\"u}bsch}},
  \bibinfo{author}{\bibfnamefont{K.~W.} \bibnamefont{Becker}},
  \bibinfo{author}{\bibfnamefont{G.}~\bibnamefont{Wellein}}, and
  \bibinfo{author}{\bibfnamefont{H.}~\bibnamefont{Fehske}},
  \bibinfo{year}{2005}, \bibinfo{journal}{Phys. Rev. B}
  \textbf{\bibinfo{volume}{71}}, \bibinfo{pages}{045112}.

\bibitem[{\citenamefont{Tal-Ezer and Kosloff}(1984)}]{TK84}
\bibinfo{author}{\bibnamefont{Tal-Ezer}, \bibfnamefont{H.}}, and
  \bibinfo{author}{\bibfnamefont{R.}~\bibnamefont{Kosloff}},
  \bibinfo{year}{1984}, \bibinfo{journal}{J. Chem. Phys.}
  \textbf{\bibinfo{volume}{81}}, \bibinfo{pages}{3967}.

\bibitem[{\citenamefont{Thouless}(1974)}]{Th74}
\bibinfo{author}{\bibnamefont{Thouless}, \bibfnamefont{D.~J.}},
  \bibinfo{year}{1974}, \bibinfo{journal}{Physics Reports}
  \textbf{\bibinfo{volume}{13}}, \bibinfo{pages}{93}.

\bibitem[{\citenamefont{Turek}(1988)}]{Tu88}
\bibinfo{author}{\bibnamefont{Turek}, \bibfnamefont{I.}}, \bibinfo{year}{1988},
  \bibinfo{journal}{J. Phys. C} \textbf{\bibinfo{volume}{21}},
  \bibinfo{pages}{3251}.

\bibitem[{\citenamefont{Vijay} \emph{et~al.}(2004)\citenamefont{Vijay, Kouri,
  and Hoffman}}]{VKH04}
\bibinfo{author}{\bibnamefont{Vijay}, \bibfnamefont{A.}},
  \bibinfo{author}{\bibfnamefont{D.~J.} \bibnamefont{Kouri}}, and
  \bibinfo{author}{\bibfnamefont{D.~K.} \bibnamefont{Hoffman}},
  \bibinfo{year}{2004}, \bibinfo{journal}{J. Phys. Chem. A}
  \textbf{\bibinfo{volume}{108}}, \bibinfo{pages}{8987}.

\bibitem[{\citenamefont{de~Vries and De~Raedt}(1993)}]{VR93}
\bibinfo{author}{\bibnamefont{de~Vries}, \bibfnamefont{P.}}, and
  \bibinfo{author}{\bibfnamefont{H.}~\bibnamefont{De~Raedt}},
  \bibinfo{year}{1993}, \bibinfo{journal}{Phys. Rev. B}
  \textbf{\bibinfo{volume}{47}}, \bibinfo{pages}{7929}.

\bibitem[{\citenamefont{Wang}(1994)}]{Wa94}
\bibinfo{author}{\bibnamefont{Wang}, \bibfnamefont{L.-W.}},
  \bibinfo{year}{1994}, \bibinfo{journal}{Phys. Rev. B}
  \textbf{\bibinfo{volume}{49}}, \bibinfo{pages}{10154}.

\bibitem[{\citenamefont{Wang and Zunger}(1994)}]{WZ94}
\bibinfo{author}{\bibnamefont{Wang}, \bibfnamefont{L.-W.}}, and
  \bibinfo{author}{\bibfnamefont{A.}~\bibnamefont{Zunger}},
  \bibinfo{year}{1994}, \bibinfo{journal}{Phys. Rev. Lett.}
  \textbf{\bibinfo{volume}{73}}, \bibinfo{pages}{1039}.

\bibitem[{\citenamefont{Wei{\ss}e}(2004)}]{We04}
\bibinfo{author}{\bibnamefont{Wei{\ss}e}, \bibfnamefont{A.}},
  \bibinfo{year}{2004}, \bibinfo{journal}{Eur. Phys. J. B}
  \textbf{\bibinfo{volume}{40}}, \bibinfo{pages}{125}.

\bibitem[{\citenamefont{Wei{\ss}e} \emph{et~al.}(1999)\citenamefont{Wei{\ss}e,
  Bouzerar, and Fehske}}]{WBF99}
\bibinfo{author}{\bibnamefont{Wei{\ss}e}, \bibfnamefont{A.}},
  \bibinfo{author}{\bibfnamefont{G.}~\bibnamefont{Bouzerar}}, and
  \bibinfo{author}{\bibfnamefont{H.}~\bibnamefont{Fehske}},
  \bibinfo{year}{1999}, \bibinfo{journal}{Eur. Phys. J. B}
  \textbf{\bibinfo{volume}{7}}, \bibinfo{pages}{5}.

\bibitem[{\citenamefont{Wei{\ss}e} \emph{et~al.}(2005)\citenamefont{Wei{\ss}e,
  Fehske, and Ihle}}]{WFI05}
\bibinfo{author}{\bibnamefont{Wei{\ss}e}, \bibfnamefont{A.}},
  \bibinfo{author}{\bibfnamefont{H.}~\bibnamefont{Fehske}}, and
  \bibinfo{author}{\bibfnamefont{D.}~\bibnamefont{Ihle}}, \bibinfo{year}{2005},
  \bibinfo{journal}{Physica B} \textbf{\bibinfo{volume}{359--361}},
  \bibinfo{pages}{702}.

\bibitem[{\citenamefont{Wei{\ss}e} \emph{et~al.}(2001)\citenamefont{Wei{\ss}e,
  Loos, and Fehske}}]{WLF01a}
\bibinfo{author}{\bibnamefont{Wei{\ss}e}, \bibfnamefont{A.}},
  \bibinfo{author}{\bibfnamefont{J.}~\bibnamefont{Loos}}, and
  \bibinfo{author}{\bibfnamefont{H.}~\bibnamefont{Fehske}},
  \bibinfo{year}{2001}, \bibinfo{journal}{Phys. Rev. B}
  \textbf{\bibinfo{volume}{64}}, \bibinfo{pages}{054406}.

\bibitem[{\citenamefont{Wheeler}(1974)}]{Wh74}
\bibinfo{author}{\bibnamefont{Wheeler}, \bibfnamefont{J.~C.}},
  \bibinfo{year}{1974}, \bibinfo{journal}{Phys. Rev. A}
  \textbf{\bibinfo{volume}{9}}, \bibinfo{pages}{825}.

\bibitem[{\citenamefont{Wheeler and Blumstein}(1972)}]{WB72}
\bibinfo{author}{\bibnamefont{Wheeler}, \bibfnamefont{J.~C.}}, and
  \bibinfo{author}{\bibfnamefont{C.}~\bibnamefont{Blumstein}},
  \bibinfo{year}{1972}, \bibinfo{journal}{Phys. Rev. B}
  \textbf{\bibinfo{volume}{6}}, \bibinfo{pages}{4380}.

\bibitem[{\citenamefont{Wheeler} \emph{et~al.}(1974)\citenamefont{Wheeler,
  Prais, and Blumstein}}]{WPB74}
\bibinfo{author}{\bibnamefont{Wheeler}, \bibfnamefont{J.~C.}},
  \bibinfo{author}{\bibfnamefont{M.~G.} \bibnamefont{Prais}}, and
  \bibinfo{author}{\bibfnamefont{C.}~\bibnamefont{Blumstein}},
  \bibinfo{year}{1974}, \bibinfo{journal}{Phys. Rev. B}
  \textbf{\bibinfo{volume}{10}}, \bibinfo{pages}{2429}.

\bibitem[{\citenamefont{Wolff}(1989)}]{Wo89}
\bibinfo{author}{\bibnamefont{Wolff}, \bibfnamefont{U.}}, \bibinfo{year}{1989},
  \bibinfo{journal}{Phys. Rev. Lett.} \textbf{\bibinfo{volume}{62}},
  \bibinfo{pages}{361}.

\bibitem[{\citenamefont{Zener}(1951)}]{Ze51b}
\bibinfo{author}{\bibnamefont{Zener}, \bibfnamefont{C.}}, \bibinfo{year}{1951},
  \bibinfo{journal}{Phys. Rev.} \textbf{\bibinfo{volume}{82}},
  \bibinfo{pages}{403}.

\bibitem[{\citenamefont{Zhong} \emph{et~al.}(1994)\citenamefont{Zhong, Sorella,
  and Parola}}]{ZSP94}
\bibinfo{author}{\bibnamefont{Zhong}, \bibfnamefont{Q.}},
  \bibinfo{author}{\bibfnamefont{S.}~\bibnamefont{Sorella}}, and
  \bibinfo{author}{\bibfnamefont{A.}~\bibnamefont{Parola}},
  \bibinfo{year}{1994}, \bibinfo{journal}{Phys. Rev. B}
  \textbf{\bibinfo{volume}{49}}, \bibinfo{pages}{6408}.

\end{thebibliography}

\end{document}